\documentclass[12pt]{article}
\usepackage{epsfig}

\usepackage{graphicx}
\newcommand{\beq}{\begin{equation}}
\newcommand{\eeq}{\end{equation}}
\newcommand{\beqn}{\begin{eqnarray}}
\newcommand{\eeqn}{\end{eqnarray}}
\newcommand{\beqs}{\begin{eqnarray*}}
\newcommand{\eeqs}{\end{eqnarray*}}
\begin{document}

\title{\Large \bf BFKL Pomeron, Reggeized gluons and \\ 
Bern-Dixon-Smirnov amplitudes}
\author{\large J.~Bartels$^{1}$, L.~N. Lipatov$^{1,2}$, 
A.~Sabio~Vera$^{3}$ \bigskip \\
{\it 
$^1$~II. Institut Theoretical Physics, Hamburg University, Germany} \\
{\it  $^2$~St. Petersburg Nuclear Physics Institute, Russia}\\
{\it  $^3$~CERN, Geneva, Switzerland}}

\maketitle

\vspace{-10cm}
\begin{flushright}
$~~$\\
CERN-PH-TH/2008-027\\
DESY-08-015
\end{flushright}
\vspace{8cm}

\abstract{After a brief review of the BFKL approach to Regge processes in 
QCD and in supersymmetric (SUSY) gauge theories we propose a strategy for 
calculating 
the next-to-next-to-leading order corrections to the BFKL 
kernel. They can be obtained in terms of various cross-sections for 
Reggeized gluon interactions. The corresponding amplitudes can be calculated
in the framework of the effective action for high energy scattering. 
In the case of 
$N=4$ SUSY it is also possible to use the Bern-Dixon-Smirnov (BDS) ansatz. 
For this purpose the analytic properties of the BDS 
amplitudes at high energies are investigated, in order to verify their 
self-consistency. It is found that, for the number of
external particles being larger than five, these amplitudes, beyond one loop, 
are not 
in agreement with the BFKL approach which predicts the existence of Regge 
cuts in some physical channels.}    

\section{Introduction}

The elastic scattering amplitude in QCD at high energies for particles with 
color indices $A,B$ and helicities $\lambda _A, \lambda _B$ in the leading 
logarithmic approximation (LLA) has the Regge form~\cite{BFKL}
\begin{equation}
A_{2\rightarrow 2}=2\,g\delta _{\lambda _{A}\lambda _{A'}}
T_{AA'}^c\frac{s^{1+\omega (t)}}{t}\,g\,T_{BB'}^c
\,\delta _{\lambda _{B}\lambda _{B'}}\,,\,\,t=-\vec{q}^{{2}}.
\end{equation}
The gluon Regge trajectory, $j(t)=1+\omega (t)$, reads
\begin{equation}
\omega (-\vec{q}^{2})=-\frac{\alpha_{s} N_c}{(2\pi )^2}\,
(2\pi \mu )^{2\epsilon}\,\int 
d^{2-2\epsilon }k
\,\frac{
\vec{q}^{2}}{\vec{k}^{2}(\vec{q}-{k})^{2}}\approx 
-\,a\,\left(\ln
\frac{\vec{q}^{2}}{\mu ^2}-\frac{1}{\epsilon}\right)\,,
\end{equation}
where we have introduced dimensional regularization with $D=4-2\,\epsilon$
and the renormalization point $\mu$ for the 't Hooft coupling 
constant
\begin{equation}
a=\frac{\alpha _{s}\,N_c}{2\pi }\,\left(4\pi e^{-\gamma}\right)^\epsilon \,.
\end{equation}
The gluon trajectory is also known in the next-to-leading approximation in 
QCD~\cite{trajQCD} and in SUSY gauge models~\cite{trajN4}.

\begin{figure}[ht]
\centerline{\epsfig{file=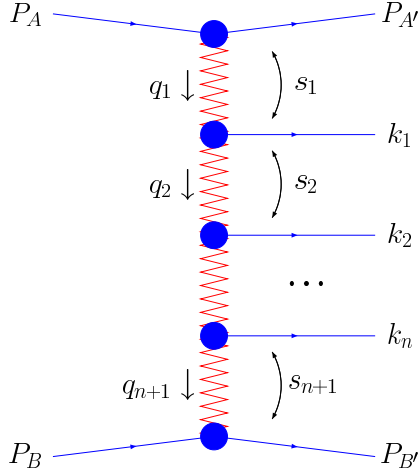,width=6cm,bbllx=93,bblly=402,bburx=429,
        bbury=783,clip=}}
\caption{Multi-Regge kinematics} 
\label{multi}
\end{figure}
In LLA gluons are produced in the multi-Regge kinematics (see 
Fig.~\ref{multi}). In this kinematics the gluon production amplitude in LLA 
has the factorized form
\begin{eqnarray}
&&\hspace{-1cm}A_{2\rightarrow 2+n} ~=~ \nonumber\\
&&\hspace{-0.8cm}-2\,s\,
g \, \delta _{\lambda _A\lambda _{A'}}\,T^{c_1}_{AA'}
\frac{s_1^{\omega (-\vec{q}_1^2)}}{\vec{q}_1^2}gC_{\mu}(q_2,q_1)
e^*_\mu (k_1)T^{d_1}_{c_2c_1}\frac{s_2^{\omega (-\vec{q}_2^2)}}{\vec{q}_2^2}
...\frac{s_{n+1}^{\omega (-\vec{q}_{n+1}^2)}}{\vec{q}_{n+1}^2}
g \, \delta _{\lambda _B\lambda _{B'}}\,T^{c_{n+1}}_{BB'}\,,
\label{MRKcinematica}
\end{eqnarray}
where 
\begin{equation}
s=(p_A+p_B)^2\gg s_r=(k_r+k_{r-1})^2\gg \vec{q}_r^2\,,\,\,k_r=q_{r+1}-q_r\,.
\end{equation}
The matrices $T^{a}_{b c}$ are the generators of the $SU(N_c)$ gauge 
group in the adjoint representation and $C_\mu (q_2,q_1)$ are the 
effective Reggeon-Reggeon-gluon vertices. In the case when the polarization 
vector $e_{\mu}(k_1)$ corresponds to a produced gluon with a definite helicity 
one can obtain~\cite{effmult}
\begin{equation}
C\equiv C_\mu 
(q_2,q_1)\,e^*_{\mu}(k_1)=\sqrt{2}\,\frac{q_2^*q_1}{k^*_1}\,,
\label{helicityproduction}
\end{equation}
where the complex notation $q=q_x+iq_y$ for the two-dimensional transverse 
vectors has been used. 

The elastic scattering amplitude with the vacuum quantum numbers in the 
$t$-channel can be calculated with the use of $s$-channel 
unitarity~\cite{BFKL}. In this approach the Pomeron appears as a composite 
state of two Reggeized gluons. It is also convenient to use  
transverse coordinates in a complex form together with their canonically 
conjugated momenta as
\begin{equation}
\rho
_{k}=x_{k}+iy_{k}\,,\,\,\rho
_{k}^{\ast
}=x_{k}-iy_{k}\,,\,\,p_{k}=i
\frac{\partial }{\partial \rho
_{k}}\,,\,\,p_{k}^{\ast }=
i\frac{\partial }{\partial \rho
_{k}^{\ast }}\,.
\end{equation}
In the coordinate representation the Balitsky-Fadin-Kuraev-Lipatov (BFKL) 
equation for the Pomeron wave function can then be written as 
follows~\cite{BFKL}
\begin{equation}
E\,\Psi
(\vec{\rho}_{1},\vec{\rho}_{2})=
H_{12}\,\Psi (\vec{\rho}_{1},\vec{%
\rho}_{2})\;,\,\,\Delta
=-\frac{\alpha _{s}N_{c}}{2\pi
}\,\min \,E\,,
\end{equation}  
where $\Delta$ is the Pomeron intercept. The BFKL Hamiltonian has the 
rather simple representation~\cite{int1}
\begin{equation}  
H_{12}=\ln
\,|p_{1}p_{2}|^{2}+\frac{1}{p_{1}p_{2}^{\ast
}}(\ln \,|\rho _{12}|^{2})\,p_{1}p_{2}^{\ast
}+\frac{1}{p_{1}^{\ast
}p_{2}}(\ln \,|\rho
_{12}|^{2})\,p_{1}^{\ast
}p_{2}-4\psi (1)\,,
\end{equation}
with $\rho _{12}=\rho _1-\rho
_2$. The kinetic energy is proportional to the sum of two gluon Regge
trajectories $\omega (-|p|_i^2)$ ($i=1, 2$).   
The potential energy $\sim \ln \,|\rho_{12}|^{2}$ is related to the product of
two gluon production vertices $C_\mu$. This Hamiltonian is invariant under the 
M\"{o}bius transformation~\cite{moeb}
\begin{equation}
\rho _{k}\rightarrow
\frac{a\rho _{k}+b}{c\rho
_{k}+d}\,,
\end{equation}
where $a,b,c$ and $d$ are complex numbers. The eigenvalues of the 
corresponding Casimir operators are expressed in terms of the conformal weights
\begin{equation}
m=\frac{1}{2}+i\nu  
+\frac{n}{2}\,,\,\,\widetilde{m}=\frac{1}{2}+i\nu
-\frac{n}{2}
\end{equation}
for the unitary principal series representation of $SL(2,C)$, with $\nu$ being 
real and $n$ integer.

The BFKL Hamiltonian can be iterated in the $s$--channel to account for the 
exchange 
of an arbitrary number of Reggeized gluons. This iteration is described by the 
Bartels-Kwiecinski-Praszalowicz
(BKP) equation~\cite{BKP} for the $n$-gluon colorless composite state. 
In the $N_c\rightarrow \infty$ limit the Hamiltonian has the property of 
holomorphic separability~\cite{separ} in the form 
\begin{equation}
H=\frac{1}{2}\sum _kH_{k,k+1}=\frac{1}{2}
(h+h^*)\,,\,\,[h,h^*]=0\,.
\end{equation}
The holomorphic
Hamiltonian can be written as 
\begin{equation}
h=\sum _kh_{k,k+1}\,,\,\,h_{12}=\ln
(p_{1}p_{2})+\frac{1}{p_{1}}\,(\ln
\rho _{12})\,p_{1}+
\frac{1}{p_{2}}\,(\ln \rho
_{12})\,p_{2}-2\psi (1)\,,
\end{equation}
where $\psi (x)=(\ln \Gamma (x))'$. 
Consequently, the wave function $\Psi $ fulfills holomorphic
factorization~\cite{separ} and there exists the remarkable duality 
symmetry under the transformation~\cite{dual}
\begin{equation}
p_i\rightarrow \rho _{i,i+1}\rightarrow p_{i+1}\,.
\end{equation}
Moreover, in the holomorphic and anti-holomorphic
sectors, there are integrals of motion
commuting among themselves and with $h$~\cite{int1, int}:
\begin{equation}
q_{r}=\sum_{k_{1}<k_{2}<...<k_{r}}\rho
_{k_{1}k_{2}}\rho
_{k_{2}k_{3}}...\rho
_{k_{r}k_{1}}\,p_{k_{1}}p_{k_{2}}...
p_{k_{r}}\,,\,\,[q_{r},h]=0\,.
\end{equation}
The integrability of BFKL dynamics was demonstrated
in~\cite{int} and it is   
related to the fact that $h$ coincides with the local Hamiltonian of the
Heisenberg spin model \cite{LiFK}.
In the LLA the Pomeron intercept 
is $\Delta =4\;\frac{\alpha
_{s}}{\pi }N_{c}\,\ln 2>0$
and the Froissart
bound $\sigma _t<c\ln ^2s$
for the total cross-section $\sigma _t\sim s^\Delta$ 
is violated~\cite{BFKL}.

In the next-to-leading logarithmic approximation
the integral kernel for the BFKL equation was
constructed in Refs.~\cite{trajN4,FL}.
Due to its M\"{o}bius
invariance, the solution of the BFKL 
equation can be classified by the anomalous
dimension $\gamma =
\frac{1}{2}+i\nu$ of twist-2 operators and
the conformal spin $|n|$, which 
coincides with the number of
transverse indices of the local operators $O_{\mu _{1}...\mu _j}$. 

The eigenvalue of the BFKL kernel in the next-to-leading approximation
has the form~\cite{FL}
\begin{equation}
\omega =4 \, \hat{a}\,
\left[2\psi (1)-\psi \left(\gamma +\frac{|n|}{2}\right)-
\psi \left(1-\gamma +\frac{|n|}{2}\right)
\right]+4\,\,\hat{a}^{2}\,\Delta
(n,\gamma
)\,,\,\,\hat{a}=g^{2}\frac{N_{c}}{16\pi^{2}}\,.
\end{equation}
In QCD, the next-to-leading contribution $\Delta (n,\gamma )$ is
a non-analytic function of the
conformal spin $|n|$ because it contains some terms depending on
the 
Kronecker symbols $\delta _{n,0}$ and $\delta _{n,2}$. However, in $N=4$ SUSY 
this dependence is cancelled and we obtain the following
hermitially separable expression~\cite{trajN4, KL}
\begin{equation}
\Delta (n,\gamma )=\phi
(M)+\phi (M^{\ast })-\frac{
\rho (M)+ \rho  
(M^{\ast
})}{2\hat{a}/\omega}\,,\,M=\gamma
+\frac{|n|}{2}\,,
\end{equation}
\begin{equation}
\rho (M)=\beta ^{\prime
}(M)+\frac{1}{2}%
\zeta (2)\,,\,\beta ^{\prime  
}(z)=\frac{1}{4}\Biggl[\Psi
^{\prime
}\Bigl(\frac{z+1}{2}\Bigr)-\Psi
^{\prime
}\Bigl(\frac{z}{2}\Bigr)\Biggr],
\end{equation}
where
\begin{equation}
\phi (M)=3\zeta (3)+\psi
^{^{\prime \prime }}(M)-2\Phi
(M)+ 2\beta
^{^{\prime }}(M)\Bigl(\psi
(1)-\psi (M)\Bigr),
\end{equation}  
and
\begin{equation} 
\Phi (M)=\sum_{k=0}^{\infty   
}\frac{\beta ^{\prime }(k+1)}{k+M}
+ \sum_{k=0}^{\infty
}\frac{(-1)^{k}}{k+M}\left(
\psi ^{\prime
}(k+1)~- ~\frac{\psi (k+1)-\psi 
(1)}{k+M}\right).
\end{equation}
Very importantly, all these contributions have the property of maximal 
transcendentality~\cite{KL}. The behaviour of the 4-point Green function
corresponding to this NLO kernel in $N=4$ SUSY was investigated 
in~\cite{GGFN4}. The NLO conformal spins affect azimuthal angle 
decorrelations in jet physics as it was originally 
suggested in~\cite{Vera:2006un}.

In a different context, the one-loop anomalous 
dimension matrix for twist-2 operators
in $N=4$ SUSY can be easily calculated since it is completely
fixed by superconformal invariance. Its 
eigenvalue is proportional to $\psi (1)-\psi (j-1)$,
which is related to the integrability of the evolution
equation for the quasi-partonic operators in
this model~\cite{L4}. The integrability of $N=4$ SUSY
has also been established for other operators and in higher 
loops~\cite{MZ, BS}.

The maximal transcendentality principle suggested
in Ref.~\cite{KL} made it  possible to extract the 
universal anomalous dimension up to three loops in $N=4$
SUSY~\cite{KLV, KLOV} from the QCD results~\cite{VMV}. 
This principle was also helpful for finding closed integral equations 
for the cusp anomalous 
dimension in this model~\cite{ES, BES} based on the AdS/CFT
correspondence~\cite{Malda, GKP, W}. In the framework of the
asymptotic Bethe ansatz the maximal transcendentality principle 
helped to fix the anomalous
dimension at four loops~\cite{KLRSV}. However, the obtained results 
contradict the predictions stemming from 
the BFKL equation~\cite{trajN4, KL}. The origin of this discrepancy is 
related to the onset of wrapping 
effects~\cite{KLRSV}. In this framework it is, therefore, crucial  
to obtain more information from the BFKL side through the calculation 
of its 
higher order corrections to the integral kernel. 
We would like to point out that the intercept of the
BFKL Pomeron at large 't Hooft coupling constant in $N=4$ SUSY was
found in Refs.~\cite{KLOV, Polch}. 

In the present paper we want to formulate a program to calculate 
the three loop corrections to the BFKL kernel in the 't Hooft coupling. 
Our approach is based on the use of the high energy effective 
action developed in~\cite{eff, last} for the construction of the various 
Reggeized gluon couplings, and on the BDS ansatz~\cite{BDS} for scattering
amplitudes in the $N=4$ super Yang-Mills theory. We begin with a short review 
of the effective action (section 2) and then turn to an analysis of the 
BDS formula in the Regge limit for the amplitudes up to six external 
gluons (section 3). An interpretation based upon known results of the high 
energy limit of scattering amplitudes in QCD is given in section 4. An outlook 
is presented in the concluding section. Some details of the calculations are 
presented in several appendices.        
 
\section{Effective action for Reggeized gluons}      

Initially calculations of scattering amplitudes in 
Regge kinematics were performed by an iterative method based on analyticity, 
unitarity and renormalizability of the theory~\cite{BFKL}. The $s$-channel
unitarity was incorporated partly in the form of bootstrap equations for the 
amplitudes 
generated by Reggeized gluons exchange. But later it turned out that 
for this purpose one can also use an effective field theory for
Reggeized gluons~\cite{eff, last}. 

We shall write below the effective action valid at high energies for 
interactions of particles inside each cluster having their
rapidities $y$ in a certain interval 
\begin{equation}
\,\,y=\frac{1}{2}\ln \frac{\epsilon _{k}+|k|}{\epsilon _{k}-|k|}\,,\,\,
|y-y_{0}|<\eta \,,\,\,\eta <<\ln \,s\,.
\end{equation}
The corresponding gluon and quark fields are
\begin{equation}
v_{\mu }(x)=-iT^{a}v_{\mu }^{a}(x)\,,\,\,\psi (x)\,,\,\,\bar{\psi}%
(x)\,,\,\,[T^{a},T^{b}]=if_{abc}T^{c}\,.
\end{equation}
In the case of the supersymmetric models one can take into account also 
the fermion and scalar fields with known Yang-Mills and Yukawa interactions.
Let us introduce now the fields describing the production and
annihilation of Reggeized gluons~\cite{eff}:
\begin{equation}
A_{\pm }(x)=-iT^{a}A_{\pm }^{a}(x)\,.
\end{equation}
Under the global color group rotations the fields are transformed in the
standard way
\begin{equation}
\delta v_{\mu }(x)=[v_{\mu }(x),\chi ],\,\,\delta \psi (x)=-\chi \,\psi
(x),\,\,\delta A(x)=[A(x),\chi ]\,,
\end{equation}
but under the local gauge transformations with $\chi (x)\rightarrow 0$ 
at $x\rightarrow \infty$ we have
\begin{equation}
\delta v_{\mu }(x)=\frac{1}{g}[D_{\mu },\chi (x)],\,\,\delta \psi (x)=-\chi
(x)\,\psi (x),\,\,\delta A_{\pm }(x)=0\,.
\end{equation}

In quasi-multi-Regge kinematics particles are produced in groups
(clusters) with fixed masses. These groups have significantly different
rapidities corresponding to the multi-Regge asymptotics. In this case  
one obtains the following
kinematical constraint
on the reggeon fields
\begin{equation}
\partial _{\mp }\,A_{\pm }(x)=0\,,\,\,\partial _{\pm}=n_{\pm}^{\mu}
\partial _{\mu}\,,
\end{equation}
$n_\pm ^{\mu}=\delta_0^\mu  \pm \delta_3^\mu$.
For QCD the corresponding effective
action local in the rapidity $y$ has the form ~\cite{eff}
\begin{equation}
S=\int d^{4}x\left( L_{0}+L_{ind}\right) \,,
\end{equation}
where $L_0$ is the usual Yang-Mills Lagrangian 
\begin{equation}
L_{0}=i\bar{\psi}\hat{D}\psi +\frac{1}{2}Tr\,G_{\mu \nu }^{2},\,D_{\mu
}=\partial _{\mu }+gv_{\mu },\,G_{\mu \nu }=\frac{1}{g}[D_{\mu },D_{\nu }]
\end{equation}
and the induced contribution is given by
\begin{equation}
L_{ind}=Tr\,(L_{ind}^{k}+L_{ind}^{GR})\,,\,\,L_{ind}^{k}=-\partial _{\mu
}A_{+}^{a}\partial _{\mu }A_{-}^{a}\,.
\end{equation}
Here the Reggeon-gluon interaction can be presented in terms
of Wilson $P$-exponents
\begin{eqnarray}
L_{ind}^{GR}&=&-\frac{1}{g}\partial _{+}\,P\exp \left( -g\frac{1}{2}
\int_{-\infty }^{x^{+}}v_{+}(x^{\prime })d(x^{\prime })^{+}\right)
\,\partial _{\sigma }^{2}A_{-} \nonumber\\
&&\hspace{0.4cm}-\frac{1}{g}\partial _{-}\,P\exp \left( 
-g\frac{1}{2}\int_{-\infty }^{x^{-}}v_{-}(x^{\prime })d(x^{\prime})^{+}\right)
\,\partial _{\sigma }^{2}A_{+}\nonumber\\
&=&\left( v_{+}-gv_{+}\frac{1}{\partial _{+}}v_{+}+g^{2}v_{+}
\frac{1}{\partial _{+}}v_{+}\frac{1}{\partial _{+}}v_{+}-...\right) 
\partial _{\sigma}^{2}A_{-}\nonumber\\
&&\hspace{0.4cm}
+\left( v_{-}-gv_{-}\frac{1}{\partial _{-}}v_{-}+g^{2}v_{-}\frac{1%
}{\partial _{-}}v_{-}\frac{1}{\partial _{-}}v_{-}-...\right)
\partial _{\sigma}^{2}A_{+}\,.
\end{eqnarray}

One can formulate the Feynman rules directly in momentum space~\cite{last}.
For this purpose
it is needed to take into account the gluon momentum conservation for induced
vertices
\begin{equation}
k_{0}^{\pm}+k_{1}^{\pm}+...+k_{r}^{\pm}=0\,.
\end{equation}
Some simple examples of induced Reggeon-gluon vertices are
\begin{equation}
\Delta _{a_0c}^{\nu _0+}=\vec{q}_{\perp }^{2}\,\delta _{a_0c}\,(n^+)^{\nu _0}
\,,\,\,\Delta _{a_{0}a_{1}c}^{\nu _{0}\nu _{1}+}=\vec{q}_{\perp
}^{2}\,T_{a_{1}a_{0}}^{c}\,(n^{+})^{\nu _{1}}\frac{1}{k_{1}^{+}}%
(n^{+})^{\nu _{0}},
\end{equation}
\begin{equation}
\Delta _{a_{0}a_{1}a_{2}c}^{\nu _{0}\nu _{1}\nu _{2}+}=\vec{q}_{\perp
}^{\,2}\,(n^{+})^{\nu _{0}}(n^{+})^{\nu _{1}}(n^{+})^{\nu _{2}}\left( \frac{%
T_{a_{2}a_{0}}^{a}\,T_{a_{1}a}^{c}}{k_{1}^{+}k_{2}^{+}}+\frac{%
T_{a_{2}a_{1}}^{a}\,T_{a_{0}a}^{c}}{k_{0}^{+}k_{2}^{+}}\right)\,.
\end{equation}
In the general case these vertices factorize in the form 
\begin{equation}
\Delta _{a_{0}a_{1}...a_{r}c}^{\nu _{0}\nu _{1}...\nu _{r}+}=
(-1)^r\vec{q}_{\perp
}^{\,2}\prod_{s=0}^{r}(n^{+})^{\nu _{s}}\,2\,{\rm Tr} \left(
T^{c}G_{a_{0}a_{1}...a_{r}}\right),
\end{equation}
where $T^c$ are the color generators in the fundamental representation.
In more detail, $G_{a_{0}a_{1}...a_{r}}$ can be written as~\cite{last}
\begin{equation}
G_{a_{0}a_{1}...a_{r}}=\sum_{\{i_{0},i_{1},...,i_{r}\}}\frac{%
T^{a_{i_{0}}}T^{a_{i_{1}}}T^{a_{i_{2}}}...T^{a_{i_{r}}}}{%
k_{i_{0}}^{+}(k_{i_{0}}^{+}+k_{i_{1}}^{+})...(k_{i_{0}}^{+}+
k_{i_{1}}^{+}+...+k_{i_{r-1}}^{+})%
}\,.
\end{equation}
These vertices satisfy the following recurrent relations 
(Ward identities)~\cite{eff}
\begin{eqnarray}
k_{r}^{+}\,\Delta _{a_{0}a_{1}...a_{r}c}^{\nu _{0}\nu _{1}...\nu
_{r}+}(k_{0}^{+},...,k_{r}^{+}) \nonumber\\
&&\hspace{-4.6cm}=
-(n^+)^{\nu _r}\sum_{i=0}^{r-1}if_{aa_{r}a_{i}}\Delta
_{a_0...a_{i-1}aa_{i+1}...a_{r-1}c}^{\nu _{0}...
\nu_{r-1}+}(k_{0}^{+},...,k_{i-1}^{+},k_{i}^{+}+k_{r}^{+},k_{i+1}^{+}
,...,k_{r-1}).
\end{eqnarray}

With the use of this effective theory one can calculate the tree amplitude 
for the production of a cluster of three
gluons, or a gluon and a pair of fermions 
or scalar particles (in the case of an extended supersymmetric model) in the 
collision of two Reggeized gluons~\cite{last} (see also earlier calculations 
of this amplitude in~\cite{Del Duca:1999ha}). The square of the amplitude 
for three particle production integrated  
over the momenta of these particles is the main ingredient to construct 
the corresponding contribution to the BFKL
kernel in the next-to-next-to-leading approximation 
using the methods of~\cite{NextLead}. 
One can go to the helicity
basis of produced gluons or fermions~\cite{helic24}. In
principle it is also possible to calculate the loop corrections to
the above Reggeon-particle vertices with the use of the effective action, 
however, in the present paper, we will use for this purpose the results 
for $N=4$ SUSY amplitudes presented by Bern, Dixon and Smirnov in~\cite{BDS}.

\section{BDS amplitudes in multi-Regge kinematics}

As we have already remarked in the previous section, to find the 
next-to-next-to-leading corrections to the BFKL kernel in $N=4$ SUSY we need 
to calculate, apart from the 
amplitude for the transition of two Reggeized gluons to three particles, also 
the three loop correction to the gluon Regge trajectory, the two loop 
correction to the Reggeon-Reggeon-gluon vertex, and the one loop correction 
to the amplitude for
the transition of two Reggeized gluons to two gluons or their superpartners.
In this section we consider, as a first step, the corrections to the Regge 
trajectory and corrections to the Reggeon-Reggeon-gluon vertex 
(valid up to one loop) which can be obtained from the multi-Regge
asymptotics of the amplitude with the maximal helicity violation, calculated 
by Bern, Dixon and Smirnov (BDS)~\cite{BDS}. We also investigate the six point 
amplitudes $2 \to 4$ and $3 \to 3$ in multi-Regge kinematics, thus preparing 
the comparison with QCD calculations to be carried out in the following   
section.

The BDS formula determines the logarithm of the scattering amplitude 
(to be more precise: after the Born amplitude has been removed).
In our analysis of the BDS formula we will, thoughout our paper, restrict 
ourselves, in the logarithm of the amplitudes, to those terms which are singular or constant in $\epsilon$, 
i.e. we do not (yet) consider corrections of order $\epsilon$ or $\epsilon^2$ in the 
logarithm of the amplitude. As a consequence, 
all results for the scattering amplitude are correct up to relative 
corrections of the order $\epsilon$, i.e all results should be multiplied 
by a factor of the form $\left(1+ {\cal O}(\epsilon)\right)$. Details of 
our analysis of the BDS formula are outlined in several appendices.  
 
\begin{figure}[ht]
\centerline{\epsfig{file=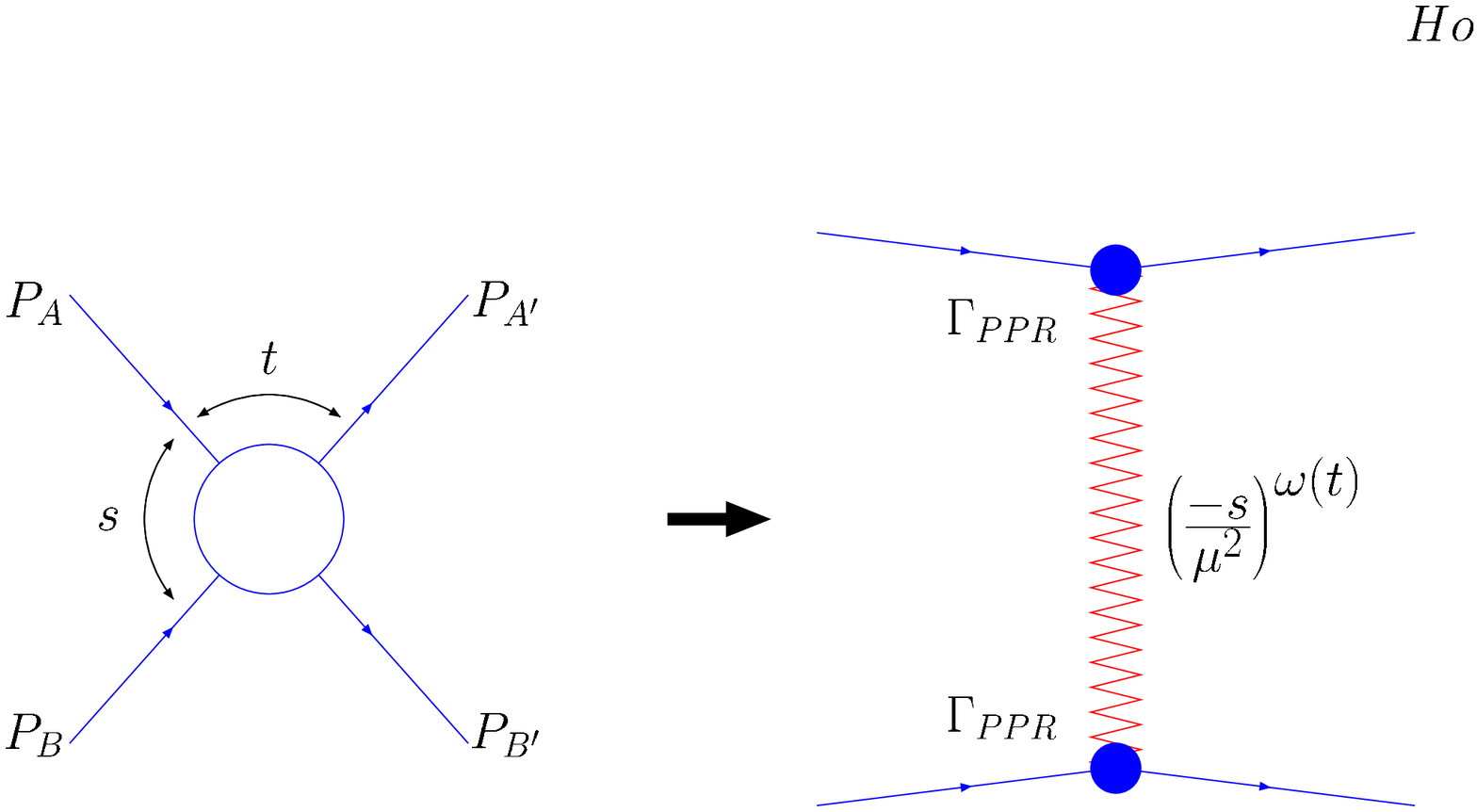,width=10cm,angle=0,bbllx=0,bblly=390,bburx=612,
        bbury=689,clip=}}
\caption{Elastic scattering in the Regge asymptotics} 
\label{elast}
\end{figure}
According to Ref.~\cite{BDS}, in the case of maximal helicity violation
the amplitude $A_n$ with $n$ legs in the large-$N_c$ limit is factorized in the 
product of the tree result (including the corresponding
color structure) and the simple scalar quantity $M_n$. In the Regge limit
$s\gg (-t)$ the expression for $M_4$ having the singularities in $s$ and 
$t$-channels
can be simplified as follows (see Fig.\ref{elast} and 
Appendix A)\footnote{As we have said before, the factor 
$\left(1+ {\cal O}(\epsilon)\right)$ 
on the rhs is present in all our results for scattering amplitudes, and 
it will be omitted in the following.   
For example, for the calculation of the vertex function $\Gamma$, from the BDS 
formula, our neglect of order-$\epsilon$ corrections in the 
logarithm has consequences: in order to determine the vertex function 
beyond the one loop approximation,   
it is necessay to compute, in the logarithm, also the higher order terms.
However, such a computation is not the aim of this paper, and we will 
not go beyond the one loop approximations for the logarithm of vertex functions, 
$\ln \Gamma(t)$ or $\ln \Gamma(t_2,t_1,\ln \kappa)$ (see below). 
We will have to come back to this question 
when calculating the full higher order corrections to the BFKL equation.
We thank V. Del Duca for discussions on this point.}
\begin{eqnarray}
M_{2\rightarrow 2}&=&\Gamma (t)\,\left(\frac{-s}{\mu ^2}\right)^{\omega (t)}\,
\Gamma (t)\,\cdot\, \left(1+ {\cal O}(\epsilon)\right),
\label{M2a2Regge}
\end{eqnarray}  
where $\mu ^2$ is the renormalization point,
\begin{eqnarray}
\omega (t)&=&-\frac{\gamma (a)}{4}\,\ln \frac{-t}{\mu ^2}+\int _0^a
\frac{da'}{a'}\left(\frac{\gamma (a')}{4\epsilon}+\beta (a')\right)\,,
\label{gluontrajectory}
\end{eqnarray}
is the all-order gluon Regge trajectory, as obtained from the BDS 
formula~\cite{DKS,NS} (for a verification by comparison with explicit 
calculations see discussions below), and
\begin{eqnarray}
\ln \Gamma (t)&=&\ln  \frac{-t}{\mu ^2}\,\int _0^a   
\frac{da'}{a'}\left(\frac{\gamma (a')}{8\epsilon}+\frac{\beta (a')}{2}\right)
+\frac{C(a)}{2}+\frac{\gamma (a)}{2}\,\zeta _2 \nonumber\\
&-& \int _0^a \frac{da'}{a'}
\ln \frac{a}{a'}\,\left(\frac{\gamma (a')}{4\epsilon ^2}+
\frac{\beta (a')}{\epsilon} +\delta (a')\right),
\label{ReggeonParticlevertex}
\end{eqnarray}
is the vertex for the coupling of the Reggeized gluon to the external 
particles.

The 't Hooft coupling is defined as in eq.(3):
\begin{equation}
a=\frac{\alpha N_c}{2\pi}\,\left(4\pi e^{-\gamma}\right)^
\epsilon 
\end{equation}
and the small parameter $\epsilon$ is related to the dimensional regularization
$4\rightarrow 4-2\epsilon$.
The cusp anomalous dimension $\gamma (a)$ is known to all 
loops~\cite{KLOV,Bern:2006ew,BES}

\begin{equation}
\gamma (a)=4a-4\zeta _2\,a^2+22\zeta_4\,a^3+...\,,
\end{equation}
and the functions $\beta (a)$, $\delta (a)$ and $C(a)$ read~\cite{BDS}
\begin{eqnarray}
\beta (a) &=& - \zeta_3 \, a^2 +
(6 \,\zeta _5 + 5 \, \zeta _2 \, \zeta _3)\,a^3+...\,,\nonumber\\
\delta (a) &=& - \, \zeta _4 \, a^2+...\,,\nonumber\\
C(a) &=& - \frac{\zeta _2^2}{2}\,a^2+...\,,
\end{eqnarray}
where $\zeta (n)$ is the Riemann $\zeta$-function
\begin{equation}
\zeta (n)=\sum _{k=1}^\infty k^{-n}\,.
\end{equation}  
Written as in Eq.~(\ref{M2a2Regge}) we can see that the asymptotic behavior of 
the $M_{2 \rightarrow 2}$ BDS amplitude corresponds to the Regge ansatz with 
the gluon trajectory $j=1+\omega (t)$ given by the perturbative expansion
\begin{eqnarray}
\omega (t)&=&\left(-\ln \frac{-t}{\mu ^2}+\frac{1}{\epsilon}\right)a+
\left[\zeta _2\left(\ln \frac{-t}{\mu ^2}-\frac{1}{2\epsilon}\right)-
\frac{\zeta _3}{2}\right]a^2 \nonumber\\
&+&\left[-\frac{11}{2}\zeta _4\left(\ln \frac {-t}{\mu ^2}-
\frac{1}{3\epsilon}\right)+\frac{6\zeta _5+5\zeta _2\zeta _3}{3}\right]a^3
+...\,.
\label{expandedtrajectory}
\end{eqnarray}
The first two terms in this expansion are in agreement with the predictions 
in Refs.~\cite{BFKL, trajN4}. Note that in Ref.~\cite{trajN4}, where the 
BFKL kernel at NLO was calculated in $N=4$ SUSY, initially the 
$\overline{\rm MS}$-scheme was used, and only later, in Ref.~\cite{KL} 
the final result was also presented in the dimensional reduction scheme 
(DRED). The NLO terms in the BDS expression for $\omega (t)$ 
can be obtained from Ref.~\cite{trajN4} by converting it
to the DRED scheme, where, apart from
the finite renormalization of the coupling constant,  
one should also take into account in the loop the additional 
number $2\epsilon$ of scalar particles (for details see 
Appendix A and the recent paper~\cite{FF})\footnote{
We thank A.~V.~Kotikov and E.~M.~Levin for helpful discussions regarding 
these redefinitions.}.
The ${\cal O}(a^3)$ term in $\omega(t)$, extracted from the BDS 
amplitude~\cite{DKS}), corresponds to the three-loop correction to 
the gluon Regge trajectory needed when calculating the 
next-to-next-to-leading corrections to the BFKL kernel in this model. 
Strictly speaking the Regge asymptotics of scattering amplitudes corresponds
to a different order of taking  two limits $\epsilon \rightarrow 0$ and
$s \rightarrow \infty$, but it is probable that they can be interchanged.

It is noteworthy to point out that the expression for 
$M_{2\rightarrow 2}$, derived in the Regge kinematics, in fact, is 
valid also outside the Regge limit. That is to say that, when analysing 
the BDS formula for the logarithm of the $2 \to 2$ amplitude (see Appendix A), 
we do not make use of the high energy limit.  
In particular, the amplitude can also be written in the dual form
\begin{equation}
M_{2\rightarrow 2}=\Gamma (s)\,\left(\frac{-t}{\mu ^2}\right)^{\omega (s)}\,
\Gamma (s) \,.  
\end{equation}

\begin{figure}[ht]
\centerline{\epsfig{file=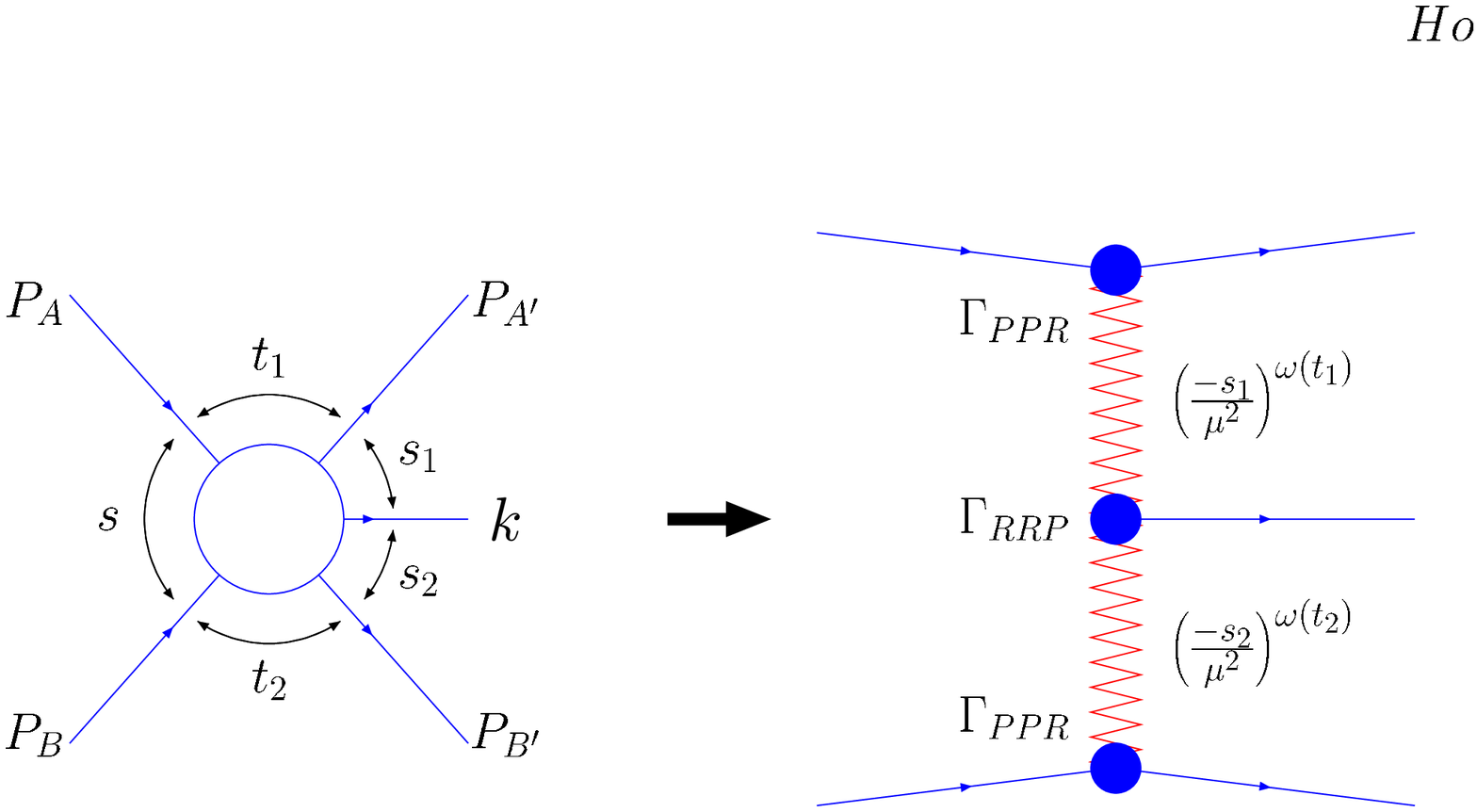,width=11cm,angle=0,bbllx=0,bblly=400,bburx=610,
        bbury=686,clip=}}
\caption{Production amplitude in the multi-Regge regime} 
\label{prod1}
\end{figure}
After having described 
the elastic scattering amplitude we now focus on the BDS 
production amplitude (see Fig.~\ref{prod1}). The analysis of 
$\ln M_{2 \to 3}$, described in 
Appendix B shows that the amplitude (as before, up to the corrections 
$\left(1+ {\cal O}(\epsilon)\right)$: see also the footnote on p.7) 
can be written in the factorized form
\begin{eqnarray}
M_{2\rightarrow 3} &=& \Gamma (t_1)\,
\left(\frac{-s_1}{\mu ^2}\right)^{\omega (t_1)}\,\Gamma (t_2,t_1, \ln -\kappa )
\,\left(\frac{-s_2}{\mu ^2}\right)^{\omega (t_2)}\,\Gamma (t_2)\,,
\label{M2a3mregge}
\end{eqnarray}
with
\begin{eqnarray}
-\kappa &=&\frac{(-s_1)(-s_2)}{(-s)}\,.
\end{eqnarray}
Since Eq.~(\ref{M2a3mregge}) is exact it is also valid 
in the multi-Regge kinematics
\begin{equation}
-s\gg -s_1,-s_2\gg -t_1\sim -t_2 \sim -\kappa\,,
\end{equation}
where all invariants $s, s_1, s_2, t_1$ and $t_2$ are negative. 
Due to the correct factorization properties of this amplitude 
the Reggeon--particle--particle 
vertex $\Gamma(t)$ in Eq.~(\ref{M2a3mregge}) is exactly the same as in the 
elastic amplitude in Eq.~(\ref{ReggeonParticlevertex}). The gluon Regge 
trajectory $\omega(t)$ in Eq.~(\ref{M2a3mregge}) also coincides with the 
one discussed above. The new component is the Reggeon-Reggeon-particle 
vertex. Its logarithm is given by
\begin{eqnarray}
\ln \Gamma (t_2,t_1, \ln -\kappa )&=&-\frac{\gamma (a)}{16}\,
\ln ^2\frac{-\kappa }{\mu ^2}-\frac{1}{2}\int _0^a \frac{da'}{a'}
\ln \frac{a}{a'}\,\left(\frac{\gamma (a')}{4\epsilon ^2}+\frac{\beta (a')}{\epsilon}
+\delta (a')\right)\nonumber\\
&&\hspace{-4cm} -\frac{\gamma (a)}{16}\ln ^2\frac{-t_1}{-t_2}-
\frac{\gamma (a)}{16}\zeta _2-\frac{1}{2}
\left(\omega (t_1)+\omega (t_2)-
\int _0^a\frac{da'}{a'}\,\left(\frac{\gamma (a')}{4\epsilon}+\beta(a')\right)\right)
\ln \frac{-\kappa }{\mu ^2}
\,.
\label{Gammavertex}
\end{eqnarray}

\begin{figure}[ht]
\centerline{
\epsfig{file=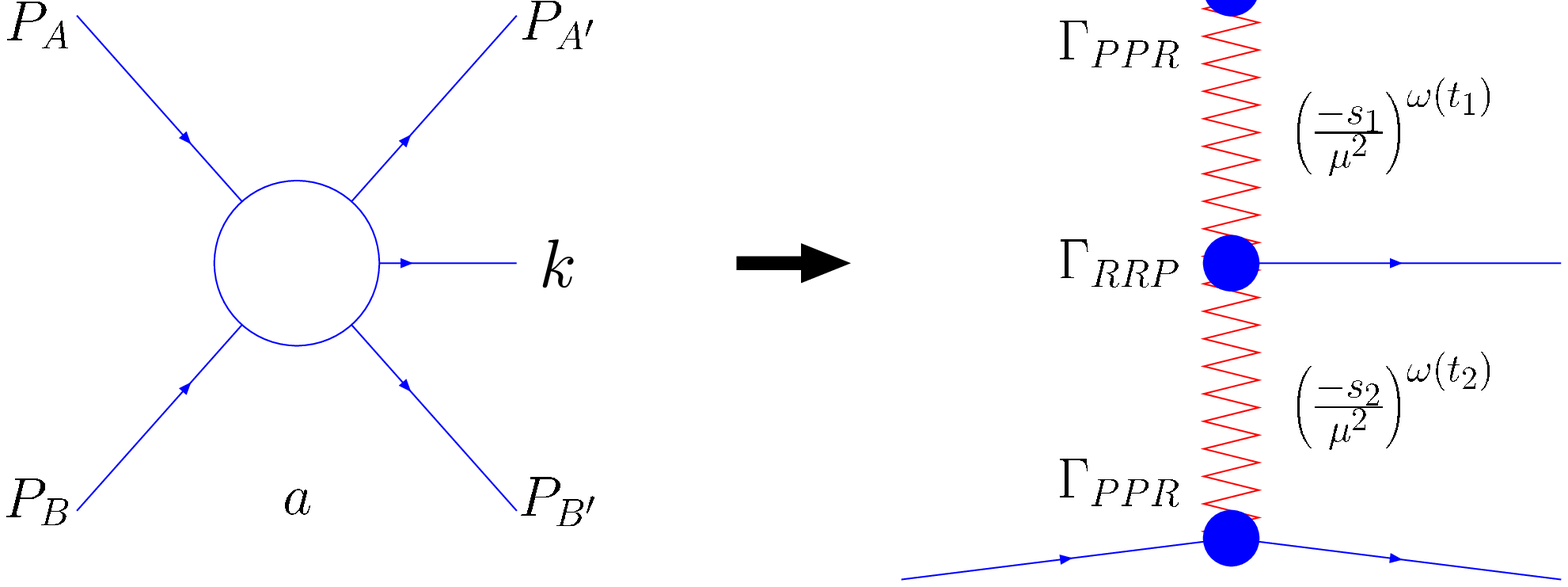,width=5cm,angle=0,bbllx=0,bblly=584,bburx=250,
        bbury=815,clip=}\epsfig{file=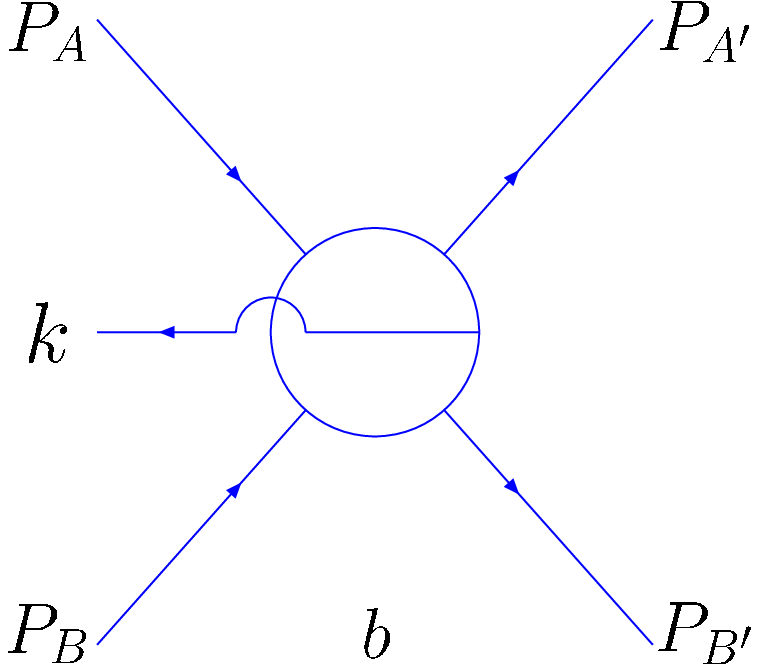,width=5cm,angle=0,bbllx=0,bblly=584,bburx=250,
        bbury=810,clip=}}
\caption{Physical channels for the one particle production amplitude} 
\label{phys23}
\end{figure}

It is now possible to analytically continue this $2 \rightarrow 3$ production 
amplitude to the physical region where the invariants $s$, $s_1$ and $s_2$ are 
positive (see Fig. \ref{phys23}a)
\begin{equation}
M_{2\rightarrow 3}=\Gamma (t_1)\,
\left(\frac{-s_1-i\epsilon}{\mu ^2}\right)^{\omega (t_1)}\,\Gamma (t_2,t_1, 
\ln \kappa -i\pi )
\,\left(\frac{-s_2-i\epsilon }{\mu ^2}\right)^{\omega (t_2)}\,\Gamma (t_2)\,.
\label{dosatres}
\end{equation}

A similar continuation to another physical region can be performed in the case 
when $s$ is positive but $s_1$ and $s_2$ are negative 
(see Fig. \ref{phys23}b):
\begin{equation}
M_{2\rightarrow 3}=\Gamma (t_1)\,
\left(\frac{-s_1}{\mu ^2}\right)^{\omega (t_1)}\,\Gamma (t_2,t_1,
\ln \kappa +i\pi )
\,\left(\frac{-s_2}{\mu ^2}\right)^{\omega (t_2)}\,\Gamma (t_2)\,.
\end{equation}

\begin{figure}[ht]
\centerline{\epsfig{file=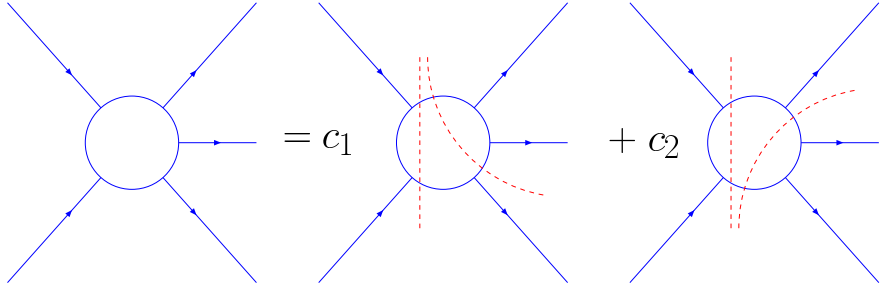,width=12cm,angle=0,bbllx=0,bblly=601,bburx=580,
        bbury=800,clip=}}
\caption{Dispersion representation for $M_{2\rightarrow 3}$, exhibiting its analytic 
structure} 
\label{disper23}
\end{figure}
Using a `dispersive' representation illustrated in
Fig. \ref{disper23} (in the following, 
we will refer to this type of representation as `analytic' representation), 
it can be easily 
verified that in all physical regions (including the crossing channels with
$s,s_1<0,s_2>0$ and $s,s_2<0,s_1>0$)
the amplitude can be written as follows 
\begin{eqnarray}
\frac{M_{2\rightarrow 3}}{\Gamma (t_1)\Gamma (t_2)} ~=~  
\left(\frac{-s_1}{\mu^2}\right)^{\omega (t_1)-\omega (t_2)}
\left(\frac{-s\kappa }{\mu^4}\right)^{\omega (t_2)}c_1+
\left(\frac{-s_2}{\mu^2}\right)^{\omega (t_2)-\omega (t_1)}
\left(\frac{-s\kappa }{\mu^4}\right)^{\omega (t_1)}c_2
\label{Mdosatres}
\end{eqnarray}
with the real coefficients $c_i$
\begin{eqnarray}
c_1&=&|\Gamma (t_2,t_1, \ln {-\kappa} )|\,
\frac{\sin \pi (\omega (t_1)-\phi _\Gamma )}{\sin \pi (\omega (t_1)-\omega (t_2))}\,,
\label{ces1}\\
c_2&=&|\Gamma (t_2,t_1, \ln {-\kappa} )|\,
\frac{\sin \pi (\omega (t_2)-\phi _\Gamma )}{\sin \pi (\omega (t_2)-\omega (t_1))}\,,
\label{ces2}
\end{eqnarray}
where $\phi _\Gamma$ is the phase of $\Gamma$, {\it i.e.}
\begin{equation}
\Gamma (t_2,t_1, \ln \kappa -i\pi )=|\Gamma (t_2,t_1, \ln -\kappa )|\,
e^{i\pi \phi _\Gamma}\,. 
\label{phigamma}
\end{equation}
In this dispersion-type representation  for all physical channels we use  
the on-mass shell constraint for the produced gluon momentum 
\begin{equation}
\kappa = \vec{k}^2_\perp =(\vec{q}_1-\vec{q}_2)^2\,,
\label{kappaonshell}
\end{equation}
where $\vec{k}_\perp$ is its transverse component 
($k_\perp p_A =k_\perp p_B=0$). In this case the amplitude $M_{2-3}$ does not
have simultaneous singularities in the overlapping channels ($s_1, s_2$), in an 
agreement with the condition of the gluon stability 
(this will be discussed further in section 4.2). 

The fact that there exists a solution for the coefficients $c_1$ and $c_2$ 
proves that the scattering amplitude, derived from the BDS formula, has 
the correct analytic structure in all physical regions. In particular, 
it satisfies the Steinmann relations  
(a somewhat more detailed discussion of analyticity properties will be 
presented in the following section). We therefore conclude that 
the BDS amplitude for the production of one particle  
in multi-Regge kinematics has the correct multi-Regge form.
This is encouraging, and we proceed now to study the production of two particles 
in multi-Regge kinematics, for which we use the $M_{2 \rightarrow 4}$ 
BDS scattering amplitudes.

\begin{figure}[ht]
\centerline{\epsfig{file=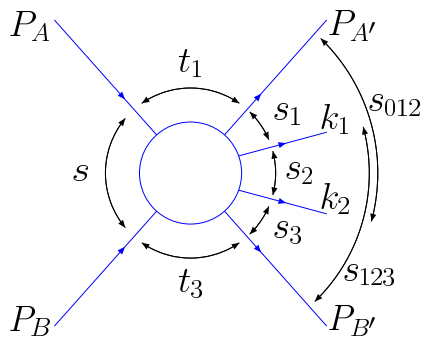,width=6cm,angle=0,bbllx=0,bblly=575,bburx=265,
        bbury=810,clip=}}
\caption{Production of two particles} 
\label{prod24}
\end{figure}
We have first checked that the planar BDS amplitude for two particle 
production having singularities only at positive values of the invariants 
$s,s_1,s_2, s_3, t_1, t_2, t_3$ has the correct multi-Regge form
in the multi-Regge kinematics in the region where all invariants
$s,s_1,s_2, s_3, t_1, t_2,t_3$ are negative (see Fig. \ref{prod24}
and Appendix C)
\begin{eqnarray}
\frac{M_{2\rightarrow 4}}{\Gamma (t_1)\Gamma (t_3)} =
\left(\frac{-s_1}{\mu ^2}\right)^{\omega (t_1)}\,\Gamma (t_2,t_1, 
\ln -\kappa _{12})
\,\left(\frac{-s_2}{\mu ^2}\right)^{\omega (t_2)}\,
\Gamma (t_3,t_2, \ln -\kappa _{23})
\left(\frac{-s_3}{\mu^2}\right)^{\omega (t_3)}\hspace{-0.3cm}
\end{eqnarray} 
and the quantities
\begin{equation}
-\kappa _{12}=\frac{(-s_1)(-s_2)}{-s_{012}}\,,\,\,\,\,
-\kappa _{23}=\frac{(-s_2)(-s_3)}{-s_{123}}\,.
\end{equation}
are fixed together with $t_i$.
The invariants $s_{012}$ and $s_{123}$ are the squared masses of
the corresponding three particles in their center-of-mass system.

\begin{figure}[ht]
\centerline{\epsfig{file=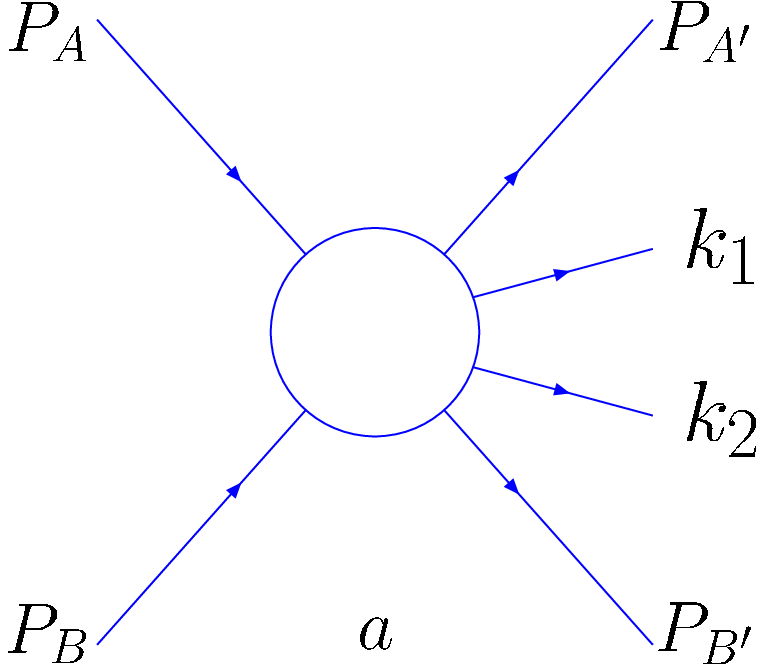,width=5cm,angle=0,bbllx=0,bblly=575,bburx=255,
        bbury=810,clip=}\epsfig{file=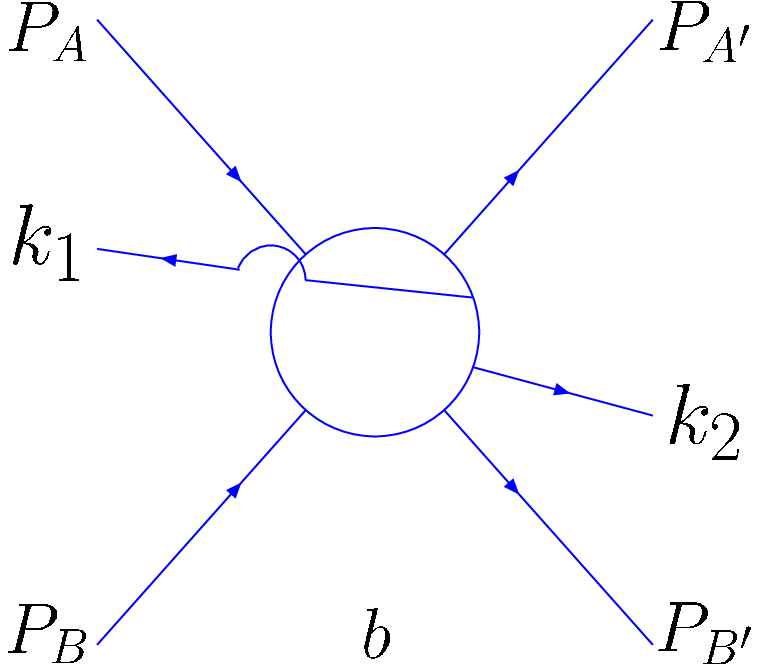,width=5cm,angle=0,bbllx=0,bblly=575,bburx=255,
        bbury=810,clip=}}
\centerline{\epsfig{file=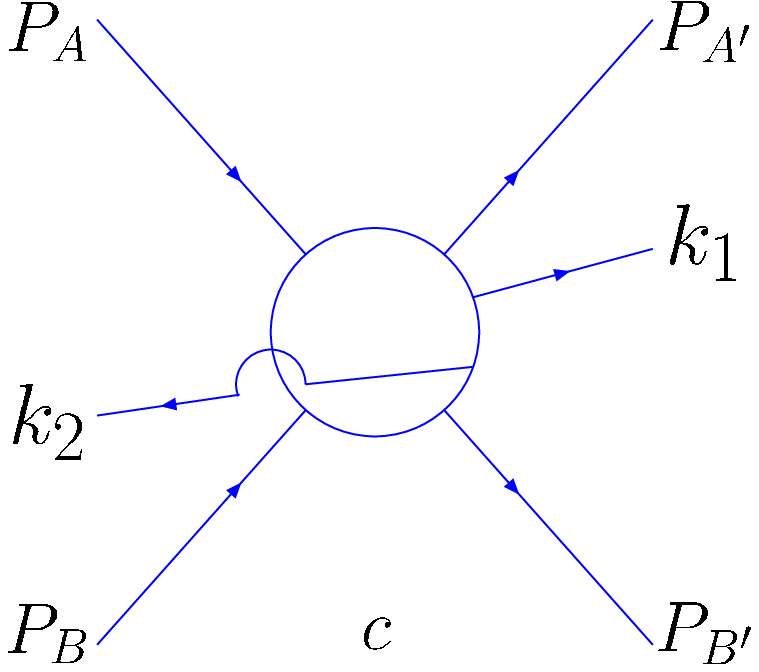,width=5cm,angle=0,bbllx=0,bblly=575,bburx=255,
        bbury=810,clip=}\epsfig{file=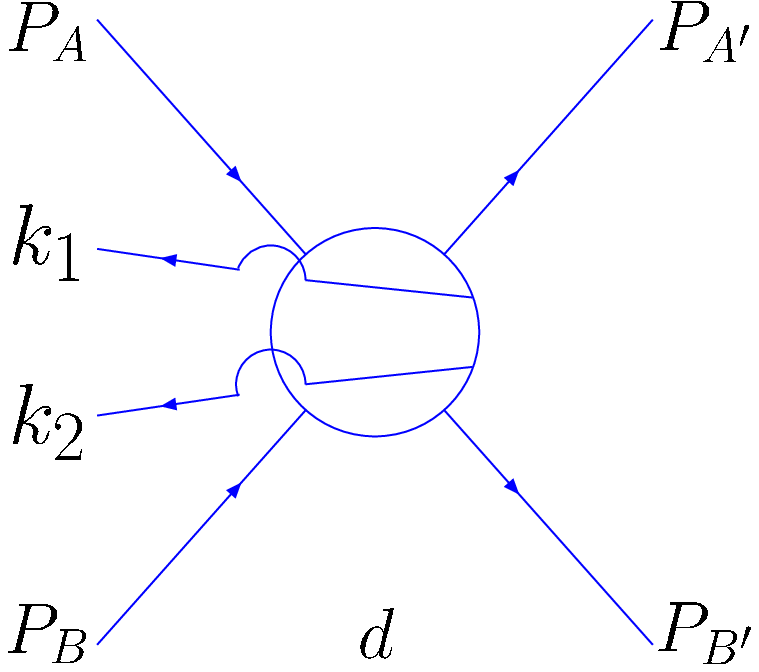,width=5cm,angle=0,bbllx=0,bblly=575,bburx=255,
        bbury=810,clip=}}
\caption{Physical channels for two particle production} 
\label{phys24}
\end{figure}
In a similar way to what we did in the $2 \rightarrow 3$ case, the BDS  
$2 \rightarrow 4$ amplitude can be analytically continued to several
physical channels, each of them corresponding to different signs of
the invariants $s, s_{012}, s_{123}, s_1, s_2, s_3$. To begin with,
in the region (see Fig.~\ref{phys24}a)
\begin{eqnarray}
s,  s_{012}, s_{123}, s_1, s_2, s_3 >0
\end{eqnarray}
the amplitude has the form
\begin{eqnarray}
&&\hspace{-1cm}
\frac{M_{2\rightarrow 4}}{\Gamma (t_1)\Gamma (t_3)} ~=~ \nonumber\\
&&\hspace{-0.8cm}
\left(\frac{-s_1}{\mu ^2}\right)^{\omega (t_1)}\,\Gamma (t_2,t_1,
\ln \kappa _{12}-i\pi )  
\,\left(\frac{-s_2}{\mu ^2}\right)^{\omega (t_2)}\,
\Gamma (t_3,t_2, \ln \kappa _{23}-i\pi )
\left(\frac{-s_3}{\mu
^2}\right)^{\omega (t_3)},
\end{eqnarray}  
where
we can replace $\kappa _{12}$ and $\kappa _{23}$ by their values
on the mass shell
\begin{equation}
\kappa _{12}\rightarrow \frac{s_1s_2}{s_{012}}=\vec{k_1}_\perp^2\,,\,\,
\kappa _{23}\rightarrow \frac{s_2s_3}{s_{123}}=\vec{k_2}_\perp^2\,.
\end{equation}
Here $k_1=q_1-q_2$ and $k_2=q_2-q_1$ are the momenta of the 
produced particles (see Fig.~\ref{phys24}).

In the physical region, represented in Fig.~\ref{phys24}b, where 
\begin{equation}
s,s_{012}, s_3 >0\,\,,\,\,s_1, s_2, s_{123} <0,
\end{equation}
one can obtain, for the multi-Regge asymptotics of the BDS amplitude, the 
following expression
\begin{eqnarray}
&&\hspace{-1cm}\frac{M_{2\rightarrow 4}}{\Gamma (t_1)\Gamma (t_3)} ~=~ 
\nonumber\\
&&\hspace{-0.8cm}
\left(\frac{-s_1}{\mu ^2}\right)^{\omega (t_1)}\,\Gamma (t_2,t_1,
\ln \kappa _{12}+i\pi )
\,\left(\frac{-s_2}{\mu ^2}\right)^{\omega (t_2)}\,
\Gamma (t_3,t_2, \ln \kappa _{23}-i\pi )
\left(\frac{-s_3}{\mu
^2}\right)^{\omega (t_3)}.
\end{eqnarray}
In a similar way, in the region (see Fig. \ref{phys24}c)
\begin{equation}
s,s_{123}, s_1 >0\,\,,\,\,\,s_3, s_2, s_{012} <0
\end{equation} 
we have
\begin{eqnarray}
&&\hspace{-1cm}\frac{M_{2\rightarrow 4}}{\Gamma (t_1)\Gamma (t_3)} ~=~ \nonumber\\
&&\hspace{-0.8cm}
\left(\frac{-s_1}{\mu ^2}\right)^{\omega (t_1)}\,\Gamma (t_2,t_1,
\ln \kappa _{12}-i\pi )
\,\left(\frac{-s_2}{\mu ^2}\right)^{\omega (t_2)}\,
\Gamma (t_3,t_2, \ln \kappa _{23}+i\pi )
\left(\frac{-s_3}{\mu
^2}\right)^{\omega (t_3)}.
\label{antesdeds}
\end{eqnarray}

\begin{figure}[ht]
\centerline{\epsfig{file=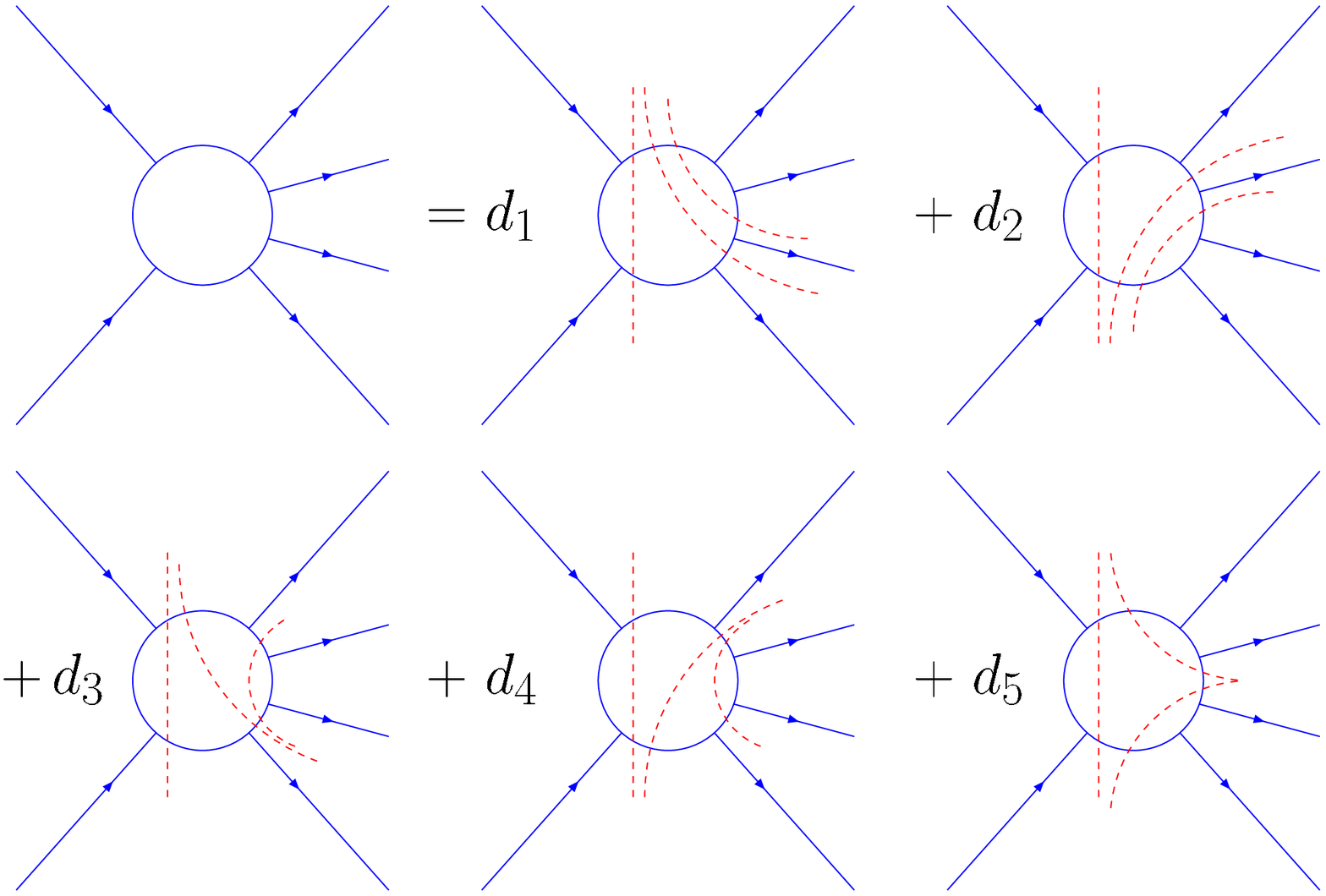,width=12cm,angle=0,bbllx=0,bblly=390,bburx=580,
        bbury=800,clip=}}
\caption{Analytic representation of the amplitude $M_{2\rightarrow 4}$} 
\label{disper24}
\end{figure}

We can now attempt to write $M_{2\rightarrow 4}$ in all physical regions
in terms of
the following dispersion ansatz (represented in Fig.~\ref{disper24})
\begin{eqnarray}
\frac{M_{2\rightarrow 4}}{\Gamma (t_1)\Gamma (t_3)} &=&
\left(\frac{-s_1}{\mu^2}\right)^{\omega (t_1)-\omega (t_2)}
\left(\frac{-s_{012}\kappa _{12}}{\mu^4}\right)^{\omega (t_2)-\omega (t_3)}
\left(\frac{-s\kappa_{12} \kappa_{23}}{\mu^6}\right)^{\omega (t_3)}\,
d_1 \nonumber\\
&+&\left(\frac{-s_3}{\mu^2}\right)^{\omega (t_3)-\omega (t_2)}
\left(\frac{-s_{123}\kappa _{23}}{\mu^4}\right)^{\omega (t_2)-\omega (t_1)}
\left(\frac{-s\kappa_{12} \kappa_{23}}{\mu^6}\right)^{\omega (t_1)}\,d_2
\nonumber\\
&+&\left(\frac{-s_2}{\mu^2}\right)^{\omega (t_2)-\omega (t_1)}
\left(\frac{-s_{012}\kappa _{12}}{\mu^4}\right)^{\omega (t_1)-\omega (t_3)}
\left(\frac{-s\kappa_{12} \kappa_{23}}{\mu^6}\right)^{\omega (t_3)}\,
d_3  \nonumber\\ 
&+&\left(\frac{-s_2}{\mu^2}\right)^{\omega (t_2)-\omega (t_3)}
\left(\frac{-s_{123}\kappa _{23}}{\mu^4}\right)^{\omega (t_3)-\omega (t_1)}
\left(\frac{-s\kappa_{12} \kappa_{23}}{\mu^6}\right)^{\omega (t_1)}\,d_4   
\nonumber\\
&+&\left(\frac{-s_3}{\mu^2}\right)^{\omega (t_3)-\omega (t_2)}
\left(\frac{-s_1}{\mu^2}\right)^{\omega (t_1)-\omega (t_2)}
\left(\frac{-s\kappa_{12} \kappa_{23}}{\mu^6}\right)^{\omega (t_2)}\,d_5
\label{Mdosacuatro}
\end{eqnarray}
with the real coefficients $d_{i=1,2,3,4,5}$. Here
\begin{eqnarray}
\kappa _{12}=(\vec{q}_1-\vec{q}_2)^2\,,\,\,
\kappa _{23}=(\vec{q}_2-\vec{q}_3)^2\,.
\end{eqnarray}

By comparing the above
`dispersive' 
representation with the previous expressions for the BDS amplitude 
in the three physical regions Figs~\ref{phys24}a,b,c
it is possible to extract the coefficients. 
They read 
\begin{eqnarray}
d_1 &=& c_1(\kappa _{12})\,c_1(\kappa _{23})\,, \nonumber\\
d_2 &=& c_2(\kappa _{12})\,c_2(\kappa _{23})\,, \nonumber\\
d_3 + d_4 &=& c_2 (\kappa _{12})\,c_1(\kappa _{23})\,, \nonumber\\
d_5 &=& c_1 (\kappa _{12})\,c_2(\kappa _{23})\,,
\label{dcoefficients}
\end{eqnarray}
where, in fact, $c_1(\kappa )$ and $c_2(\kappa )$ were defined 
in Eqs.~(\ref{ces1}), (\ref{ces2}):
\begin{eqnarray}
c_1(\kappa _{12}) &=& |\Gamma _{12} |\frac{\sin \pi (\omega (t_1)
-\phi _{\Gamma _{12}})}{\sin \pi (\omega (t_1)-\omega (t_2))}\,,\\
c_2(\kappa _{12}) &=& |\Gamma _{12}|
\frac{\sin \pi (\omega (t_2)-\phi _{\Gamma _{12}})}{\sin \pi 
(\omega (t_2)-\omega (t_1))}\,,\\
c_1(\kappa _{23})&=&|\Gamma _{23} |
\frac{\sin \pi (\omega (t_2)-\phi _{\Gamma _{23}})}{\sin \pi 
(\omega (t_2)-\omega (t_3))}\,, \\
c_2(\kappa _{23})&=&|\Gamma _{23}|
\frac{\sin \pi (\omega (t_3)-\phi _{\Gamma _{23}})}{\sin 
\pi (\omega (t_3)-\omega (t_2))}\,,
\end{eqnarray}
with (cf.(\ref{Gammavertex}))
\begin{equation}
\Gamma _{12}=\Gamma (t_2,t_1,\ln -\kappa _{12})\,,\,\,
\Gamma _{23}=\Gamma (t_3,t_2,\ln -\kappa _{23})\,.
\end{equation}

We note that for the coefficients $d_3$ and $d_4$ only their sum can be 
determined from the three physical regions previously discussed. However, 
an attempt to fix separately these two coefficients from the multi-Regge 
asymptotics in the physical region (see Fig. \ref{phys24}d)
\[
s,s_2>0\,,\,\,s_1,s_3,s_{012},s_{123}<0
\]  
leads to a disaster: the corresponding equations do not have any solution.
The reason for this is that the BDS amplitude in this region does not have 
the correct Regge factorization (see the discussion in section 4). 
According to Appendix C its asymptotics here is
\begin{eqnarray}
&&\hspace{-1cm}
\frac{M_{2\rightarrow 4}}{\Gamma (t_1)\Gamma (t_3)} ~=~ \nonumber\\
&&\hspace{-0.8cm}C \left(\frac{-s_1}{\mu ^2}\right)^{\omega (t_1)}
\Gamma (t_2,t_1,\ln \kappa _{12}-i\pi )
\left(\frac{-s_2}{\mu ^2}\right)^{\omega (t_2)}
\Gamma (t_3,t_2, \ln \kappa _{23}-i\pi )
\left(\frac{-s_3}{\mu^2}\right)^{\omega (t_3)},
\label{24mixedregion}
\end{eqnarray}
where the coefficient $C$ is given by 
\begin{equation}
C=\exp \left[\frac{\gamma _K(a)}{4} \,i\pi \,\left( 
\ln \frac{\vec{q}_1^2\vec{q}_3^2}{(\vec{k}_1+\vec{k}_2)^2\mu ^2}
-\frac{1}{\epsilon}\right)\right].
\label{coeffC}
\end{equation} 
The fact that, for this region, we find no solution for the coefficients
$d_i$ indicates that, in this region, the BDS amplitude does not have the 
correct analytic structure. In section 4 we will show, by comparing with 
explicit calculations of the high energy limit of scattering amplitudes, 
that in the BDS formula a piece is missing. This piece belongs to a Regge 
cut singularity, which - apart from the one-loop approximation -  does not 
fit into the simple exponentiation of the BDS ansatz. In Appendix C we 
write down the amplitude $M_{2\rightarrow 4}$ also in the quasi-multi-Regge 
kinematics, where the variable $s_2$ is fixed.

\begin{figure}[ht]
\centerline{\epsfig{file=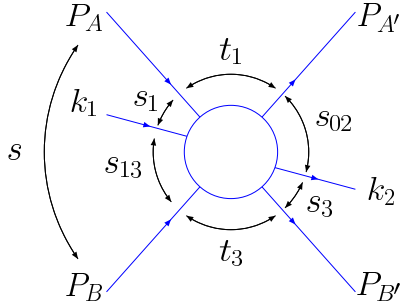,width=6cm,angle=0,bbllx=70,bblly=570,
bburx=360,bbury=815,clip=}}
\caption{Three particle transition }
\label{kin3to3}
\end{figure}
To continue our analysis of the BDS six point amplitude we now discuss the 
asymptotics of $M_{3\rightarrow 3}$ (see Fig.~\ref{kin3to3}). According to 
Appendix D in the multi-Regge region where all invariants 
$s, s_{1}, s_{3}, s_{13}, s_{02}, s_2 \equiv t'_2$ are large and negative,
its asymptotics is similar to the corresponding asymptotics of the 
$M_{2\rightarrow 4}$ amplitude, {\it i.e.}
\begin{eqnarray}
&&\hspace{-1cm}\frac{M_{3\rightarrow 3}}{\Gamma (t_1)\Gamma (t_3)} =
\nonumber\\
&&\hspace{-0.8cm}\left(\frac{-s_1}{\mu ^2}\right)^{\omega (t_1)} 
\Gamma (t_2,t_1,\ln -\kappa _{12})
\left(\frac{-s_2}{\mu ^2}\right)^{\omega (t_2)}
\Gamma (t_3,t_2, \ln -\kappa _{23})
\left(\frac{-s_3}{\mu^2}\right)^{\omega (t_3)}.
\end{eqnarray}
\begin{figure}[ht]
\centerline{\epsfig{file=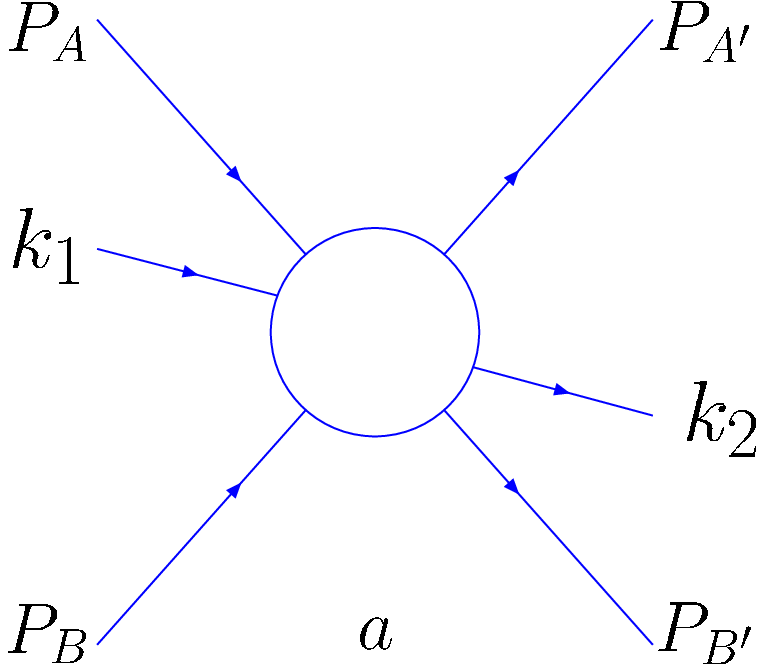,width=5cm,angle=0,bbllx=0,
bblly=575,bburx=255,bbury=810,clip=}\epsfig{file=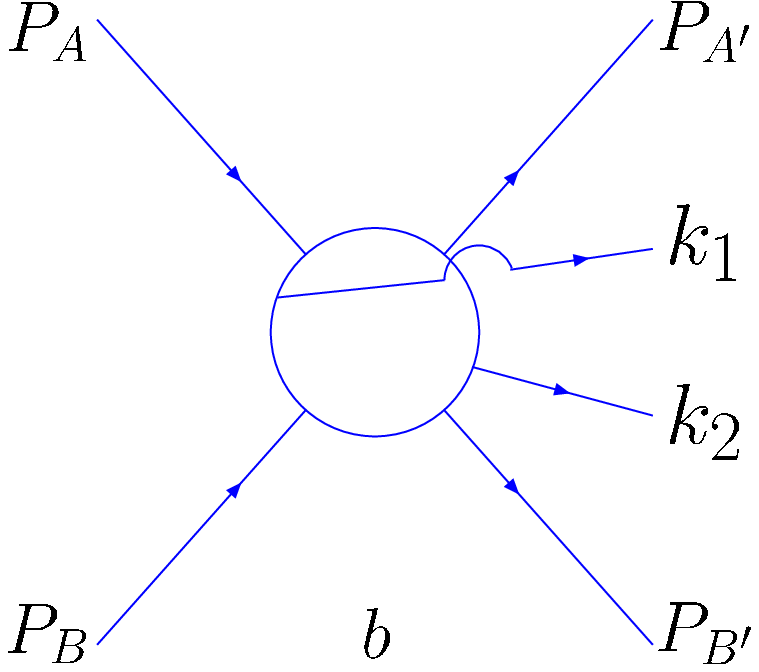,
width=5cm,angle=0,bbllx=0,bblly=575,bburx=255,bbury=810,clip=}}
\centerline{\epsfig{file=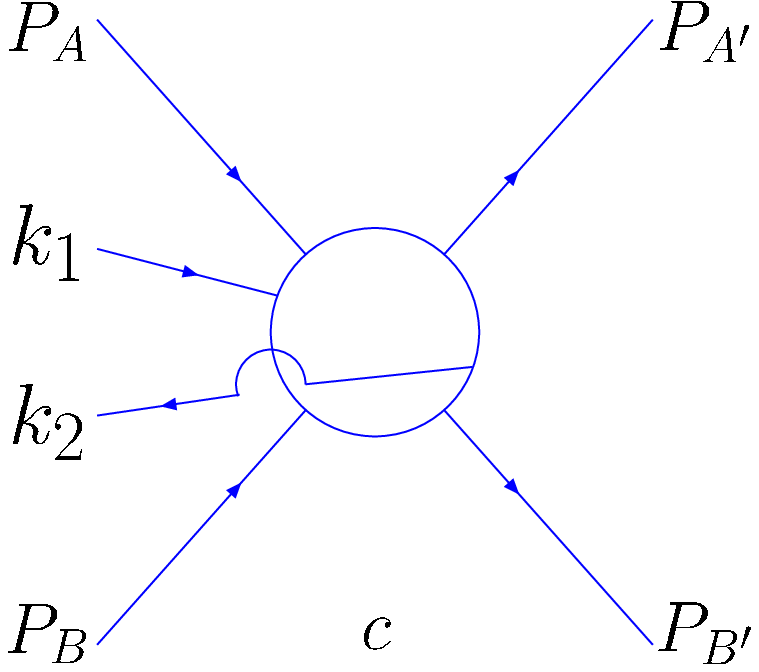,width=5cm,angle=0,bbllx=0,   
bblly=575,bburx=255,bbury=810,clip=}\epsfig{file=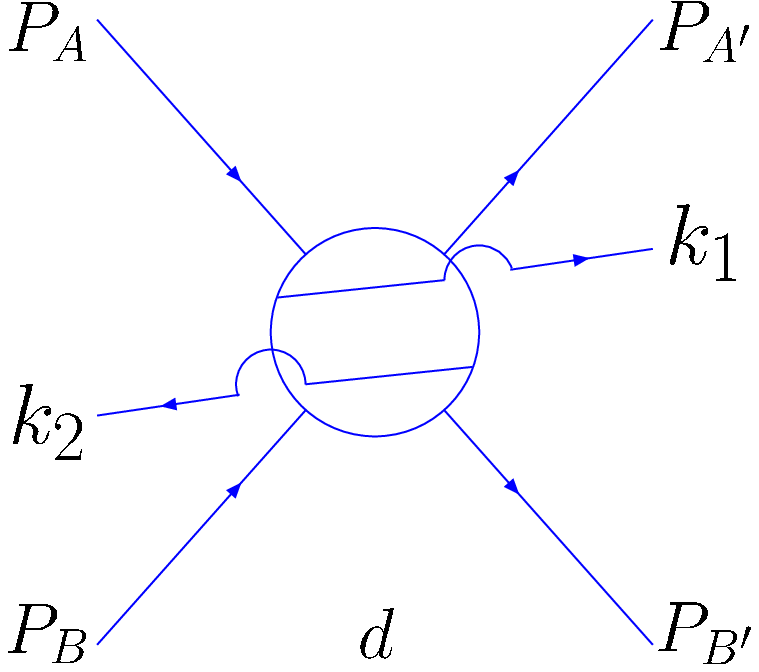,
width=5cm,angle=0,bbllx=0,bblly=575,bburx=255,bbury=810,clip=}}
\caption{Physical regions for the amplitude $M_{3\rightarrow 3}$}
\label{Phys3to3}
\end{figure}
This BDS amplitude can be now analytically continued to the physical region 
where the invariants $s, s_{1}, s_{3}, s_{12}, s_{02}, t'_2$ are positive
(see Fig. \ref{Phys3to3}a). The resulting amplitude can be written as 
\begin{eqnarray}
&&\hspace{-1cm}\frac{M_{3\rightarrow 3}}{\Gamma (t_1)\Gamma (t_3)}~=~
\nonumber\\
&&\hspace{-0.8cm}
\left(\frac{-s_1}{\mu ^2}\right)^{\omega (t_1)}
\Gamma (t_2,t_1,\ln \kappa _{12}-i\pi )
\left(\frac{-s_2}{\mu ^2}\right)^{\omega (t_2)}
\Gamma (t_3,t_2, \ln \kappa _{23}-i\pi )
\left(\frac{-s_3}{\mu^2}\right)^{\omega (t_3)},
\end{eqnarray}
Similarly, the analytic continuation to the region where 
$s_1, s_{12}, t'_2<0$ and $s, s_{3}, s_{02}>0$ (see Fig. \ref{Phys3to3}b) is 
of the form
\begin{eqnarray}                                   
&&\hspace{-1cm}\frac{M_{3\rightarrow 3}}{\Gamma (t_1)\Gamma (t_3)} ~=~
\nonumber\\
&&\hspace{-0.8cm}
\left(\frac{-s_1}{\mu ^2}\right)^{\omega (t_1)}
\Gamma (t_2,t_1,\ln \kappa _{12}+i \pi )
\left(\frac{-s_2}{\mu ^2}\right)^{\omega (t_2)}
\Gamma (t_3,t_2, \ln \kappa _{23}-i\pi )
\left(\frac{-s_3}{\mu^2}\right)^{\omega (t_3)}.
\end{eqnarray}
Finally, the continuation to the region where 
$s_3, s_{02}, t'_2<0$ and $s, s_{1}, s_{12}>0$
(see Fig. \ref{Phys3to3}c) reads
\begin{eqnarray}
&&\hspace{-1cm}\frac{M_{3\rightarrow 3}}{\Gamma (t_1)\Gamma (t_3)}~=~
\nonumber\\
&&\hspace{-0.8cm}
\left(\frac{-s_1}{\mu ^2}\right)^{\omega (t_1)}
\Gamma (t_2,t_1,\ln \kappa _{12}-i \pi )
\left(\frac{-s_2 }{\mu ^2}\right)^{\omega (t_2)}
\Gamma (t_3,t_2, \ln \kappa _{23}+i\pi )
\left(\frac{-s_3}{\mu^2}\right)^{\omega (t_3)}.
\end{eqnarray}

\begin{figure}[ht]
\centerline{\epsfig{file=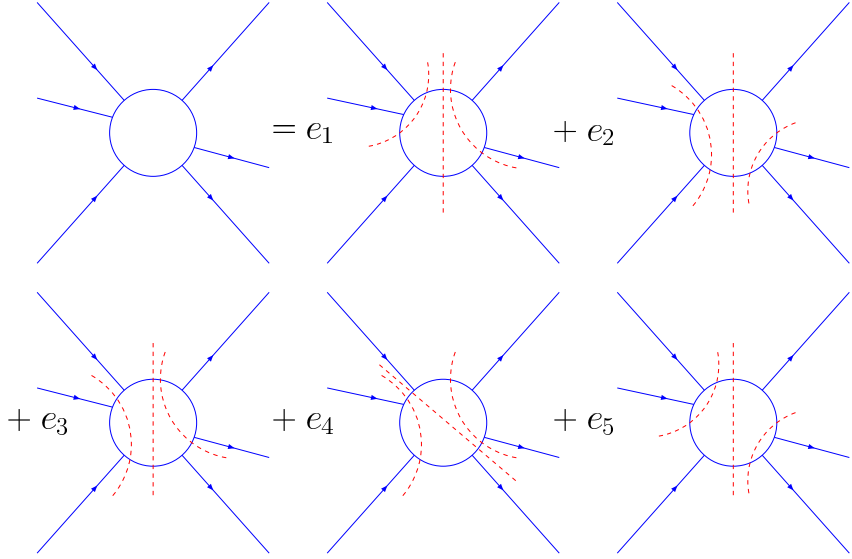,width=12cm,angle=0,bbllx=0,bblly=400,bburx=560
,bbury=800,clip=}}
\caption{Analytic representation of the amplitude $M_{3\rightarrow 3}$}
\label{Dispers3to3}
\end{figure}

As it was done in the $M_{2 \rightarrow 4}$ case one can write the 
dispersion relation for $M_{3\rightarrow 3}$ valid in these physical regions, 
which includes five 
contributions, shown in Fig.~\ref{Dispers3to3}, and calculate the real 
coefficients $e_1,e_2,e_3+e_4,e_5$. But once again, we find that it is impossible 
to fix separately the coefficients $e_3$ and $e_4$ from the BDS amplitude,
calculated in the physical region where $s_1,s_3, s_{13}, s_{02}<0$ and
$s, t'_2>0$ (see Fig. \ref{Phys3to3}d)
\begin{eqnarray}
&&\hspace{-1cm}\frac{M_{3\rightarrow 3}}{\Gamma (t_1)\Gamma (t_3)} ~=~
\nonumber\\
&&\hspace{-0.8cm}
C' \, \left(\frac{-s_1}{\mu ^2}\right)^{\omega (t_1)}
\Gamma (t_2,t_1,\ln \kappa _{12}+i \pi )
\left(\frac{-s_2}{\mu ^2}\right)^{\omega (t_2)}
\Gamma (t_2,t_1, \ln \kappa _{23}+i\pi )
\left(\frac{-s_3}{\mu^2}\right)^{\omega (t_3)},
\label{33mixedregion}
\end{eqnarray}
where the phase factor $C'$ is
\begin{equation}
C'=\exp \left[\frac{\gamma _K(a)}{4} \,(-i\pi ) \,
\ln \frac{(\vec{q}_1-\vec{q}_2)^2\,
(\vec{q}_2-\vec{q}_3)^2}{(\vec{q}_1+\vec{q}_3-\vec{q}_2)^2\,\vec{q}_2^2}  
\right]. 
\label{coeffC'}
\end{equation}
 The reason for this drawback is the 
same as before: the absence of a correct Regge factorization for the BDS 
amplitude. In the next section, using the BFKL approach, we shall discuss the 
reason for this problem. Namely, the BDS amplitude does not contain
the Mandelstam cut 
contributions
plotted in Fig.~\ref{BFKLlad}.  

\section{Regge cuts and breakdown of factorization}
\subsection{Regge pole models } 
The results of the previous section can best be understood if we confront them
with the known high energy behavior of QCD scattering amplitudes in the Regge
limit. In the LLA, the high energy behavior of the 
QCD scattering amplitudes is the same as in the supersymmetric case. 
We proceed in three steps: we first review the findings for models containing 
only Regge poles. We then summarize the results obtained in gauge theories, 
and finally compare with the scattering amplitude derived from the BDS formula.
    
A key element in analyzing the high energy limit are the Steinmann
relations~\cite{Steinmann} which forbid the existence of simultaneous energy 
discontinuities
in overlapping channels. As an illustrative example of the Steinmann
relations, consider the $2 \to 3$ amplitude shown in Fig.~\ref{disper23}:
obviously, the produced particle in the central region can form resonance 
states with particle $A'$ or with particle $B'$, but not simultaneously with
both of them. As a result, in the physical region the scattering amplitude
cannot have simultaneous discontinuities in the energy variables $s_1$ and
$s_2$. The way in which this restriction is implemented into   
scattering amplitudes is that, in the double
Regge limit, the signatured amplitude can be written as a sum of two pieces,
one of them with cuts in the $s_1$ and in the $s$ channels,
the other one in the $s_2$ and in the $s$ channels. In general, there are
cut singularities both in the right and left half energy planes, and one has to form
signatured combinations. Decompositions of this kind have first been derived from simple models
which contain only Regge poles (massive scalar
$\varphi^3$ theory~\cite{Cambridge} or the dual
Veneziano $6$ point amplitude, $B_6$~\cite{Brower}), and from
studies of dispersion relations and generalized Froissart-Gribov partial
wave representations~\cite{White}. For $2 \to 4$ or $3 \to 3$ amplitudes, we have five 
independent terms, and for scattering processes with higher number of legs the number 
of terms grows rapidly.  

For models which contain only Regge poles
the general structure of the signatured $2 \to 3$ amplitude is:
\begin{eqnarray}
\frac{A_{2\to 3}}{\beta_A(t_1) \beta_B(t_2)} &=& \nonumber\\
&&\hspace{-3cm}\left[ \left(\frac{-s_1}{\mu^2}\right)^{\alpha(t_1)-\alpha(t_2)}
      \hspace{-0.5cm}+\tau_1 \tau_2\left(\frac{s_1}{\mu^2}
\right)^{\alpha(t_1)-\alpha(t_2)} \right]
  \left[
      \left(\frac{-s}{\mu^2}\right)^{\alpha (t_2)} \hspace{-0.5cm}+
     \tau_2 \left(\frac{s}{\mu^2}\right)^{\alpha (t_2)} \right]
\tilde{V}_1(t_1,t_2,\kappa) \nonumber \\ 
&&\hspace{-3.4cm}+ 
\left[ \left(\frac{-s_2}{\mu^2}\right)^{\alpha(t_2)-\alpha(t_1)}
      \hspace{-0.5cm}+\tau_1 \tau_2 \left(\frac{s_2}{\mu^2}
\right)^{\alpha(t_2)-\alpha(t_1)} \right]
  \left[
      \left(\frac{-s}{\mu^2}\right)^{\alpha (t_1)} \hspace{-0.5cm}+
     \tau_1 \left(\frac{s}{\mu^2}\right)^{\alpha (t_1)} \right]
\tilde{V}_2(t_1,t_2,\kappa).
\end{eqnarray}
Here $\alpha(t_i)$ denotes the trajectory function of the Regge pole in the 
$t_i$ exchange channel, 
$\tau_1$ ($\tau_2$) are the
signatures of the $t_1$ ($t_2$) channels, and
as usual, $(-s)^{\alpha} = \left(|s|\right)^{\alpha}e^{-i \pi \alpha}$.
In this representation, the energy singularities are explicit, i.e.
all phase factors are contained in the energy factors, and the functions
$\beta(t)$, $V_i$ are real valued functions. 
With the abreviations 
\beq
\alpha_i = \alpha(t_i), \alpha_{ij} = \alpha(t_i) - \alpha(t_j)
\eeq
and with the signature factors
\begin{equation}  
\xi_i = e^{-i \pi \alpha_i} + \tau_i,\hspace{1cm}
\xi_{ij} = e^{-i \pi \alpha_{ij}} + \tau_i \tau_j
\label{signature1}
\end{equation}
we can rewrite the expression for $A_{2 \to3}$:
\begin{eqnarray} 
\frac{A_{2\to 3}}{\beta_A(t_1)\beta_B(t_2)} &=&
\left(\frac{|s_1|}{\mu^2}\right)^{\alpha_{12}}
      \left(\frac{|s|}{\mu^2}\right)^{\alpha_2} \xi_{12} \, \xi_2 \, 
\frac{V_1(t_1,t_2,\kappa)}{\sin \pi \alpha_{12}} \nonumber\\
&+& \left(\frac{|s_2|}{\mu^2}\right)^{\alpha_{21}}
     \left(\frac{|s|}{\mu^2}\right)^{\alpha_1} \xi_{21} \, \xi_1 \,
\frac{V_2(t_1,t_2,\kappa)}{\sin \pi \alpha_{21}},
\label{23analyticpole}
\end{eqnarray}
where the vertex function $V_i$ is proportional to $\tilde{V}_i$.  
The generalization to the signatured $2 \to 4$ amplitude 
(consisting of five different pieces) is illustrated in 
Fig.~\ref{disper24}, and from 
Eq.~(\ref{Mdosacuatro}) one easily obtains the analogue of 
(\ref{23analyticpole})~\cite{Brower}:
\begin{eqnarray}
\frac{A_{2\to4}}{\beta_A(t_1)\beta_B(t_3)} &=& \nonumber\\ 
&&\hspace{-3.6cm}\left(\frac{|s_1|}{\mu^2}\right)^{\alpha_{12}}
\left(\frac{|s_{012}|}{\mu^2}\right)^{\alpha_{23}}
\left(\frac{|s|}{\mu^2}\right)^{\alpha_3}
\xi_{12} \xi_{23} \xi_3 \frac{ V_1(t_1,t_2,\kappa_{12}) V_1(t_2,t_3,\kappa_{23})}
{\sin \pi \alpha_{12}  \sin \pi \alpha_{23}}
\nonumber \\  
&&\hspace{-3.6cm}+
\left(\frac{|s_3|}{\mu^2}\right)^{\alpha_{32}}
\left(\frac{|s_{123}|}{\mu^2}\right)^{\alpha_{21}}
\left(\frac{|s|}{\mu^2}\right)^{\alpha_1}
\xi_{32} \xi_{21} \xi_1 \frac{V_2(t_1,t_2,\kappa_{12}) V_2(t_2,t_3,\kappa_{23})}
{\sin \pi \alpha_{32} \sin \pi \alpha_{21}}
\nonumber \\  
&&\hspace{-3.6cm}+
\left(\frac{|s_2|}{\mu^2}\right)^{\alpha_{21}}
\left(\frac{|s_{012}|}{\mu^2}\right)^{\alpha_{13}}
\left(\frac{|s|}{\mu^2}\right)^{\alpha_3}
\xi_{21} \xi_{13} \xi_3 
\frac{\sin \pi \alpha_1}{\sin \pi \alpha_2}
\frac{V_2(t_1,t_2,\kappa_{12}) V_1(t_2,t_3,\kappa_{23})}
{\sin \pi \alpha_{21} \sin \pi \alpha_{13}}
\nonumber \\  
&&\hspace{-3.6cm}+
\left(\frac{|s_2|}{\mu^2}\right)^{\alpha_{23}}
\left(\frac{|s_{123}|}{\mu^2}\right)^{\alpha_{31}}
\left(\frac{|s|}{\mu^2}\right)^{\alpha_1}
\xi_{23} \xi_{31} \xi_1 
\frac{\sin \pi \alpha_3}{\sin \pi \alpha_2}
\frac{V_2(t_1,t_2,\kappa_{12}) V_1(t_2,t_3,\kappa_{23})}
{\sin \pi \alpha_{23} \sin \pi \alpha_{31}}
\nonumber  \\  
&&\hspace{-3.6cm}+
\left(\frac{|s_3|}{\mu^2}\right)^{\alpha_{32}}
\left(\frac{|s_1|}{\mu^2}\right)^{\alpha_{12}}
\left(\frac{|s|}{\mu^2}\right)^{\alpha_2}
\xi_{32} \xi_{12} \xi_2 \frac{ V_1(t_1,t_2,\kappa_{12}) V_2(t_2,t_3,\kappa_{23})}
{\sin \pi \alpha_{32}  \sin \pi \alpha_{12}}.
\label{24analyticpole}
\end{eqnarray}
The analogue for the $3\to3$ process again consists of five pieces
which are shown in Fig.~\ref{Dispers3to3}.   

From the discussions of these Regge pole models it has also been recognized
that the analytic decomposition into a sum of terms
in (\ref{23analyticpole}) and (\ref{24analyticpole}) is consistent
with a factorizing form.
For the $2\to 3$ case we can write:
\begin{eqnarray}
\frac{A_{2 \to 3}}{\beta_A(t_1)\beta_B (t_2)} &=&
\left(\frac{|s_1|}{\mu^2}\right)^{\alpha_1} \xi_1 \,
      V_{\tau_1\tau_2}(t_1,t_2,\kappa)\,
   \left(\frac{|s_2|}{\mu^2}\right)^{\alpha_2} \xi_2.
\label{23factor}
\end{eqnarray}
Here the important point to be stressed is that the new production vertex
function $V_{\tau_1 \tau_2}$ contains phases (in contrast to the real-valued
functions $V_i$ in (\ref{23analyticpole})), and it has cut singularities in the
$\kappa$-plane. Similarly for the $2 \to 4$ case we have  
\begin{eqnarray}
\frac{A_{2 \to 4}}{\beta_A(t_1)\beta_B (t_3)} =
\xi_1\,\left(\frac{|s_1|}{\mu^2}\right)^{\alpha_1} \,
      V_{\tau_1\tau_2}(t_1,t_2,\kappa_{12}) \,
   \xi_2\,\left(\frac{|s_2|}{\mu^2}\right)^{\alpha_2} 
      V_{\tau_2\tau_3}(t_2,t_3,\kappa_{23}) \,
   \xi_3 \, \left(\frac{|s_3|}{\mu^2}\right)^{\alpha_3} 
\label{24factor}
\end{eqnarray}  
with the production vertex function from (\ref{23factor}). 
As a result, for this class of Regge-pole models the production amplitudes,
in the multi-Regge limit, can be written either in the `analytic' form
(sum of terms with simple analytic properties and real-valued vertex functions
$V_i$)
or in the `factorized' form (with the production vertices $V$ containing
phases and singularities in $\kappa$).

Let us comment on the planar approximation.
Planar amplitudes have right hand cut singularities only, and in the physical 
region where all energies are positive, their phases 
follow from the signatured amplitudes in (\ref{23analyticpole}) or 
(\ref{24analyticpole}) by simply 
dropping all 'twisted' terms containing factors $\tau_i$. 
One can easily verify that, in the physical region where all 
energies are positive, these planar amplitudes can also be written in the 
factorized form (\ref{23factor}) and (\ref{24factor}) (with vertex  
functions $V(t_1,t_2,\kappa_{12})$ being slightly different from 
the signatured ones, $V_{\tau_1\tau_2}(t_1,t_2,\kappa_{12})$). 
When analytically continuing into the unphysical region, where all energy
variables are negative and
well separated from their threshold singularities, all phases inside the 
production vertex $V(t_1,t_2,\kappa_{12})$ disappear, 
the vertex function turns into a real-valued function, and the factorized 
form remains valid. However, in the physical 
region where $s, s_2 > 0$ and $s_1, s_3, s_{012}, s_{123} < 0$  
the factorized form is not valid, and the structure of the amplitude is 
more complicated.  
 
\subsection{High energy behavior in Yang Mills theories}

Let us now turn to QCD. Throughout this section we will restrict ourselves to 
scattering amplitudes with odd signature in all t-channels. 
Compared to the Regge pole models discussed in the previous subsection, 
the situation is slightly more complicated since also Regge cut pieces 
appear in some of the $t$-channels.
In the LLA the real part of the
$2 \to n$ scattering amplitude is well known to have the factorized form
of Eq.~(\ref{MRKcinematica}), and it is in agreement with our previous result 
in (\ref{23factor}) and (\ref{24factor}).
However, when turning to the imaginary parts ({\it i.e.} to the energy 
discontinuities) of the production amplitudes, a new piece appears which 
destroys the simple factorization property. The best way of understanding 
the appearance of this new piece is the use of $s$-channel unitarity in the 
physical region where all energies are positive. 
 
Starting from the analytic representation of the scattering amplitude 
$A_{2 \to n}$,
it is possible to determine, in QCD, the partial waves from energy 
discontinuities
and unitarity equations~\cite{JB1,JB2}.
As the simplest example, let us consider, in the LLA,
the $2 \to 3$ amplitude, consisting of the
two terms illustrated in Fig.~\ref{disper23}.
Anticipating that, in the $2 \to 3$ process, there are only 
Regge pole contributions, we start from the ansatz 
\begin{eqnarray} 
\frac{A_{2\to 3}}{\Gamma(t_1) \Gamma(t_2)} &=&  \frac{2s}{t_1t_2} \Big[                      
\left(\frac{|s_1|}{\mu^2}\right)^{\omega_{12}}
      \left(\frac{|s|}{\mu^2}\right)^{\omega_2} \xi_{12} \, \xi_2 \, 
\frac{V_1(t_1,t_2,\kappa)}{\sin \pi \omega_{12}}        \nonumber\\ 
& +&        \left(\frac{|s_2|}{\mu^2}\right)^{\omega_{21}}
     \left(\frac{|s|}{\mu^2}\right)^{\omega_1} \xi_{21} \, \xi_1 \,
\frac{V_2(t_1,t_2,\kappa)}{\sin \pi \omega_{21}} \Big]
\label{23analyticgeneral}
\end{eqnarray}
with (cf.(\ref{kappaonshell}))
\beq
\kappa = (\vec{q}_1-\vec{q}_2)^2
\eeq
and
\beq
\Gamma(t_1) = g \delta_{\lambda_A \lambda_{A'}}, 
\Gamma(t_2) = g \delta_{\lambda_B \lambda_{B'}}.
\eeq
Here we used $\alpha_i(t_i) = 1 + \omega (t_i)$ and 
\beq
\omega_i = \omega(t_i),\;\; \omega_{ij} = \omega(t_i) - \omega(t_j),
\eeq
and the signature factors can be written in following form:
\begin{equation}  
\xi_i = e^{-i \pi \omega_i} + 1,\hspace{0.5cm}
\xi_{ij} = e^{-i \pi \omega_{ij}} + 1.
\label{signature2}
\end{equation}
Taking the discontinuity in
$s_1$, only the first term in (\ref{23analyticgeneral}) contributes. Making use 
of the unitarity equation
and invoking, for the ladder diagrams in the $t_1$ channel, the BFKL
bootstrap condition we find the partial wave $V_1$ in the LLA. For a definite
helicity $V_1$ has the form 
\begin{equation}
V_1= g \pi C(q_2,q_1) \left( \frac{1}{2} (\omega_1 - \omega_2) -
\frac{a}{2} (\ln \frac{\kappa}{\mu ^2} - \frac{1}{\epsilon}) \right)
\label{V1}
\end{equation}
with $C(q_2,q_1)$ being the production vertex from (\ref{helicityproduction}). 
Similarly, the discontinuity in $s_2$ leads to
\begin{equation}
V_2= g \pi  C(q_2,q_1) \left( \frac{1}{2} (\omega_2 - \omega_1) - \frac{a}{2} 
(\ln \frac{\kappa}{\mu ^2} - \frac{1}{\epsilon} )\right).
\label{V2}
\end{equation}
A comment may be in place on the term $\ln \kappa$ in $V_1$ and $V_2$:
it indicates that, in contrast to massive field theories where the $V_i$'s 
are analytic functions near $\kappa=0$, in massless theories this is no longer the 
case. Therefore, when computing the discontinuity in $s_1$ or $s_2$ of 
$A_{2 \to 3}$, there is, 
at first sight, an uncertainty in handling the cut in $\kappa$. 
It turns out that the correct prescription for computing the discontinuity 
in $s_1$ or $s_2$ in the physical region, is keeping 
$\kappa = (\vec{q}_1-\vec{q}_2)^2$ fixed. 
This can be derived either from a direct analyis of Feynman diagrams where 
the Steinmann relations $disc_{s_1} disc_{s_2} A_{2 \to 3} =0$ are fulfilled 
explicitly. Alternatively, one can consider the massless  
Yang Mills theory as the zero mass limit of a nonabelian Higgs model 
where the gauge bosons are massive: before the zero mass limit is taken,
the vertex functions are analytic near $\kappa=0$ and there is no ambiguity in 
computing the energy discontinuities. As a result, in the physical region 
the singularities of in $\kappa$ of $V_1$ and $V_2$ are not related to
singularities in $s_1$ or $s_2$.   
We also mention that, in the next-to-leading approximation, the functions $V_1$ and $V_2$ 
contain an additional dependence on $\ln \kappa$, which, again, does not 
contradict the Steinmann relations~\cite{NextLead}.     

Inserting these expressions into Eq.~(\ref{23analyticgeneral}), using 
(\ref{signature2}), and restricting ourselves to the planar approximation, 
we find for the real part (apart from the color factor):  
\beq
\frac{A_{2 \to 3}}{\Gamma(t_1)\Gamma(t_2)} =\frac{2s}{t_1t_2}
(|s_1|)^{\omega_1}\; g C(q_2,q_1) \; (|s_2|)^{\omega_2}
\label{23leadinglog}
\eeq  
in agreement with (4). In particular, the infrared singular pieces in 
$V_1$ and $V_2$ cancel.
As a further test, one could also compute, from the
corresponding unitarity equation, the single discontinuity in $s$: here both
partial waves $V_1$ and $V_2$ contribute, and the result is in agreement with
(\ref{V1}) and (\ref{V2})\footnote{We emphasize that the same results are obtained 
if one starts
from the double discontinuities in $s$ and $s_1$: using unitarity conditions
and making use of generalized bootstrap conditions, one again arrives at
(\ref{V1}). This is a crucial test of the selfconsistency of this
`unitarity-based approach'.}. 

\begin{figure}[ht]
\centerline{\epsfig{file=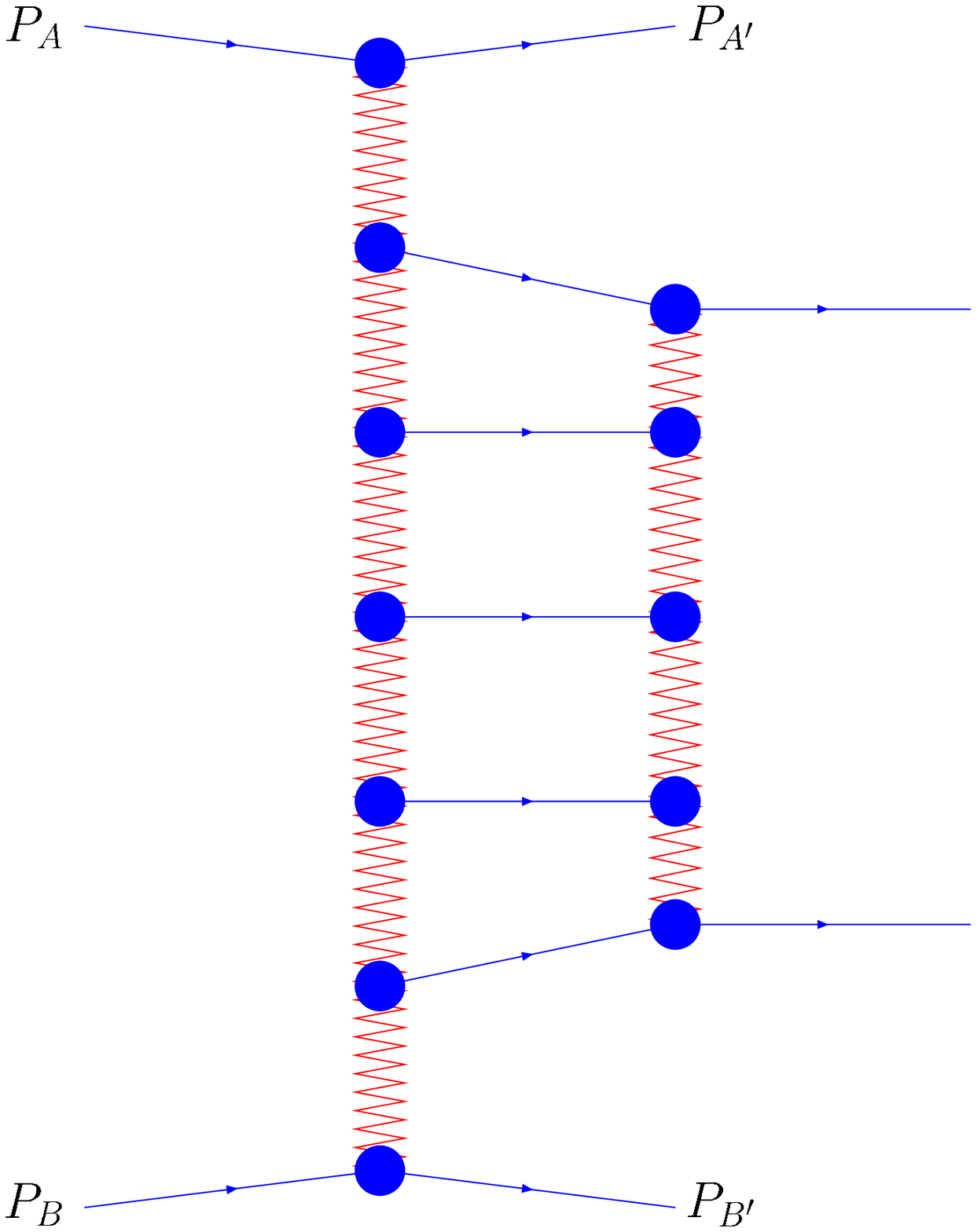,width=6cm,angle=0,bbllx=85,
bblly=240,bburx=540,bbury=800,clip=}\epsfig{file=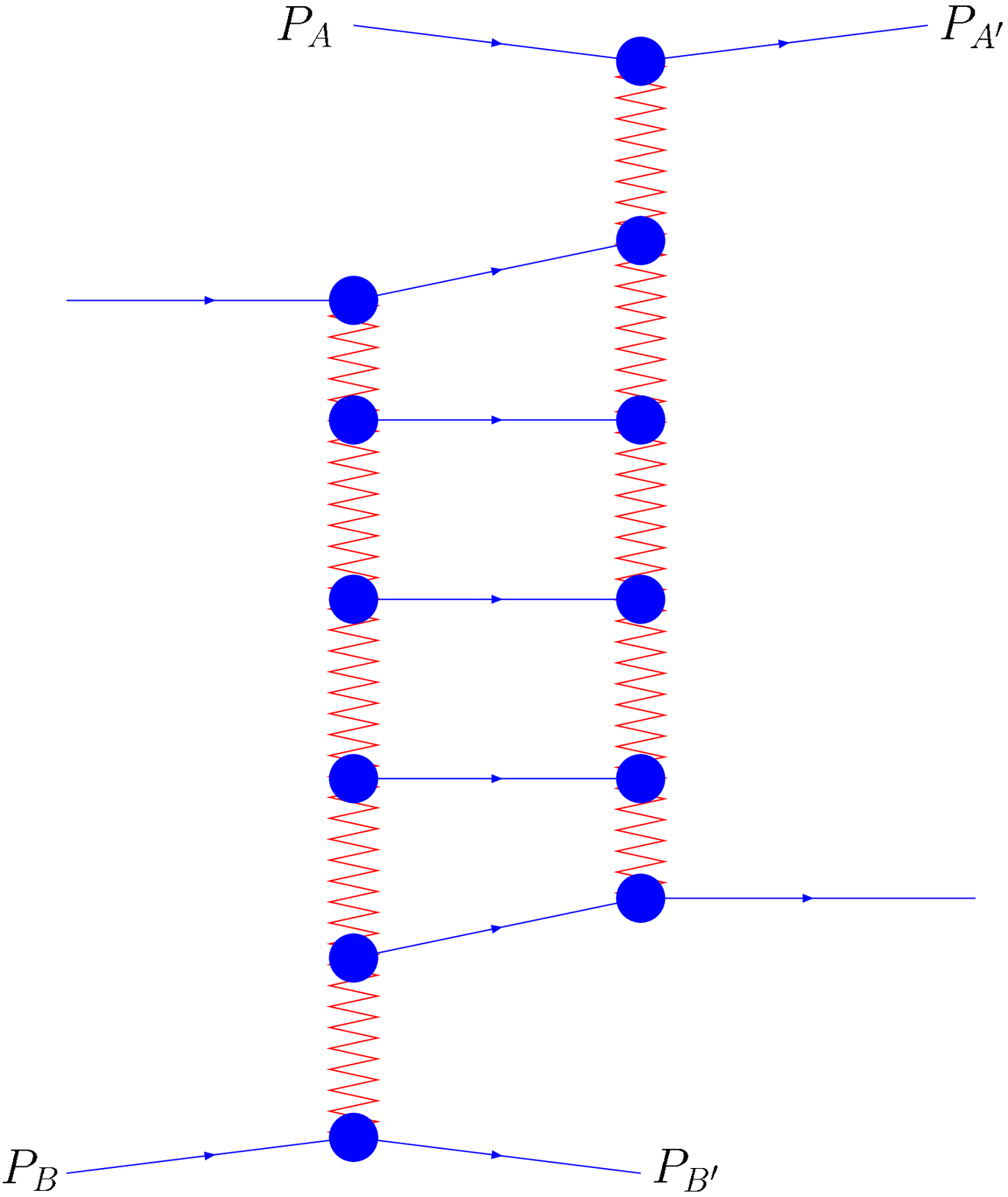,width=5.8cm,
angle=0,bbllx=0,bblly=250,bburx=430,bbury=800,clip=}}
\caption{BFKL contributions to the amplitudes $M_{2\rightarrow 4}$
and $M_{3\rightarrow 3}$}
\label{BFKLlad}
\end{figure}

For the $2 \to 4$ amplitude we start from an ansatz which is slightly more 
general than (\ref{24analyticpole}). In order to account for the Regge cut 
in the $t_2$ channel, we introduce, in the $t_2$-channel, the 
Sommerfeld-Watson integral $\int d \omega'_2/2\pi i$:
\begin{eqnarray}
\frac{A_{2\to4}}{\Gamma(t_1) \Gamma( t_3)} &=& \frac{2s}{t_1 t_2 t_3}
\int \frac{d \omega'_2}{2\pi i} \nonumber\\ 
&&\hspace{-3.6cm} \Big[\left(\frac{|s_1|}{\mu^2}\right)^{\omega_1-\omega'_2}
\left(\frac{|s_{012}|}{\mu^2}\right)^{\omega'_2-\omega_3}
\left(\frac{|s|}{\mu^2}\right)^{\omega_3}
\xi_{12'} \xi_{2'3} \xi_3 \,\,
\frac{W_1(t_1,t_2,t_3, \kappa_{12}, \kappa_{23};\omega'_2)}
{\sin \pi \omega_{12'} \sin \pi \omega_{2'3}} 
\nonumber \\  
&&\hspace{-3.6cm}+
\left(\frac{|s_3|}{\mu^2}\right)^{\omega_3-\omega'_2}
\left(\frac{|s_{123}|}{\mu^2}\right)^{\omega'_2-\omega_1}
\left(\frac{|s|}{\mu^2}\right)^{\omega_1}
\xi_{32'} \xi_{2'1} \xi_1\,\, 
\frac{W_2(t_1,t_2,t_3,\kappa_{12},\kappa_{23};\omega'_2)}
{\sin \pi \omega_{32'} \sin \pi \omega_{2'1}} 
\nonumber \\  
&&\hspace{-3.6cm}+
\left(\frac{|s_2|}{\mu^2}\right)^{\omega'_2-\omega_1}
\left(\frac{|s_{012}|}{\mu^2}\right)^{\omega_1-\omega_3}
\left(\frac{|s|}{\mu^2}\right)^{\omega_3}
\xi_{2'1} \xi_{13} \xi_3 \,\,
\frac{W_3(t_1,t_2,t_3, \kappa_{12}, \kappa_{23},\kappa_{123};\omega'_2)}
{\sin \pi \omega_{2'1} \sin \pi \omega_{13}} 
\nonumber \\  
&&\hspace{-3.6cm}+
\left(\frac{|s_2|}{\mu^2}\right)^{\omega'_2-\omega_3}
\left(\frac{|s_{123}|}{\mu^2}\right)^{\omega_3-\omega_1}
\left(\frac{|s|}{\mu^2}\right)^{\omega_1}
\xi_{2'3} \xi_{31} \xi_1 \,\,
\frac{W_4(t_1,t_2,t_3, \kappa_{12}, \kappa_{23},\kappa_{123};\omega'_2)}
{\sin \pi \omega_{2'3} \sin \pi \omega_{31}} 
\nonumber  \\  
&&\hspace{-3.6cm}+
\left(\frac{|s_3|}{\mu^2}\right)^{\omega_3-\omega'_2}
\left(\frac{|s_1|}{\mu^2}\right)^{\omega_1-\omega'_2}
\left(\frac{|s|}{\mu^2}\right)^{\omega_2}
\xi_{32'} \xi_{12'} \xi_{2'}\,\, 
\frac{W_5(t_1,t_2,t_3,\kappa_{12},\kappa_{23};\omega'_2)}
{\sin \pi \omega_{32'} \sin \pi \omega_{12'}} 
 \Big].
\label{24analyticgeneral}
\end{eqnarray}
with
the partial wave functions, denoted by $W_{i=1,2,3,4,5}$ and to be determined 
from single energy discontinuity equations. 
The partial wave functions $W_3$ and $W_4$ 
also depend upon the additional variable $\kappa_{123}=(\vec{k}_1+\vec{k}_2)^2$.    
We have the five single
discontinuities in $s_1$, $s_2$, $s_3$, $s_{012}$, and $s_{123}$
which allow to find the partial waves $W_{i=1,2,3,4,5}$.
In leading log accuracy the results are the following:
\begin{eqnarray}
W_1 &=& V_1(t_1,t_2,\kappa_{12}) \frac{1}{\omega'_2 - \omega_2} 
V_1(t_2,t_3,\kappa_{23}),
\label{W1}\\
W_2 &=& V_2(t_1,t_2,\kappa_{12}) \frac{1}{\omega'_2 - \omega_2} 
V_2(t_2,t_3,\kappa_{23}),
\label{W2} \\
W_3 &=& \frac{\sin \pi \omega_1}{\sin \pi \omega_2} V_2(t_1,t_2,\kappa_{12}) 
\frac{1}{\omega'_2 - \omega_2} 
V_1(t_2,t_3,\kappa_{23})\nonumber\\
&- &\sin \pi (\omega_2' - \omega_1)
\left(V_{\rm cut}-V_p \right),
\label{W3}\\
W_4 &=& \frac{\sin \pi \omega_3}{\sin \pi \omega_2} V_2(t_1,t_2,\kappa_{12}) 
\frac{1}{\omega'_2 - \omega_2} 
V_1(t_2,t_3,\kappa_{23}) \nonumber\\
&- & \sin \pi (\omega_2' - \omega_3) 
\left(V_{\rm cut}-V_p \right),
\label{W4}\\
W_5 &=& V_1(t_1,t_2,\kappa_{12}) \frac {1}{\omega'_2 - 
\omega_2}V_2(t_2,t_3,\kappa_{23}).
\label{W5}
\end{eqnarray}
The three amplitudes $W_1$, $W_2$ and $W_5$ are products of the
production vertices $V_i$ in (\ref{V1}) and (\ref{V2}), found in the
$2 \to 3$ case, whereas the amplitudes
$W_3$ and $W_4$ contain, in addition to the products of production
vertices $V_i$, the extra pieces, $V_{\rm cut}-V_p$ which will be defined 
in the following. The term $V_{\rm cut}$ contains
Regge cuts and cannot be written as a simple product of vertices for the
two produced gluons. It takes the form of BFKL-like ladder 
diagrams in the color octet   
channel, and it is illustrated in Fig. \ref{BFKLlad} (left figure):
\begin{eqnarray}
V_{\rm cut} &=& \frac{t_2 N_c}{2} g^4 \int \frac{d^2k d^2k'}{(2\pi)^6} 
        \frac{q_1^2}{(k+k_1)^2} C(k,k+k_1) \nonumber\\
&\times&
G^{(8_A)}(k,q_2-k;k',q_2-k';\omega_2') 
C(k'-k_2,k') \frac{q_3^2}{(k'-k_2)^2}.
\label{Vcut}
\end{eqnarray}
Here $C$ denotes the effective Reggeon-Reggeon-gluon vertex given in (6), and 
$G^{(8_A)}$ is the BFKL Green's function in the color octet channel,
satisfying the integral equation
\begin{eqnarray}
\omega_2' G^{(8_A)}(k,q-k;k',q-k';\omega_2') &=& \nonumber\\ 
&&\hspace{-6cm}\frac{(2\pi)^3 \delta^{(2)}(k-k')}{k^2 (q-k)^2} 
+ \frac{1}{k^2 (q-k)^2} \left(K\otimes G^{(8_A)} \right)(k,q-k;k',q-k'),
\label{BFKLoctet}
\end{eqnarray}
where $K$ denotes the BFKL kernel in the color octet channel, containing 
both real emission and the gluon trajectory.
In lowest order in the coupling, and for equal helicities of the two 
produced gluons, $V_{\rm cut}$ equals:
\begin{equation}
V_{\rm cut}^{(0)} = g^2 \frac{C(q_2,q_1) C(q_3,q_2)}{2 \omega'_2} \left[ \omega_1 + \omega_3
+a \left(\ln \frac{\kappa_{123}}{\mu ^2}  - \frac{1}{\epsilon}\right) \right].
\label{Vcut0}
\end{equation}
The term $V_{\rm cut}$ not only violates the simple factorization of Regge pole models, 
but also, when computed beyond the one loop approximation, will be shown to
disagree with the BDS formula. Finally, the subtraction term $V_{p}$ removes 
the Regge pole piece inside $V_{\rm cut}$, and it is of the form:
\begin{eqnarray}
\label{Vpole}
 V_p= g^2 \frac{C(q_2,q_1) C(q_3,q_2)}{4\omega_2} \left[\omega_1 +\omega_2 + 
a \left(\ln \frac{\kappa_{12}}{\mu ^2} - \frac{1}{\epsilon}\right) \right]
                      \frac{1}{\omega_2' -\omega_2}\nonumber \\
                          \cdot \left[\omega_2 +\omega_3 
+ a \left(\ln \frac{\kappa_{23}}{\mu ^2} - \frac{1}{\epsilon}\right)\right].
\end{eqnarray}

Before we compare with the BDS formula, let us remark on a few features
of these leading order QCD results (for details, see Appendix E). 
From now on, we specialize on the planar approximation, i.e. in the 
signature factors in eq.(\ref{signature2}) we only retain the phases. 
Inserting the results of
(\ref{W1}) - (\ref{W5}) into the full amplitude (\ref{24analyticgeneral})
we can derive the results for different kinematic regions. 

Beginning with 
the physical region where all energies are positive, one finds  
that the sum of the Regge pole terms can be written in the simple 
factorizing form (\ref{24factor}). In particular, the Regge cut pieces contained 
in $W_3$ and $W_4$ cancel completely, and the real part of the scattering 
amplitude coincides with (4).

Next, in the unphysical region where all energies are negative and 
all phases disappear, again, the Regge pole contributions can be written 
in a simple factorizing form, and the cut pieces in $W_3$ and $W_4$ cancel.

Finally, we go into the physical region where $s, s_2 > 0$ and 
$s_1,s_3, s_{012}, s_{123} <0$. Nonzero phases appear only in $s$ and in $s_2$.
After some algebra we obtain:
$$\frac{A_{2 \to 4}}{\Gamma(t_1) \Gamma(t_3)}=\frac{2s}{t_1 t_2 t_3} 
g^2 C(q_2,q_1) C(q_3,q_2) 
(|s_1|)^{\omega_1} (|s_3|)^{\omega_3} \cdot $$
\beqn 
e^{-i \pi \omega_2} (|s_2|)^{\omega_2} 
\Big[1+ i \frac{\pi}{2} \left(
\omega_1+\omega_2 + 
 a \left( \ln \frac{\kappa_{12}}{\mu^2} -\frac{1}{\epsilon}\right) 
+ \omega_3 + \omega_2 + 
 a \left( \ln \frac{\kappa_{23}}{\mu^2} -\frac{1}{\epsilon} \right) 
\right) \Big] \nonumber \\ 
- 2 i \pi \frac{2s}{t_1t_2t_3} \int \frac{d \omega'_2}{2 \pi i} 
(e^{-i \pi} |s_2|)^{\omega'_2} V_{cut}.
\label{A2to4phase} 
\eeqn
In the last term, $V_{cut}$, it is possible to factor out the gluon 
trajectory (details are presented in ~\cite{BLS}):
\beq
\int \frac{d \omega'_2}{2 \pi i} (e^{-i \pi} |s_2|)^{\omega'_2} V_{cut}  
= g^2 C(q_2,q_1) C(q_3,q_2) (e^{-i \pi} |s_2|)^{\omega_2} \int \frac{d \omega'_2}{2 \pi i} 
(e^{-i \pi} |s_2|)^{\omega'_2} V_{cut, reduced},
\label{Vreduction}
\eeq
where in the one loop approximation (\ref{Vcut0})
\beqn
g^2 C(q_2,q_1) C(q_3,q_2) \int \frac{d \omega'_2}{2 \pi i} 
(e^{-i \pi} |s_2|)^{\omega'_2} V_{cut, reduced} \nonumber \\
= g^2 \frac{C(q_2,q_1) C(q_3,q_2)}{2} 
\left[ a \left( \ln \frac{\kappa_{123}\mu^2}{q_1^2 q_3^2} + \frac{1}{\epsilon}\right) +
{\cal O}(a^2 \ln s_2) \right],
\eeqn
and the two loop and higher order terms of $V_{cut, reduced}$ are infrared 
finite~\cite{BLS}.

Inserting (\ref{Vreduction}) into (\ref{A2to4phase}) we see that 
all terms on the rhs of (\ref{A2to4phase}) are proportional to the 
common phase factor $ e^{-i \pi \omega_2}$:
$$\frac{A_{2 \to 4}}{\Gamma(t_1) \Gamma(t_3)}=\frac{2s}{t_1 t_2 t_3} 
(|s_1|)^{\omega_1} (|s_3|)^{\omega_3} (|s_2|)^{\omega_2} 
 g^2 C(q_2,q_1) C(q_3,q_2) e^{-i \pi \omega_2} \cdot $$
\beqn 
\left[ 1+ i \frac{\pi}{2} \left(
\omega_1+\omega_2 + 
 a \left( \ln \frac{\kappa_{12}}{\mu^2} -\frac{1}{\epsilon}\right) 
+ \omega_3 + \omega_2 + 
 a \left( \ln \frac{\kappa_{23}}{\mu^2} -\frac{1}{\epsilon} \right) 
\right)  \right.\nonumber \\ \left. 
- 2 i \pi \int \frac{d \omega'_2}{2 \pi i} (e^{-i \pi} |s_2|)^{\omega'_2} 
V_{cut,reduced} \right],
\label{A2to4phaseexp} 
\eeqn
and the coefficient in the square brackets is infrared finite.
This shows that, in LLA, the imaginary part of $A_{2 \to 4}$ 
is infrared singular, but the singularities 
are assembled in the phase factor $ e^{-i \pi \omega_2}$. 
This observation will be important when comparing with the BDS formula.
    
A completely analogous discussion applies to the case $3 \to 3$ 
(Figs.~\ref{kin3to3},~\ref{Dispers3to3}) in the 
multi-Regge region (a more detailed 
discussion is given in Appendix E). Again, the scattering 
amplitude consists of five terms, and two of them contain the Regge cut piece:
$$
\int \frac{d \omega'_2}{2 \pi i} (e^{-i \pi} |s_2|)^{\omega'_2} U_{\rm cut} 
$$
with  
\begin{eqnarray}
U_{\rm cut} &=& \frac{t_2 N_c}{2} g^4 \int \frac{d^2k d^2k'}{(2\pi)^6} 
        \frac{q_1^2}{(k-q_1)^2} C(q_2-k,q_1-k) \nonumber\\
&\times&
G^{(8_A)}(k,q_2-k;k',q_2-k';\omega_2') 
C(k'-k_2,k') \frac{q_3^2}{(k'-k_2)^2}.
\label{Ucut}
\end{eqnarray}
In lowest order (and for equal helicities of the produced gluns) 
this Regge cut piece equals:  
\begin{equation}
U_{\rm cut}^{(0)} = \frac{g^2 C(q_2,q_1) C(q_3,q_2)}{2\omega'_2}  
a \ln \frac{\kappa_{12}\kappa_{23}}{(\vec{q}_1+\vec{q}_3-\vec{q}_2)^2 q_2^2}.
\label{Ucut0}
\end{equation}

Proceeding in the same way as for the $2 \to 4$ amplitude, 
on derives results for the scattering amplitude in the different 
kinematic regions. In the region
where all energies are positive, we find the same factorization 
as for simple Regge pole models, i.e. 
the Regge cut pieces cancel. 
In the region $s,s_2>0$, $s_1,s_3,s_{13},s_{02}<0$ 
the Regge cut piece appears. First we rewrite it in the same form as in 
(\ref{Vreduction}):
 \beq
\int \frac{d \omega'_2}{2 \pi i}
(e^{-i \pi} |s_2|)^{\omega'_2} U_{cut} 
= g^2 C(q_2,q_1) C(q_3,q_2) (e^{-i \pi} |s_2|)^{\omega_2} \int \frac{d \omega'_2}{2 \pi i}
(e^{-i \pi} |s_2|)^{\omega'_2} U_{cut,reduced}
\label{Ureduction}
\eeq
with the infrared finite one loop approximation 
\beqn
 g^2 C(q_2,q_1) C(q_3,q_2) \int \frac{d \omega'_2}{2 \pi i}
(e^{-i \pi} |s_2|)^{\omega'_2} U_{cut,reduced} \nonumber \\ =
g^2 \frac{C(q_2,q_1) C(q_3,q_2)}{2} 
\left[ a \left(\ln \frac{\kappa_{12}\kappa_{23}}{(\vec{q}_1+\vec{q}_3-\vec{q}_2)^2 q_2^2}\right)
+ {\cal O}(a^2 \ln s_2) \right].
\eeqn 
As in the case of the $2 \to 4$ amplitude, the higher order corrections 
(denoted by ${\cal O}(a^2)$) are infrared finite. 
With this result, the $3 \to 3$ amplitude can be written in the form 
(cf.(\ref{A2to4phaseexp})):  
$$\frac{A_{3\to3}}{\Gamma(t_1) \Gamma( t_3)} = \frac{2s}{t_1 t_2 t_3} 
(|s_1|)^{\omega_1} (|s_2|)^{\omega_2} (|s_3|)^{\omega_3} g^2 C(q_2,q_1) C(q_3,q_2)\cdot$$
\beqn
\left[ 1+ i \frac{\pi}{2} \left(
\omega_1+ 
 a \left( \ln \frac{\kappa_{12}}{\mu^2} -\frac{1}{\epsilon}\right) 
+ \omega_3 + 
 a \left( \ln \frac{\kappa_{23}}{\mu^2} -\frac{1}{\epsilon} \right) 
\right)  \right. \nonumber \\ \left. 
- 2 i \pi \int \frac{d \omega'_2}{2 \pi i} (e^{-i \pi} |s_2|)^{\omega'_2} 
U_{cut,reduced} \right].
\label{A3to3phaseexp} 
\eeqn 
Note that, in contrast to the $2\to4$ case, there is no common 
infrared singular phase factor, $e^{-i\pi \omega_2}$, and the square bracket term on the 
rhs of (\ref{A3to3phaseexp}) is infrared finite.
This shows that the infrared structure of $A_{3 \to 3}$ is quite 
different from  $A_{2 \to 4}$.
 
\subsection{Comparison with the BDS formula}

Let us now return to the BDS amplitude discussed in the section 3,
take the leading logarithmic approximation and compare with the results 
discussed in the previous subsection.
In the leading logarithmic approximation we retain, in the exponent 
$\ln \,M_n$, only the lowest order (in powers of $a$) of the coefficients 
of the energy logarithms, and the lowest order of the real and imaginary parts 
of logarithms of the vertex functions, $\ln\, \Gamma$. 
In the case of the $2 \to 2$ scattering 
process the coefficient of $\ln s$ is given by the gluon trajectory  
function (eq.(\ref{gluontrajectory})), and the leading order coefficient
is the term proportional to $a$. The logarithm of the vertex function is 
given in (\ref{ReggeonParticlevertex}); the lowest order term is  
of order $a$, and since $t$ is negative, there is 
no imaginary part. Therefore, in the leading logarithmic approximation         
we put $\ln \Gamma$ equal to zero (note that $M_4$ multiplies the 
Born approximation which contains a reggeon-particle-particle vertex of the 
order $g$). 
 
Turning to the case $2 \to 3$ in the physical region, we use   
(\ref{M2a3mregge}) and (\ref{Gammavertex}) (see also Appendix B). 
The new element, the logarithm of the production vertex, starts with 
terms of the order $a$, and the real part can be neglected (i.e. the absolute value of 
$\Gamma(t_2,t_1,\kappa)$  can be put equal to unity). But, depending 
upon the kinematic region, terms with $\ln (-\kappa)$ may lead to imaginary 
parts of order $a$ which have to be kept. In the region where all energies are positive the 
relevant terms of order $a$ are (see (\ref{phigamma}):
\beq
\Phi_{\Gamma}= \frac{1}{2} (\omega(t_1) + \omega(t_2)) + 
\frac{a}{2} \left( \ln \frac{\kappa}{\mu^2} - \frac{1}{\epsilon} \right).
\eeq   
In order to compare with the QCD results we use (\ref{Mdosatres}), 
(\ref{ces1}), and (\ref{ces2}). In (\ref{Mdosatres}) we approximate 
the factors $\kappa ^{\omega(t_1)} \to 1$ etc, and for the real coefficients 
$c_1$ and $c_2$ we obtain:
\beq
c_1 = \frac{\frac{1}{2} \left(\omega(t_1) - \omega(t_2) 
- a (\ln \frac{\kappa}{\mu^2} -
\frac{1}{\epsilon}) \right)}{\omega(t_1) - \omega(t_2)},\,\, 
c_2 = \frac{\frac{1}{2} \left(\omega(t_2) - \omega(t_1) 
- a (\ln \frac{\kappa}{\mu^2} -
\frac{1}{\epsilon}) \right)}{\omega(t_2) - \omega(t_1)}
\eeq   
which agrees with the leading log result in (\ref{23analyticgeneral}),
(\ref{V1}), and (\ref{V2}) .
In the unphysical region where all energies are negative we have no imaginay 
parts and again find complete agreement with the results of the previous 
subsection. 

In the case of $2 \to 4$ we begin with the physical region
where all energies are positive. Using eqs.(\ref{Mdosacuatro}) - (\ref{dcoefficients}) and applying 
the same arguments as for the $2 \to 3$ case, we find 
\beqn
\frac{M_{2\to4}}{\Gamma(t_1)\Gamma(t_3)} =  (e^{-i\pi}\frac{|s_1|}{\mu^2})^{\omega_1} 
\left( c_1(\kappa_{12}) +c_2(\kappa_{12}) \right)   
(e^{-i\pi}\frac{|s_2|}{\mu^2})^{\omega_2} \nonumber \\
\left( c_1(\kappa_{23}) +c_2(\kappa_{23}) \right)
(e^{-i\pi}\frac{|s_3|}{\mu^2})^{\omega_3},
\eeqn 
quite in agreement with the LLA of the Regge pole part in (\ref{24polepart_E}). 
Since, in the QCD calculation for this kinematic region, the Regge cut pieces cancel, 
there is no conflict between the BDS formula and the leading logarithmic 
approximation obtained by direct calculations.  

Let us now turn to the region
$s,\,s_2>0$, $s_{1}\,, s_{012}\,,s_{123}\,,s_3<0$ where, in the QCD calculations,
the imaginary part contains the factorization breaking term $V_{\rm cut}$
(corresponding to a BFKL ladder with the octet quantum numbers in the $t$-channel). 
In the BDS amplitude (\ref{24mixedregion}) we have, 
compared to the physical region with only positive energies, the additional phase factor $C$ 
in (\ref{coeffC}). In the leading log approximation which we have described 
before we find (from (\ref{24mixedregion}), (\ref{coeffC}), or from (C.11)):
$$\frac{M_{2\to 4}}{\Gamma(t_1) \Gamma(t_3)} = 
\left(\frac{|s_1|}{\mu^2}\right)^{\omega_1}
\left(\frac{|s_3|}{\mu^2}\right)^{\omega_3} 
\left(e^{-i \pi}\frac{|s_2|}{\mu^2}\right)^{\omega_2} \cdot$$
\beqn
\Big[1 + i \frac{\pi}{2}\left(
\omega_1+\omega_2 + 
 a \left( \ln \frac{\kappa_{12}}{\mu^2} -\frac{1}{\epsilon}\right) 
+ \omega_3 + \omega_2 + 
 a \left( \ln \frac{\kappa_{23}}{\mu^2} -\frac{1}{\epsilon} \right) \right)
\nonumber\\
+ i \pi a \left(\ln \frac{q_1^2 q_3^2}{(k_1+k_2)^2 \mu ^2}  - \frac{1}{\epsilon}\right) ) \Big].   
\label{BDS24phase}
\eeqn  
In the imaginary part of the square brackets the $\epsilon$ poles cancel.  
Comparing this result with (\ref{A2to4phaseexp}) and (\ref{Vcut0}) we see that the BDS formula reproduces 
the lowest order term of the Regge cut contribution, $V_{cut,reduced}$, but not the
higher order terms (which are still part of the leading logarithmic approximation).
We therefore conclude that, beyond the one loop approximation, the BDS formula 
does not agree with the leading log results listed in the previous subsection.

Let us remark on the order $\cal{O}(\epsilon)$ corrections in the BDS formula.
As explained at the beginning of section 3, our analysis of the BDS formula 
(which applies to the logarithm of the scattering amplitude) does not include terms which 
vanish as $\epsilon \to 0$. Nevertheless, the comparison of (\ref{A2to4phaseexp}) and 
(\ref{BDS24phase}) shows that such corrections cannot reproduce the finite (in $\epsilon$)
terms which are missing in the BDS formula. The key point is that, in the BDS formula, 
the leading log approximation for the imaginary part of $\ln M_{2 \to 4}$  
contains terms of the order $1/\epsilon$ only inside $\omega_2$. 
Comparing (\ref{A2to4phaseexp}) with (\ref{BDS24phase}) one sees that 
the infrared divergent phase factor for the cut contribution in 
(\ref{A2to4phaseexp}) is the same as in the BDS formula.
Therefore, when going from $\ln M_{2 \to 4}$ to $M_{2 \to 4}$ it is incorrect 
to expand this infrared singular piece $e^{-i \pi a/\epsilon}$, and 
it becomes clear that terms of order $\epsilon$ in the logarithm of the 
scattering amplitude cannot produce constant (in $\epsilon$) terms in the 
scattering  amplitude. As a result, our 
conclusion concerning the validity of the BDS formula is not affected by the order 
$\cal{O}(\epsilon)$ corrections in the BDS formula for the logarithm of the 
scattering amplitude.

For the $3 \to 3$ amplitude the comparison 
between the BDS amplitude and the high energy behavior in Yang Mills theories leads 
to the same conclusion, although some details are different.    
For the kinematic region where all energies are positive the BDS formula agrees with 
the leading log calculations, and we directly turn to the region  
$s,s_2>0$, $s_1,s_3,s_{13},s_{02}<0$. The crucial element is the phase $C'$ in (\ref{coeffC'}) 
which, in contrast to $C$ in (\ref{coeffC}), is infrared finite.
Collecting, in (\ref{33mixedregion}), (\ref{coeffC'}), or in (D.89), 
within LLA, all imaginary parts in $\ln M_{3 \to 3}$ we note that all terms of the 
form $a/\epsilon$ cancel, and we arrive at:
$$\frac{M_{3\to 3}}{\Gamma(t_1) \Gamma(t_3)} = 
\left(\frac{|s_1|}{\mu^2}\right)^{\omega_1}
\left(\frac{|s_3|}{\mu^2}\right)^{\omega_3} 
\left(\frac{|s_2|}{\mu^2}\right)^{\omega_2} \cdot$$
\beqn
\Big[1 + i \frac{\pi}{2}\left(
\omega_1+ 
 a \left( \ln \frac{\kappa_{12}}{\mu^2} -\frac{1}{\epsilon}\right) 
+ \omega_3 + 
a \left( \ln \frac{\kappa_{23}}{\mu^2} -\frac{1}{\epsilon} \right) \right)
- i \pi a \left(\ln \frac{\kappa_{12}\kappa_{23}}{(\vec{q}_1+\vec{q}_3-\vec{q}_2)^2 q_2^2} \right) \Big].   
\label{BDS33phase}
\eeqn  
The square bracket expression is infrared finite. Comparison with 
(\ref{A3to3phaseexp}) shows that the BDS formula correctly reproduces the one loop 
approximation to $U_{cut,reduced}$, but not the higher order loops. Again, terms of order 
$\epsilon$ in $\ln M_{3 \to 3}$ cannot reproduce those finite (in $\epsilon$) terms which are missing in $M$.  

Discrepancies in the BDS finite pieces for six gluon amplitudes,
starting at two loops, were also hinted in~\cite{Drummond:2007bm} where 
the equivalence between Wilson loops and MHV amplitudes was assumed.
In a particular kinematic configuration, and for a very large number of
external gluons at strong 't Hooft coupling, the finite pieces of the BDS
ansatz failed when compared to the results of~\cite{Alday:2007he}.

\section{Conclusions}

\begin{figure}[ht]
\centerline{\epsfig{file=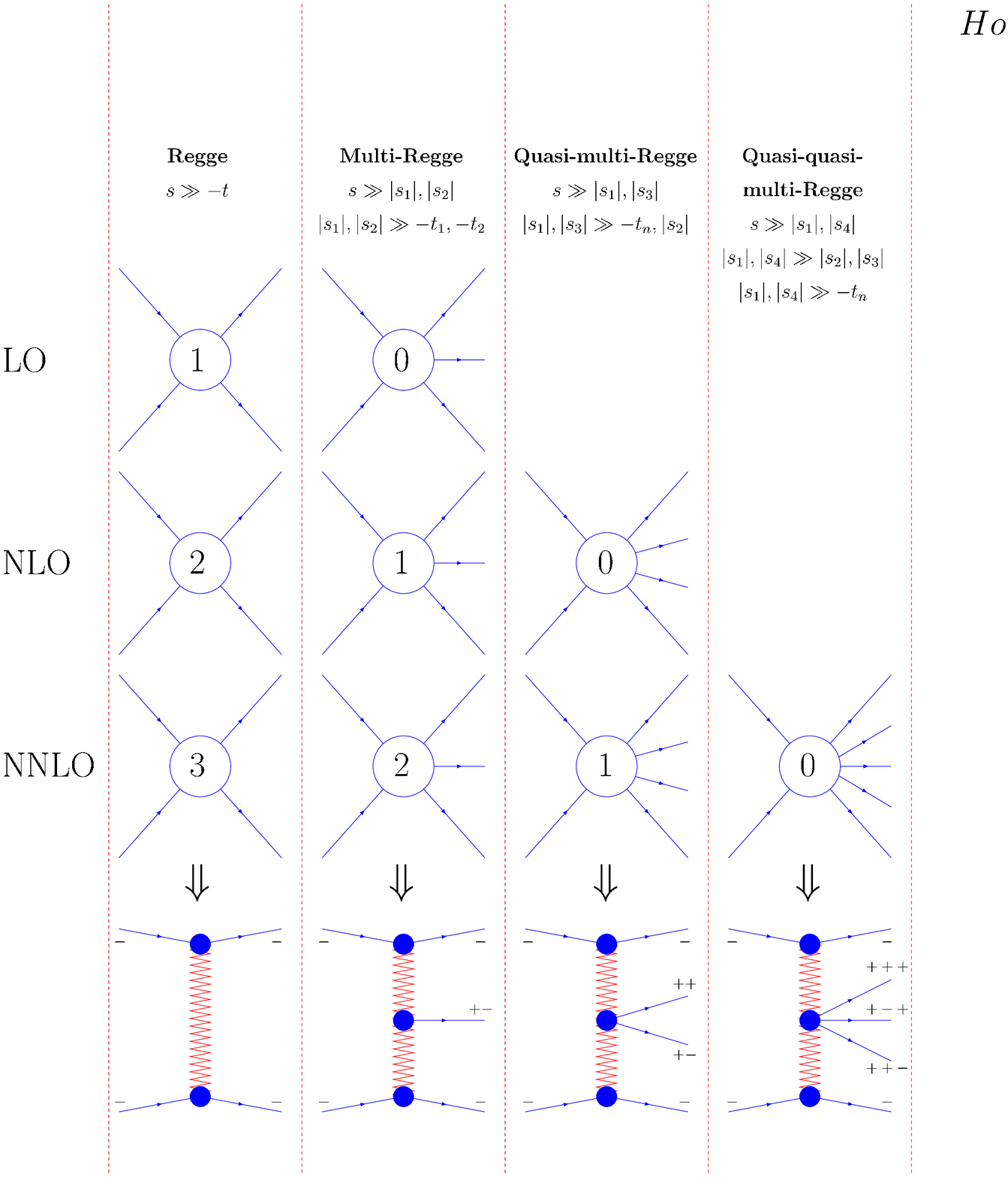,width=14cm,angle=0,bbllx=0,
bblly=50,bburx=570,bbury=700,clip=}}
\caption{Diagrams contributing to the BFKL kernel in NNLLA}
\label{Alldia}
\end{figure}

In this paper we have assembled the ingredients needed for the
three loop corrections (NNLO) to the BFKL kernel in $N=4$ SYM theory at
large $N_c$. Following earlier calculations
we can obtain the kernel from unitarity sums, {\it i.e.} by computing squares
of production amplitudes, keeping in mind that at large $N_c$ the contributing diagrams 
belong to the
cylinder topology. Figure \ref{Alldia} illustrates the production vertices which enter
the three loop
calculation. 

Elements in the first two lines are known, whereas the
building blocks in the third line are new: they can be (and partly have been)
computed from the effective action summarized in section 2.

For most of the cases, also the BDS formula can be used.
The NNLO gluon trajectory function follows from the $2 \to 2$ scattering 
amplitude (first column); details are described in Appendix A.
For the $2\to3$ case (second column), the two loop approximation to the gluon
production vertex can be read off from the analysis presented in section
3 (c.f. Eq.~(\ref{Gammavertex})). In this case one should take into account the
$\epsilon$-corrections to the BDS amplitude.  In column 
3, we should take into account the  
the Reggeon + Reggeon $\to 2$ gluon vertex
in the one loop approximation for fixed invariant masses of produced gluons. 
Based upon 
the analysis carried out
in sections 3 and 4 we trust that the BDS formula for the maximal  
helicity violating case can be used (with $\epsilon$-corrections). For the 
non-maximal violating cases
in column 3 one can use the results of~\cite{BDK}. Finally, in column 4
we encounter the Born vertex: Reggeon + Reggeon $\to 3$ gluons for
the fixed invariant mass of the gluons. This vertex
has been obtained in~\cite{last} by means of the effective action 
(see also Ref.~\cite{Del Duca:1999ha}).

We have shown that the BDS amplitude $M_{2\rightarrow 3}$ in the multi-Regge
kinematics satisfies the
dispersive representation, which is valid in all physical regions and is 
compatible with the Steinmann relations and gluon reggeization. For the case of the 
gluon transitions $2\rightarrow 4$ and $3\rightarrow 3$, in the 
multi-Regge kinematics and in the physical 
region where $s,s_2>0$ and $s_1,s_3<0$, the Regge factorization of the
BDS amplitude is badly violated. In the one loop aproximation the BDS result
in this region coincides with the direct QCD calculations, but in
higher loops we have shown that these amplitudes should contain the
Mandelstam Regge cut in the $t_2$-channel. It was demonstrated, that this cut 
is absent in the BDS expression and cannot be reproduced by the
${\cal O}(\epsilon)$-corrections to this expression. 

\begin{figure}[ht]
\centerline{\epsfig{file=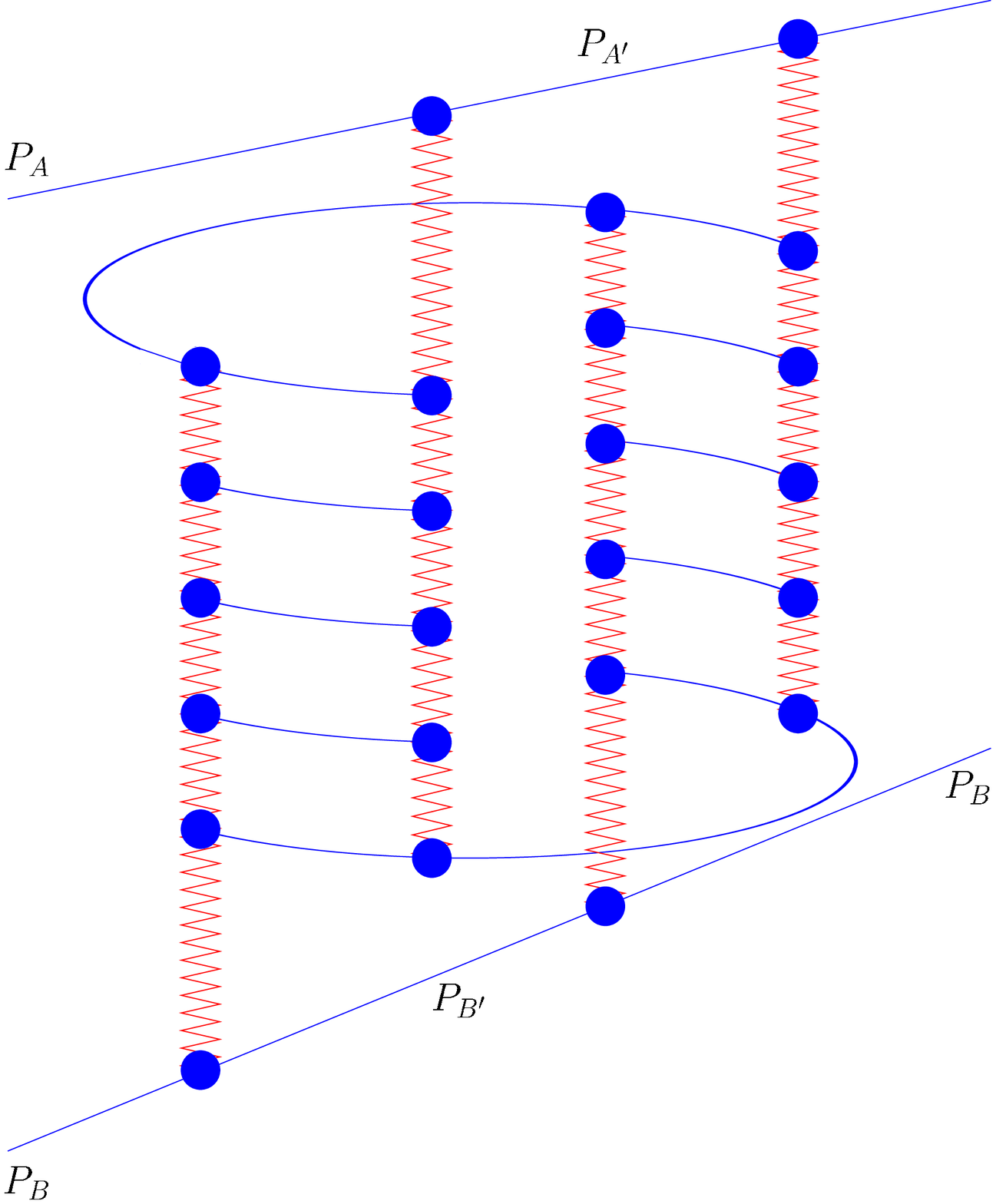,width=7cm,angle=0,bbllx=1,bblly=150,bburx=590,bbury=840,
clip=}\epsfig{file=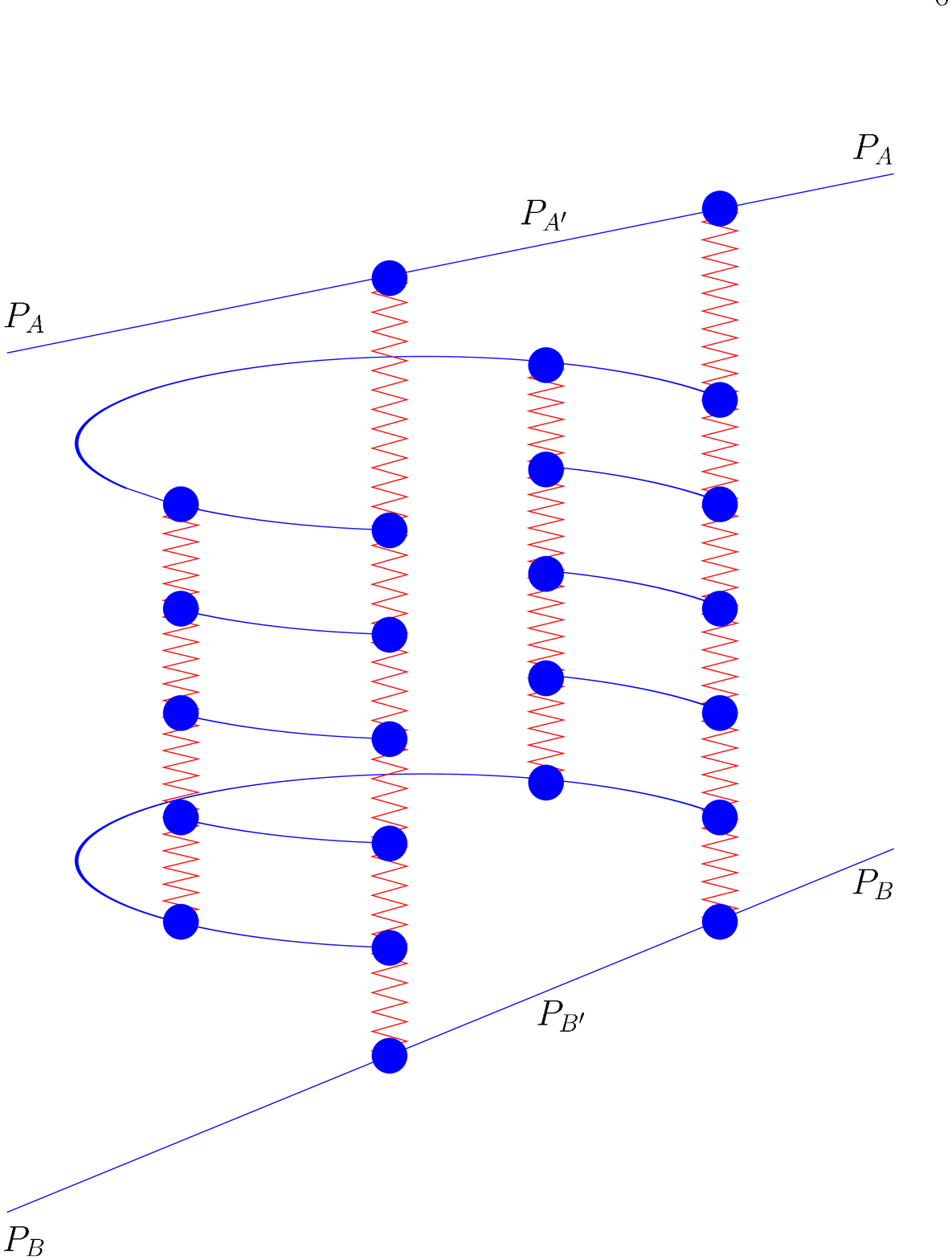,width=7cm,angle=0,bbllx=0,bblly=40,bburx=580,bbury=750,clip=}}
\caption{Cylinder-type topologies in the unitarity sums for 
the total cross section: the intermediate states (discontinuity cuts) 
are obtained by slicing the cylinders in all possible ways across the intermediate 
momenta $p_{A'}$ 
and $p_{B'}$}
\label{Cylinder}
\end{figure}

A remark is in place on the Regge-cut contribution
illustrated in Fig.~\ref{BFKLlad} and discussed in section 4. In
addition to the corrections to the production vertex functions
which are illustrated in Fig.~\ref{Alldia}, 
we still have to take into account those   
corrections to the production amplitude in the multi-Regge limit which
do not fall into the class of loop corrections to the production vertices:
in NLO these are just the Regge-cut contributions to the imaginary part  
in the $2 \to 4$ and the $3 \to 3$ cases which we have discussed in the
previous section. The diagrams
contributing to the BFKL Pomeron in the large $N_c$ limit belong to the
cylinder topology: two examples are illustrated in Fig.~\ref{Cylinder}, 
and, to begin with, we consider the discontinuity due to the 4-particle 
intermediate state.
In the left figure, on both sides of the discontinuity cut, we have the 
$3 \to 3$ production amplitudes continued into the physical region of a $2 \to 4$ process 
(c.f. rhs of Fig.~\ref{BFKLlad}), and in the right 
hand figure we recognize a configuration where the $2 \to 4$ amplitude has to 
be evaluated in a region with negative energies.  
As discussed before, in the latter case the non-factorizing pieces of the $2 \to 4$  
production amplitude do not cancel.  
If these contributions would survive in the total cross section, the NNLO BFKL
Pomeron would receive a four-reggeon cut contribution, and the simple ladder
structure would be lost. There are, however, reasons to expect that,
in the large $N_c$ limit, the sum of these contributions might cancel 
in the total cross section. Namely, in
addition to the contribution of the 4-particle intermediate state, we also need      
other cuts, which, for example, run across one of the ladders or along one of the reggeized gluons 
above or below the cylinder. These different cuts provide similar contributions, but 
they come with different signs. 
It is likely that, similar to the AFS cancellation of Regge cuts in planar amplitudes, the 
four reggeon cut contributions cancel in the sum. We will study
this in the subsequent part of our investigation.\\ \\
{\bf Acknowledgements:} We wish to thank V.~S.~Fadin, M.~Strassler, and C.~I.~Tan 
for very helpful discussions. 
Part of this work has been done while one of us 
(ASV) has been visiting the II.~Institut f.~Theoretische Physik, 
University Hamburg. The hospitality is gratefully acknowledged.
One of us (LNL) wishes to thank the Issac Newton Institute for the 
invitation to participate in the program 
``String Fields, Integrability and Strings''. LNL is supported by the  
RFBR grants 06-02-72041-MNTI-a, 07-02-00902-a, andRSGSS-5788.2006.2.\\ \\
{\bf Note to be added:} After our paper had been submitted, another study 
appeared which, in some parts, parallels our investigation~\cite{Brower2}. Like 
ours, it studies several different Regge limits of the 
BDS amplitudes. In the unphysical region (negative energies),  
the results on the energy dependence are fully consistent with ours. 
In contrast to our paper, however, in~\cite{Brower2} the continuation 
into physical regions has not been investigated, and the conflict with 
QCD calculations was not found. Recently another paper of these 
authors~\cite{Brower3} has appeared. We completely disagree with 
the statement in their section 4.3, saying that the Steinmann relations for
the BDS amplitude $M_5$ are violated. In our view, the authors compute
the discontinuity of the BDS amplitude (4.16) in an incorrect way: for example,
the discontinuity in $s_1$ in the 
physical regions should be defined at fixed $\kappa$  
(this follows already from a simple one loop calculation in QCD where,
in physical regions, the presence of simultaneous singularities 
in the overlapping channels $s_1$ and $s_2$ would contradict the gluon 
stability). 
We also find their section 4.6 very confusing. First, contrary to their
statement, in our paper the dispersive representation was not used to prove
the absence of the Regge factorization of the BDS amplitudes 
$M_{2\rightarrow 4}$ and $M_{3\rightarrow 3}$ in the physical region
with $s,s_2>0$ and $s_1,s_3<0$. We simply analytically continued the
BDS formula to this region and compared with the QCD results. 
Such a continuation is absent in the paper~\cite{Brower3}. 
Second, the ${\cal O}(\epsilon)$ terms in the 
BDS formula do not affect any of our conclusions 
(this is explained in some detail in our section 4.2). Next, the discussion around (4.29) is very misleading:
the 'unwanted piece' in (4.29) has been derived, via the correct analytic continuation, 
from BDS, it certainly has the correct infrared properties. Comparison with 
high energy QCD calculations shows agreement with the infrared divergent 
piece, and the disagreement beyond one loop comes in when expanding, in (4.29), the finite 
term of $\ln M_6$. Finally, it was shown in our Appendices
C and D that the finite parts of the factors $C$ and $C'$ appear
just from the analytic continuations of the dilogarithm functions 
$Li_2$, and they depend upon the conformal invariant cross ratios
$\Phi$ and $\Phi '$. So what do the authors criticize?

Recently, the paper~\cite{DelDuca:2008pj} appeared where the authors 
calculated the 
three-loop Regge trajectory and three loop coefficient functions. Further,
the breakdown of the BDS ansatz for the 6-point amplitude in two loops 
was found by direct calculations~\cite{BDKRSVV} in agreement with 
the predictions from the Wilson-loop calculations~\cite{Drumlast}.\\ \\
{\bf Second note to be added:} Recently a new paper on the high energy 
behavior of the BDS formula appeared~\cite{DelDuca:2008jg}.
We do not agree with the main result of the most recent version 3 of 
this work, stating that in the multi-Regge kinematics
the special functions appearing in the BDS ansatz are not important.
To be more precise, the authors argue that the two limits: energy
$s_2 \to \infty$ and $\epsilon \to 0$ do not commute. According to appendix C,
in the region $s,s_2>0$, $s_{345}, s_{456}<0$ the sequence of limits:
$ \lim_{\epsilon \to 0} \lim_{s_2 \to \infty} F $
implies that the special functions do not contribute, whereas the opposite
order $ \lim_{s_2 \to \infty} \lim_{\epsilon \to 0}  F $
leads to our result with the special functions being important.
We disagree with this `non-commutativity', since the first part of the
argument is based on a simple arithmetic mistake. Namely, starting
from eq.(C.16), the multiplication of the factor $(-P^2)^{-\epsilon}$ with
$(1-{\tilde \Phi})^{-\epsilon}$ in Eq. (C.22), in the limit $s_2 \to \infty$,
gives the {\it finite} expression 
$({p_4}_\perp+{p_5}_\perp)^{-2\epsilon}$, in agreement with Eq. (B.11).
In this way the dependence on $s_2$ cancels out, 
both sequences of limits lead to
the same answer (contrary to what is stated after Eq. (C.25)),
and our result has been confirmed: in the BDS formula, the special
functions are important in the multiregge kinematics, and their presence
implies that the multiregge factorization is violated. 

\appendix

\setcounter{equation}{0}

\renewcommand{\theequation}{A.\arabic{equation}}
\section{The $2\rightarrow 2$ amplitude}

Let us write the BDS amplitude for the general case of $n$ legs
(see \cite{BDS}):
\beq
\ln M_n= \sum _{l=1}^\infty a^l\left(f^{(l)}(\epsilon )\,
\left(\hat{I}_n^{(1)}(l\epsilon )+F_n^{(1)}(0)\right)+C^{(l)}+E_n^{(l)}(\epsilon )
\right),
\eeq
where $E_n^{(1)}(\epsilon )$ can be neglected for $\epsilon \rightarrow 0$, 
the values of the constants are 
\begin{eqnarray}
C^{(1)}&=&0, \\ 
C^{(2)}&=&-\zeta _2^2/2, \\ 
f^{(l)}(\epsilon )&=&f_0^{(l)}+\epsilon f_1^{(1)}+\epsilon ^2 f_2^{(l)},\\
f_0^{(l)}&=&\frac{1}{4}\gamma _K^{(l)}, \\
f_1 &=& -a \zeta _3/2+a^2(2\zeta_5+5\zeta_2\zeta_3/3), 
\end{eqnarray}
$\gamma _K$ is the cusp anomalous dimension~\cite{KR}, 
\beq
\hat{I}_n^{(1)}(\epsilon )=-\frac{1}{2\epsilon ^2}
\sum_{i=1}^n\left(\frac{\mu ^2}{-s_{i,i+1}}\right)^\epsilon\,, 
\eeq
and the finite remainders $F_n^{(1)}$ are expressed in terms 
of logarithms and dilogarithms. For the elastic scattering 
amplitude case we have
\beq
\hat{I}_4^{(1)}(\epsilon )=-\frac{2}{\epsilon ^2}+
\frac{1}{\epsilon }\ln \frac{(-s)(-t)}{\mu ^4}   
-\frac{1}{2}\left(\ln ^2\frac{-s}{\mu ^2}
+\ln ^2\frac{-t}{\mu ^2}\right)\,,
\eeq
\beq
F_4^{(1)}=-\frac{1}{2}\ln ^2\frac{-s}{-t}+4\zeta _2\,.
\eeq
Therefore
\beq
\hat{I}_4^{(1)}(\epsilon )+F_4^{(1)}=-\frac{2}{\epsilon ^2}
+\ln (-t)\frac{1}{\epsilon}+
\ln (-s)\left(\frac{1}{\epsilon}-\ln \frac{-t}{\mu ^2}\right)
+4\zeta _2\,.
\eeq 
As a result we obtain for $M_4$ Regge-type behaviour, as already 
discussed in the main part of our paper, with the gluon Regge trajectory 
given by
\beq
\omega (t)=a\,\left(\frac{1}{\epsilon}-\ln \frac{-t}{\mu ^2}\right)+
a^2\left(-\zeta _2\left(\frac{1}{2\epsilon}-\ln \frac{-t}{\mu ^2}\right)
-\frac{\zeta _3}{2}\right)+...\,.
\eeq
Note that this result at two loops is in agreement with the direct
calculations~\cite{trajQCD, trajN4} based on the BFKL 
approach~\cite{BFKL}. Indeed, in Ref.~\cite{trajN4} the following
expression for the gluon Regge trajectory was obtained in the 
$\overline{\rm MS}$-scheme (using the same notations):
\beq
\omega _{\overline{\rm MS}}(t)=
a\,\left(\frac{1}{\epsilon}-\ln \frac{-t}{\mu ^2}\right)+
a^2\left[\left(\frac{1}{6}-\zeta _2\right)
\left(\frac{1}{2\epsilon}-\ln \frac{-t}{\mu ^2}\right)
+\frac{2}{9}-\frac{\zeta _3}{2}\right]\,.
\eeq  
The contribution of the scalar loop to this trajectory is
proportional to the contribution of the fermion loop~\cite{trajN4}
\beq
\omega ^{s}_{\overline{\rm MS}}(t)=\frac{n_s}{4n_q}
\frac{1}{1-\epsilon}
\,\omega ^{q}_{\overline{\rm MS}}(t)=
n_s\,\frac{a^2}{24}
\left[\frac{1}{\epsilon ^2}-\ln \frac{-t}{\mu ^2}-\frac{8}{3}
\left(\frac{1}{\epsilon}+2\ln \frac{-t}{\mu ^2}\right)-
\frac{52}{9}\right]\,,
\eeq
where $n_s$ is the number of scalar fields transforming according
to the adjoint representation of the gauge group.
For the transition from the $\overline{\rm MS}$-scheme to the dimensional 
reduction (DRED) scheme, which respects $N=4$ supersymmetry, one should 
first increase the number of scalar fields
\beq
n_s\rightarrow 6+2\epsilon \,,
\eeq
because, in the pure gluonic contribution, $\Delta n=-2\epsilon$
for the gluon fields was taken into account after performing the dimensional 
regularization $4\rightarrow 4-2\epsilon$.
This gives the additional contribution to $\omega _{\overline{\rm MS}}(t)$
\beq
\Delta \omega _{\overline{\rm MS}}(t)=\frac{a^2}{12}
\left(\frac{1}{\epsilon }-\frac{8}{3}\right)\,.
\eeq
After that the subsequent finite renormalization of the coupling constant
needed for the transition between the $\overline{\rm MS}$ and DRED schemes
\beq
a\rightarrow a-\frac{1}{6}a^2
\eeq
leads to the above result for the trajectory
\beq
\omega _{\overline{\rm MS}}(t)\rightarrow \omega (t)=
a\,\left(\frac{1}{\epsilon}-\ln \frac{-t}{\mu ^2}\right)+
a^2\left[-\zeta _2\left(\frac{1}{2\epsilon}-\ln \frac{-t}{\mu ^2}\right)
-\frac{\zeta _3}{2}\right]\,.
\eeq

Concerning the residues $\Gamma (t)$ of the Regge pole, they 
have been calculated in the one-loop approximation in QCD \cite{NextLead}.
In supersymmetric models the helicity non-conserving contribution 
of each of the colliding gluons is cancelled, 
in accordance with the BDS ansatz.

\section{The $2\rightarrow 3$ amplitude}
\setcounter{equation}{0}

\renewcommand{\theequation}{B.\arabic{equation}}

For the $2 \to 3$ production amplitude we have (see Fig.\ref{prod1})
\begin{eqnarray}
\hat{I}_5^{(1)}(\epsilon ) &=& -\frac{5}{2\epsilon ^2}+
\frac{1}{2\epsilon }
\ln\frac{(-s)(-s_1)(-s_2)(-t_1)(-t_2)}{\mu ^{10}} \nonumber\\
&-& \frac{1}{4}\left(\ln ^2\frac{-s}{\mu ^2}+\ln ^2\frac{-s_1}{\mu ^2}
+\ln ^2\frac{-s_2}{\mu ^2}+\ln ^2\frac{-t_1}{\mu ^2}
+\ln ^2\frac{-t_2}{\mu ^2}\right)\,,\\  
F_5^{(1)} &=& -\frac{1}{4}\ln \frac{-s}{-s_1}\ln \frac{-t_2}{-s_2}
-\frac{1}{4}\ln \frac{-t_2}{-t_1}\ln \frac{-s_2}{-s_1}
-\frac{1}{4}\ln \frac{-s_2}{-s}\ln \frac{-s_1}{-t_1} \nonumber\\
&-&\frac{1}{4}\ln \frac{-s_1}{-t_2}\ln \frac{-t_1}{-s}
-\frac{1}{4}\ln \frac{-t_1}{-s_2}\ln \frac{-s}{-t_2}+
\frac{15}{4}\zeta_2\,.
\end{eqnarray}
Thus the total contribution in multi-Regge kinematics
is
\beqs
I_5^{(1)}(\epsilon )+F_5^{(1)}=-\frac{5}{2\epsilon ^2}
\eeqs
\beqs
+\ln \frac{-s_1}{\mu ^2}
\left(\frac{1}{\epsilon}-
\ln \frac{-t_1}{\mu ^2}\right)
+\ln \frac{-s_2}{\mu ^2}\left(\frac{1}{\epsilon}-
\ln \frac{-t_2}{\mu ^2}\right)+ 
\frac{1}{2 \epsilon} \ln{(-t_1)(-t_2) \over \mu^4}
\eeqs
\beq
-\frac{1}{4}
\ln ^2\frac{-\kappa }{\mu ^2}+
\frac{1}{2}\ln \frac{-\kappa }{\mu ^2}
\left(\ln \frac{(-t_1)(-t_2)}{\mu ^4}-
\frac{1}{\epsilon }\right)
-\frac{1}{4}\ln ^2\frac{-t_1}{-t_2}+
\frac{15}{4}\zeta _2 \, .
\eeq
In this way we obtain the Regge factorization of the production
amplitudes, discussed in the main text. Let us note that, formally, this 
result is exact and the amplitude can be written in this factorized form in 
all five channels obtained by the cyclic transmutation of the invariants
$s,t_1,s_1,s_2,t_2$. 

In the one-loop
approximation in QCD the Reggeon-Reggeon-gluon vertex
contains, apart from the Born structure proportional 
to the vector $C(q_2,q_1)$, also the contribution
proportional to the gauge-invariant vector 
$\frac{p_A}{s_1}-\frac{p_B}{s_2}$~\cite{NextLead}.  
In the supersymmetric theories this contribution is 
cancelled, in agreement with the BDS ansatz.

One can calculate also the production amplitude in
the quasi-elastic kinematics, where 
$s\sim s_1\gg s_2\sim t_1,t_2, k_\perp ^2$.
The amplitude here has the usual Regge factorization.

\section{The $2\rightarrow 4$ amplitude}
\setcounter{equation}{0}
\renewcommand{\theequation}{C.\arabic{equation}}

In the case 
of the $2\rightarrow 4$ transition we have (see Fig.~\ref{prod24}) 
\begin{eqnarray}
\hat{I}_6^{(1)}(\epsilon )&=&-\frac{3}{\epsilon ^2}+
\frac{1}{2\epsilon }\ln 
\frac{(-s)(-s_1)(-s_2)(-s_3)(-t_1)(-t_3)}{\mu ^{12}}\nonumber\\
&-& \frac{1}{4}\left(\ln ^2\frac{-s}{\mu ^2}+\ln ^2\frac{-s_1}{\mu ^2}
+\ln ^2\frac{-s_2}{\mu ^2}+
\ln ^2\frac{-s_3}{\mu ^2}+\ln ^2\frac{-t_1}{\mu ^2}
+\ln ^2\frac{-t_3}{\mu ^2}\right)\,,\\
F_6^{(1)}&=&-\frac{1}{2}\ln \frac{-s}{-s_{012}}\ln \frac{-t_3}{-s_{012}}
-\frac{1}{2}\ln \frac{-t_3}{-t_2}\ln \frac{-s_3}{-t_2}
-\frac{1}{2}\ln \frac{-s_3}{-s_{123}}\ln \frac{-s_2}{-s_{123}}\nonumber\\
&-&\frac{1}{2}\ln \frac{-s_2}{-s_{012}}\ln \frac{-s_1}{-s_{012}}   
-\frac{1}{2}\ln \frac{-s_1}{-t_2}\ln \frac{-t_1}{-t_2}
-\frac{1}{2}\ln \frac{-t_1}{-s_{123}}\ln \frac{-s}{-s_{123}}\nonumber\\
&-&\frac{1}{2}Li_2\left(1-\frac{ss_2}{s_{012}s_{123}}\right)-
\frac{1}{2}Li_2\left(1-\frac{t_3s_1}{t_2s_{012}}\right)-
\frac{1}{2}Li_2\left(1-\frac{t_1s_3}{t_2s_{123}}\right)\nonumber\\
&+&\frac{1}{4}\left(\ln \frac{-t_2}{-s_{012}}\right)^2
+\frac{1}{4}\left(\ln \frac{-t_2}{-s_{123}}\right)^2
+\frac{1}{4}\left(\ln \frac{-s_{123}}{-s_{012}}\right)^2
+\frac{9}{2}\,\zeta _2\,,
\end{eqnarray}
where the dilogarithm function is defined as 
\beq
Li_2(z)=-\int _0^z\frac{dt}{t}\ln (1-t)\,.
\eeq

In multi-Regge kinematics it is natural to introduce the independent variables
\beq
s_1,s_2,s_3\,,\,\, -\kappa _{12}=\frac{(-s_1)(-s_2)}{(-s_{012})}\,,
\,\, -\kappa _{23}=\frac{(-s_2)(-s_3)}{(-s_{123})}\,,\,\,
\Phi =\frac{(-s)(-s_2)}{(-s_{012})(-s_{123})}\,.
\eeq
Note that the variable $\Phi $ is unity in the region where all 
above invariants are negative, but $\Phi =\exp (-2\pi i)$ in
the physical region where $s,s_2>0,\,s_{012},s_{123}< 0$. In 
the multi-Regge kinematics we obtain the following general result:
\begin{eqnarray}
I_6^{(1)}(\epsilon )+F_6^{(1)}&=&-\frac{3}{\epsilon ^2}
-\frac{1}{4}\,\ln^2 \Phi-\frac{1}{2}\,\ln \Phi
\left(\ln \frac{(-t_1)(-t_3)}{(-s_2)\mu^2}-\frac{1}{\epsilon}\right)-
\frac{1}{2}\,Li_2(1-\Phi )\nonumber\\
&&\hspace{-3cm}+\ln \frac{-s_1}{\mu ^2}
\left(\frac{1}{\epsilon}-
\ln \frac{-t_1}{\mu ^2}\right)
+\ln \frac{-s_2}{\mu ^2}\left(\frac{1}{\epsilon}-
\ln \frac{-t_2}{\mu ^2}\right)
+\ln \frac{-s_3}{\mu ^2}
\left(\frac{1}{\epsilon}-
\ln \frac{-t_3}{\mu ^2}\right)\nonumber\\
&&\hspace{-3cm}-\frac{1}{4}
\left(\ln ^2\frac{-\kappa _{12}}{\mu ^2}+
\ln ^2\frac{-\kappa _{23}}{\mu ^2}\right)+
\frac{1}{2}\ln \frac{-\kappa _{12}}{\mu ^2}
\left(\ln \frac{(-t_1)(-t_2)}{\mu ^4}-
\frac{1}{\epsilon }\right)+ \frac{1}{2 \epsilon} \ln{(-t_1)(-t_3) \over \mu^4}
\nonumber\\
&&\hspace{-3cm}+\frac{1}{2}\ln \frac{-\kappa _{23}}{\mu ^2}
\left(\ln \frac{(-t_2)(-t_3)}{\mu ^4}-
\frac{1}{\epsilon }\right)
-\frac{1}{4}\left(\ln ^2\frac{-t_1}{-t_2}+
\ln ^2\frac{-t_3}{-t_2}\right)+\frac{7}{2}\zeta _2\,.
\end{eqnarray}

At first sight the arguments of the dilogarithm functions
in the multi-Regge kinematics are either 0 or 1, and we can 
use the relations
\beq
Li_2(0)=0\,,\,\,Li _2(1)=\zeta _2\,.
\eeq
However, in the physical region $s,s_2>0$, $s_1,s_3,s_{012},s_{123}<0$ 
it is needed to be cautious: we should analytically continue the expression
\beq
f(\Phi )=Li_2(1-\Phi )\,,\,\,\Phi =\frac{ss_2}{s_{012}s_{123}}
\eeq
from the region $\Phi \approx 1 $ to the region $\Phi \approx e^{-2\pi i}$
along a unit circle. In multi-Regge kinematics we have 
\beq
s_2\approx \frac{s_{012}s_{123}}{s}-
\left(\vec{k}_1+\vec{k}_2\right)^2\
\eeq
and
\beq
s_2(1-\Phi )\Phi^{-1}=\left(\vec{k}_1+\vec{k}_2\right)^2.
\eeq
Therefore $1-\Phi>0$, and 
after the analytic continuation we obtain
\beq
f(\Phi )=-\int _0^{1-\Phi}\frac{dt}{t}\ln (1-t)
+2\pi i \int _1^{1-\Phi}\frac{dt}{t}
\approx 2\pi i\ln (1-\Phi )\,
\eeq
with $\ln (1-\Phi)$ being real valued. We obtain the following result in the physical region $s,s_2>0$, 
$s_1,s_3,s_{012},s_{123}<0$:
\begin{eqnarray}
I_6^{(1)}(\epsilon )+F_6^{(1)}&=& -\frac{3}{\epsilon ^2}+\pi i \left(
\ln \frac{(-t_1)(-t_3)}{\left(\vec{k}_1+\vec{k}_2\right)^2\mu ^2} 
-\frac{1}{\epsilon}\right)
+\ln \frac{-s_1}{\mu ^2}\left(\frac{1}{\epsilon}-
\ln \frac{-t_1}{\mu ^2}\right)\nonumber\\
&&\hspace{-2.5cm}+
\ln \frac{-s_2}{\mu ^2}\left(\frac{1}{\epsilon}-
\ln \frac{-t_2}{\mu ^2}\right)
+\ln \frac{-s_3}{\mu ^2}
\left(\frac{1}{\epsilon}-
\ln \frac{-t_3}{\mu ^2}\right)
-\frac{1}{4}
\left(\ln ^2\frac{-\kappa _{12}}{\mu ^2}+
\ln ^2\frac{-\kappa _{23}}{\mu ^2}\right)\nonumber\\
&&\hspace{-2.5cm}+
\frac{1}{2}\ln \frac{-\kappa _{12}}{\mu ^2}
\left(\ln \frac{-t_1}{\mu ^2}+
\ln \frac{-t_2}{\mu ^2}-\frac{1}{\epsilon }\right)
+\frac{1}{2}\ln \frac{-\kappa _{23}}{\mu ^2}
\left(\ln \frac{-t_2}{\mu ^2}+
\ln \frac{-t_3}{\mu ^2}-\frac{1}{\epsilon }\right)\nonumber\\
&&\hspace{-2.5cm}-\frac{1}{4}\left(\ln ^2\frac{-t_1}{-t_2}+
\ln ^2\frac{-t_3}{-t_2}\right)
+ \frac{1}{2 \epsilon} \ln{(-t_1)(-t_3) \over \mu^4}+\frac{7}{2}\zeta _2 \, .
\end{eqnarray}

It is possible to derive, from the BDS amplitude, an expression for 
$M_{2\rightarrow 4}$ in the one-loop approximation for
the quasi-multi-Regge kinematics, where
$s>>s_1,s_3>>s_2\sim t_1,t_2,t_3$. In this case it is convenient to
introduce Sudakov variables for the momenta of the two produced
particles
\beq
k_r=\beta _rp_A+\alpha _rp_B+k_r^\perp \,,\,\,(k_r^\perp )^2=-\vec{k}_r^2,
\eeq
where
\beq
1\gg \beta _1\sim \beta _2 \gg \frac{\mu ^2}{s}\,,\,\,
1\gg \alpha _1\sim \alpha _2 \gg \frac{\mu ^2}{s}\,,\,\,
s\alpha _r\beta _r=\vec{k}_r^2 \sim \vec{q}_1^2 \sim \vec{q}_2^2  
\sim \vec{q}_3^2 \sim \mu ^2\,.
\eeq
We can express various invariants in terms of these variables
\beq
s_2\approx
s(\beta _1+\beta _2)(\alpha _1+\beta _2)-
(\vec{k}_1+\vec{k}_2)^2\,,
\eeq
\beq
s_1\approx \alpha _1s\,,\,\,s_3 \approx \beta _2s\,,\,\,
s_{012}\approx (\alpha _1+\alpha _2)s\,,\,\,s_{123}
\approx (\beta _1+\beta _2)s\,.
\eeq
The expression for the function 
$ f(\epsilon )=I_6^{(1)}(\epsilon )+F_6^{(1)}$ (for $\Phi =1$) 
in the quasi-multi-Regge kinematics can be obtained by 
adding an aditional term
\beq
f(\epsilon )\rightarrow f(\epsilon )+\Delta f\,,
\eeq 
where 
\begin{eqnarray}
\Delta f &=& -\frac{1}{2}\ln \frac{s\kappa _{12}\kappa _{23}}{s_1s_2s_3} 
\ln \frac{st_1t_3}{s_{012}s_{123}\mu ^2}
+\frac{1}{2}\ln \frac{s_{012}\kappa _{12}}{s_1s_2}
\ln \frac{t_3s_1s_2s}{t_2s_{012}^2s_{123}} \nonumber\\
&&+\frac{1}{2}\ln \frac{s_{123}\kappa _{23}}{s_3s_2}
\ln \frac{t_1s_3s_2s}{t_2s_{012}s_{123}^2}-\frac{1}{4}
\ln ^2\frac{s\kappa _{12}\kappa _{23}}{s_1s_2s_3}
-\frac{1}{2}\ln ^2\frac{s_{012}\kappa _{12}}{s_1s_2}
-\frac{1}{2}\ln ^2\frac{s_{123}\kappa _{23}}{s_3s_2}\nonumber\\
&&-\frac{1}{2}\ln \frac{s\kappa _{12}\kappa _{23}}{s_1s_2s_3}
\ln \frac{s_{012}s_{123}\kappa _{12}\kappa _{23}}{s_1s_3s^2_2}
-\frac{1}{2}\ln \frac{s_{012}\kappa _{12}}{s_1s_2}
\ln \frac{s_{123}\kappa _{23}}{s_3s_2}+\zeta _2 \nonumber\\
&&-
\frac{1}{2}Li_2\left(1-\frac{ss_2}{s_{012}s_{123}}\right)-
\frac{1}{2}Li_2\left(1-\frac{t_3s_1}{t_2s_{012}}\right)-
\frac{1}{2}Li_2\left(1-\frac{t_1s_3}{t_2s_{123}}\right)\,.
\end{eqnarray}
Here the signs $-1$ are implied to be in front of all invariants $s_i,t_i$.
Note that the expression for $\Delta f$ in the quasi-multi-Regge kinematics 
does not contain large logarithms, because
the arguments of all logarithms and dilogarithm functions 
are of the order of unity. It is proportional to the logarithm
of the amplitude for the transition of two Reggeized gluons into
two particles with the same helicity. Similar to the case of 
$M_4$ and $M_5$ the expression for $M_{2\rightarrow 4}$ in the 
quasi-multi-Regge kinematics coincides with the exact BDS amplitude.
The transition of two reggeons to particles with 
opposite helicity in the one-loop
approximation can be found in Ref.~\cite{BDK}. These transition
amplitudes are needed for the calculation of the next-to-next-to
leading corrections to the BFKL equation.

\section{The $3\rightarrow 3$ amplitude}
\setcounter{equation}{0}
\renewcommand{\theequation}{D.\arabic{equation}}
Here we consider the BDS amplitude $M_6$ in the 
channel corresponding to the transition $3\rightarrow 3$ with
the following invariants (see Fig. \ref{kin3to3}):
\beq
s=(p_A+k_1+p_B)^2\,,\,\,
s_1=(p_A+k_1)^2\,,\,s_3=(p_{B'}+k_2)^2\,,
\eeq
\beq
s_{13}=(k_1+p_B)^2\,,\,\,s_{02}=(p_{A'}+k_2)^2\,,
t'_2=(p_{A'}+k_2-p_A)^2\,,
\eeq
\beq
t_1=(p_{A'}-p_A)^2\,,\,\,t_3=(p_{B'}-p_{B})^2\,,\,\,
t_2=(p_{A'}-p_A-k_1)^2\,.\,
\eeq
The functions $\hat{I}_6^{(1)}(\epsilon )$
and $F_6^{(1)}$ in this case are given by~\cite{BDS}:
\begin{eqnarray}
\hat{I}_6^{(1)}(\epsilon )&=&-\frac{3}{\epsilon ^2}+
\frac{1}{2\epsilon }\ln
\frac{(-s_1)(-s_{13})(-s_3)(-s_{02})(-t_1)(-t_3)}{\mu ^{12}}\nonumber\\
&&\hspace{-1.5cm}-\frac{1}{4}\left(\ln ^2\frac{-s_1}{\mu ^2}+
\ln ^2\frac{-s_{13}}{\mu ^2}
+\ln ^2\frac{-s_3}{\mu ^2}+
\ln ^2\frac{-s_{02}}{\mu ^2}+\ln ^2\frac{-t_1}{\mu ^2}
+\ln ^2\frac{-t_3}{\mu ^2}\right)\,,\\
F_6^{(1)} &=& -\frac{1}{2}\ln \frac{-s_1}{-s}\ln \frac{-s_{13}}{-s}
-\frac{1}{2}\ln \frac{-s_{13}}{-t'_2}\ln \frac{-t_3}{-t'_2}
-\frac{1}{2}\ln \frac{-t_3}{-t_2}\ln \frac{-s_3}{-t_2}\nonumber\\
&-&\frac{1}{2}\ln \frac{-s_3}{-s}\ln \frac{-s_{02}}{-s}
-\frac{1}{2}\ln \frac{-s_{02}}{-t'_2}\ln \frac{-t_1}{-t'_2}
-\frac{1}{2}\ln \frac{-t_1}{-t_2}\ln \frac{-s_1}{-t_2}\nonumber\\
&-&\frac{1}{2}Li_2\left(1-\frac{s_1s_3}{st_2}\right)-
\frac{1}{2}Li_2\left(1-\frac{s_{13}s_{02}}{t'_2s}\right)-
\frac{1}{2}Li_2\left(1-\frac{t_1t_3}{t'_2t_2}\right)\nonumber\\
&+&\frac{1}{4}\left(\ln \frac{-t'_2}{-s}\right)^2
+\frac{1}{4}\left(\ln \frac{-t'_2}{-t_2}\right)^2
+\frac{1}{4}\left(\ln \frac{-t_2}{-s}\right)^2+
\frac{9}{2}\,\zeta _2\,.
\end{eqnarray}

In multi-Regge kinematics
\beq
-s\gg -s_1,-s_3,-t'_2\gg -t_1,-t_2,-t_3>0
\eeq
it is helpful to use the definitions
\beq
-\kappa _{12}=\frac{(-s_1)(-t'_2)}{-s_{02}}\,,\,\,
-\kappa _{23}=\frac{(-s_3)(-t'_2)}{-s_{13}}\,,
\,\,\Phi ' =\frac{(-s_{13})(-s_{02})}{(-t'_2)(-s)}\,,  
\eeq
which allows us to simplify the above expressions
\begin{eqnarray}
I_6^{(1)}(\epsilon )+F_6^{(1)}&=&-\frac{3}{\epsilon ^2}
-\frac{1}{2}\,\ln^2 \Phi' -\frac{1}{2}\,\ln \Phi'
\,\ln \frac{(-\kappa _{12})(-\kappa _{23})}{(-t'_2)(-t_2)}-
\frac{1}{2}\,Li_2(1-\Phi ' )\nonumber\\
&&\hspace{-3cm}+\ln \frac{-s_1}{\mu ^2}
\left(\frac{1}{\epsilon}-
\ln \frac{-t_1}{\mu ^2}\right)
+\ln \frac{-t'_2}{\mu ^2}\left(\frac{1}{\epsilon}-
\ln \frac{-t_2}{\mu ^2}\right)
+\ln \frac{-s_3}{\mu ^2}
\left(\frac{1}{\epsilon}-
\ln \frac{-t_3}{\mu ^2}\right)\nonumber\\
&&\hspace{-3cm}-\frac{1}{4}
\left(\ln ^2\frac{-\kappa _{12}}{\mu ^2}+
\ln ^2\frac{-\kappa _{23}}{\mu ^2}\right)+
\frac{1}{2}\ln \frac{-\kappa _{12}}{\mu ^2}
\left(\ln \frac{(-t_1)(-t_2)}{\mu ^4}-
\frac{1}{\epsilon }\right)+ \frac{1}{2 \epsilon} \ln{(-t_1)(-t_3) \over \mu^4}
\nonumber\\
&&\hspace{-3cm}+\frac{1}{2}\ln \frac{-\kappa _{23}}{\mu ^2}
\left(\ln \frac{(-t_2)(-t_3)}{\mu ^4}-
\frac{1}{\epsilon }\right)
-\frac{1}{4}\left(\ln ^2\frac{-t_1}{-t_2}+
\ln ^2\frac{-t_3}{-t_2}\right)+\frac{7}{2}\zeta _2\,.   
\end{eqnarray}
In the physical region, where $s,t'_2>0$ and $s_1,s_3,s_{02},s_{13}<0$,
one has $\Phi' =\exp (2\pi i)$, i.e. we have to continue in $\Phi'$ along 
the unit circle. The relation  
\beq
- t'_2(1-\Phi ')\approx (\vec{q}_1+\vec{q}_3-\vec{q}_2 )^2\,
\eeq
implies that, after continuation, $1-\Phi' <0$. Therefore, $Li_2$ becomes 
\beq
f(\Phi')=-\int _0^{1-\Phi'}\frac{dt}{t}\ln (1-t)
- 2\pi i \int _1^{1-\Phi'}\frac{dt}{t}
\approx 2 \pi^2 - 2\pi i\ln |1-\Phi' |\,,
\eeq
which allows to obtain the extra phase factor $C'$ violating the Regge 
factorization in this physical region.

\section{High energy scattering amplitudes in the \\ 
leading logarithmic approximation}
\setcounter{equation}{0}

\renewcommand{\theequation}{E.\arabic{equation}}

In this appendix we briefly summarize results for the high energy  
$2 \to 3$, $2 \to 4$, and $3 \to 3$ scattering amplitudes in 
Yang-Mills theories in the leading logarithmic approximation.  

For the $2 \to 3$ case most of the results have already been 
listed in section 4.2. We only quote, for the physical region 
where all energies are positive, the factorized form:   
$$\frac{A_{2 \to 3}}{\Gamma(t_1) \Gamma(t_2)} =\frac{2s}{t_1t_2}\cdot$$
\beqn (e^{-i \pi}|s_1|)^{\omega_1}\; 
\frac{(e^{-i \pi}\kappa_{12})^{-\omega_2} V_1(t_1,t_2,\kappa_{12}) 
- (e^{-i \pi} \kappa_{12})^{-\omega_1} V_2(t_1,t_2,\kappa_{12})}{\sin \pi(\omega_1 - \omega_2)}
\;(e^{-i \pi}|s_2|)^{\omega_2}.   
\label{23factorized_E}
\eeqn
Here the trajectory functions $\omega_i$ and the production vertices 
$V_1$, $V_2$ have been computed in LLA and NLO, whereas the phases and $\sin$ 
factors are part of the analytic representation, and do not need to be 
expanded in powers of $g^2$. 
However, since in this paper we restrict ourselves to the LLA, 
we can put $(e^{-i \pi} \kappa_{12})^{-\omega_1} \approx 1$. For the real part we find 
the result (\ref{23leadinglog}) which coincides with (4).  

For the $2 \to 4$ amplitude we start from the ansatz (\ref{24analyticgeneral}):
\begin{eqnarray}
\frac{A_{2\to4}}{\Gamma(t_1) \Gamma( t_3)} &=& \frac{2s}{t_1 t_2 t_3}\int \frac{d \omega'_2}{2\pi i} \nonumber\\ 
&&\hspace{-3.6cm} \Big[\left(\frac{|s_1|}{\mu^2}\right)^{\omega_1-\omega'_2}
\left(\frac{|s_{012}|}{\mu^2}\right)^{\omega'_2-\omega_3}
\left(\frac{|s|}{\mu^2}\right)^{\omega_3}
\xi_{12'} \xi_{2'3} \xi_3 \,\,
\frac{W_1(t_1,t_2,t_3, \kappa_{12}, \kappa_{23};\omega'_2)}
{\sin \pi \omega_{12'} \sin \pi \omega_{2'3}}
\nonumber \\  
&&\hspace{-3.6cm}+
\left(\frac{|s_3|}{\mu^2}\right)^{\omega_3-\omega'_2}
\left(\frac{|s_{123}|}{\mu^2}\right)^{\omega'_2-\omega_1}
\left(\frac{|s|}{\mu^2}\right)^{\omega_1}
\xi_{32'} \xi_{2'1} \xi_1\,\, 
\frac{W_2(t_1,t_2,t_3,\kappa_{12},\kappa_{23};\omega'_2)}
{\sin \pi \omega_{32'} \sin \pi \omega_{2'1}}
\nonumber \\  
&&\hspace{-3.6cm}+
\left(\frac{|s_2|}{\mu^2}\right)^{\omega'_2-\omega_1}
\left(\frac{|s_{012}|}{\mu^2}\right)^{\omega_1-\omega_3}
\left(\frac{|s|}{\mu^2}\right)^{\omega_3}
\xi_{2'1} \xi_{13} \xi_3 \,\,
\frac{W_3(t_1,t_2,t_3, \kappa_{12}, \kappa_{23},\kappa_{123};\omega'_2)}
{\sin \pi \omega_{2'1} \sin \pi \omega_{13}}
\nonumber \\  
&&\hspace{-3.6cm}+
\left(\frac{|s_2|}{\mu^2}\right)^{\omega'_2-\omega_3}
\left(\frac{|s_{123}|}{\mu^2}\right)^{\omega_3-\omega_1}
\left(\frac{|s|}{\mu^2}\right)^{\omega_1}
\xi_{2'3} \xi_{31} \xi_1 \,\,
\frac{W_4(t_1,t_2,t_3, \kappa_{12}, \kappa_{23},\kappa_{123};\omega'_2)}
{\sin \pi \omega_{2'3} \sin \pi \omega_{31}}
\nonumber  \\  
&&\hspace{-3.6cm}+
\left(\frac{|s_3|}{\mu^2}\right)^{\omega_3-\omega'_2}
\left(\frac{|s_1|}{\mu^2}\right)^{\omega_1-\omega'_2}
\left(\frac{|s|}{\mu^2}\right)^{\omega_2}
\xi_{32'} \xi_{12'} \xi_{2'}\,\, 
\frac{W_5(t_1,t_2,t_3,\kappa_{12},\kappa_{23};\omega'_2)}
{\sin \pi \omega_{32'} \sin \pi \omega_{12'}}
 \Big].
\label{24analyticgeneral_E}
\end{eqnarray}

The partial wave functions $W_{i=1,2,3,4,5}$ have been listed in section 
4.2. They have been obtained from the five single energy discontinuity 
equations, and use has been made of the BFKL bootstrap equations 
in the color octet channel.  

Inserting these partial waves into the ansatz  (\ref{24analyticgeneral}) or 
(\ref{24analyticgeneral_E}),
we can study the full amplitude in the different kinematic regions.
From now on we will specialize on the planar approximation, 
i.e. in the 
signature factors in eq.(\ref{signature2}) we only retain the phases. 

Beginning with 
the physical region where all energies are positive, we first collect  
the Regge pole terms in all five partial waves $W_i$. Their sum can 
be written in the simple factorizing form:
$$\frac{A_{2 \to 4, pole}}{\Gamma(t_1) \Gamma(t_3)} = \frac{2s}{t_1t_2t_3} \cdot $$
\beqn
(e^{-i\pi}|s_1|)^{\omega_2}
\frac{(e^{-i\pi} \kappa_{12})^{-\omega_2} V_1(t_1,t_2,\kappa_{12}) 
- (e^{-i\pi} \kappa_{12})^{-\omega_1} V_2(t_1,t_2,\kappa_{12})}
{\sin\pi\omega_{12}} 
(e^{-i\pi}|s_2|)^{\omega_2} \cdot \nonumber \\ 
\frac{(e^{-i\pi} \kappa_{23})^{-\omega_3} V_1(t_2,t_3,\kappa_{32}) - 
(e^{-i\pi} \kappa_{23})^{-\omega_2}
V_2(t_2,t_3,\kappa_{32})}{\sin \pi \omega_{23}} (e^{-i\pi}|s_3|)^{\omega_3}. 
\label{24polepart_E}
\eeqn  
In order to arrive at this result, we have  
combined, in (\ref{W3}) and (\ref{W4}), the Regge pole contributions 
of $W_3$ and $W_4$ (together with the signature factors), and we have used 
the identity:
\beq
\frac{\sin \pi \omega_{23}}{\sin \pi \omega_{13}}\,\cdot\, 
\frac{\sin \pi \omega_1}{\sin \pi \omega_2}\, 
+\,
\frac{\sin \pi \omega_{21}}{\sin \pi \omega_{31}}\, \cdot\,
\frac{\sin \pi \omega_3}{\sin \pi \omega_2}\, =\, 1.
\eeq  
The production vertices are the same as in the $2\to 3$ case, (\ref{23factorized_E}).  
As in the $2 \to 3$ case, in the leading order approximation we put,
in (\ref{24polepart_E}),
$(e^{-i \pi} \kappa)^{-\omega} \approx 1$. 
For the real part the terms 
proportional to $\ln (\kappa/\mu^2) - \frac{1}{\epsilon}$ 
in the production vertices $V_1$ and $V_2$ cancel, and we 
are back to the factorizing form in (4). 
As to the addititional Regge cut pieces contained in $W_3$ and $W_4$, 
they cancel completely:
$$\frac{A_{2 \to 4, cut}}{\Gamma(t_1) \Gamma(t_3)} = $$
\beq 
\frac{-2s}{t_1t_2t_3} \cdot  (|s_1|)^{\omega_1}
\int \frac{d \omega_2'}{2\pi i} (e^{-i\pi}|s_2|)^{\omega_2'}
\left( \frac{1}{\sin \pi \omega_{13}} + \frac{1}{\sin \pi\omega_{31}}\right) 
\left(V_{cut} - V_p \right) 
 (|s_3|)^{\omega_3} 
\label{2to4physical_E}
\eeq 
$$=0.$$ 
It is instructive to study the cancellation of the imaginary part of this 
Regge cut piece in more detail: from the representation 
(\ref{24analyticgeneral}) which 
shows the energy phase factors explicitly it is straightforward to compute 
the single discontinuities in $s_2$, $s_{012}$, $s_{123}$, and in $s$.
When summing 
these single discontinuities (i.e. when computing the full imaginary part),
we find complete cancellation of the Regge cut piece. This cancellation of Regge cut 
contributions in the planar amplitude is nothing else but the 
Mandelstam mechanism ~\cite{Mandelstam} of the cancellation of the Amati-Fubini-Stanghellini 
Regge cut ~\cite{AFS} in planar diagrams.

The unphysical region where all energies are negative 
can be obtained from (\ref{24polepart_E}) and (\ref{2to4physical_E})
by simply putting the phase factors equal to unity:
the factorizing form of the Regge pole contributions is preserved, 
and the cut pieces in $W_3$ and $W_4$ cancel.

Most interesting is the physical region where $s, s_2 > 0$ and 
$s_1,s_3, s_{012}, s_{123} <0$. Nonzero phases appear only in $s$ and in $s_2$.
After some algebra we rewrite (\ref{24analyticgeneral_E}) in the following form:  
\beq
\frac{A_{2 \to 4}}{\Gamma(t_1) \Gamma(t_3)}=\frac{2s}{t_1 t_2 t_3} 
(|s_1|)^{\omega_1} (|s_3|)^{\omega_3} \cdot 
\label{A2to4_total_E}
\eeq 
\beqn
 e^{-i \pi \omega_2} (|s_2|)^{\omega_2} \left[
\left( \frac{V_1}{\sin \pi \omega_{12}} 
   + e^{-i \pi \omega_{12}} \frac{V_2}{\sin \pi \omega_{21}} \right)  
\left( e^{-i \pi \omega_{32}} \frac{V_1}{\sin \pi \omega_{23}} 
+  \frac{V_1}{\sin \pi \omega_{32}} \right) \right.
           \nonumber\\ \left. 
\left( e^{-i \pi \omega_{31}} 
         \frac{\sin \pi \omega_1 \sin \pi \omega_{23}}
          {\sin \pi \omega_2 \sin \pi \omega_{13}} 
      +e^{-i \pi \omega_{13}}
          \frac{\sin \pi \omega_3 \sin \pi \omega_{21}}
          {\sin \pi \omega_2 \sin \pi \omega_{31}} 
      - e^{-i \pi (\omega_{12} +\omega_{32}} \right) 
   \frac{V_2 V_1}{\sin \pi \omega_{21} \sin \pi \omega_{23}} \right] 
\nonumber\\ 
-  \frac{2s}{t_1 t_2 t_3} 
(|s_1|)^{\omega_1} (|s_3|)^{\omega_3} 
\int \frac{d \omega'_2}{2 \pi i} (e^{-i \pi} |s_2|)^{\omega'_2} 
 \left( \frac{e^{-i \pi \omega_{31}}}{\sin \pi \omega_{13}}+
       \frac{e^{-i \pi \omega_{13}}}{\sin \pi \omega_{31}}\right)
       \left( V_{cut} - V_p) \right). \nonumber   
\eeqn
It is important to study the infrared singularities of the phases of this 
expression. First we note that the prefactor $e^{-i \pi \omega_2}$ contains,
in lowest order in $a$, the $1/\epsilon$ singularity of the gluon trajectory 
(\ref{expandedtrajectory}). All other phase factors contain differences of trajectory 
functions and are finite as $\epsilon \to 0$. Expanding the square brackets 
in powers of $a$ we arrive at (\ref{A2to4phase}):
$$\frac{A_{2 \to 4}}{\Gamma(t_1) \Gamma(t_3)}=\frac{2s}{t_1 t_2 t_3} 
g^2 C(q_2,q_1) C(q_3,q_2) 
(|s_1|)^{\omega_1} (|s_3|)^{\omega_3} \cdot $$
\beqn 
e^{-i \pi \omega_2} (|s_2|)^{\omega_2} 
\left[1+ i \frac{\pi}{2} \left(
\omega_1+\omega_2 + 
 a \left( \ln \frac{\kappa_{12}}{\mu^2} -\frac{1}{\epsilon}\right) 
+ \omega_3 + \omega_2 + 
 a \left( \ln \frac{\kappa_{23}}{\mu^2} -\frac{1}{\epsilon} \right) 
\right) \right] \nonumber \\ 
- 2 i \pi \frac{2s}{t_1t_2t_3} \int \frac{d \omega'_2}{2 \pi i} 
(e^{-i \pi} |s_2|)^{\omega'_2} V_{cut}.
\label{A2to4phase_E} 
\eeqn
Following the steps described in section 4.2 we factor out the gluon 
trajectory (details are presented in ~\cite{BLS}):
\beq
\int \frac{d \omega'_2}{2 \pi i} (e^{-i \pi} |s_2|)^{\omega'_2} V_{cut}  
= g^2 C(q_2,q_1) C(q_3,q_2) (e^{-i \pi} |s_2|)^{\omega_2} \int \frac{d \omega'_2}{2 \pi i} 
(e^{-i \pi} |s_2|)^{\omega'_2} V_{cut, reduced}.
\label{reduction_E}
\eeq
Insertion into (\ref{A2to4phase_E}) leads:
$$\frac{A_{2 \to 4}}{\Gamma(t_1) \Gamma(t_3)}=\frac{2s}{t_1 t_2 t_3} 
(|s_1|)^{\omega_1} (|s_3|)^{\omega_3} (|s_2|)^{\omega_2} 
 g^2 C(q_2,q_1) C(q_3,q_2) e^{-i \pi \omega_2} \cdot $$
\beqn 
\left[1+ i \frac{\pi}{2} \left(
\omega_1+\omega_2 + 
 a \left( \ln \frac{\kappa_{12}}{\mu^2} -\frac{1}{\epsilon}\right) 
+ \omega_3 + \omega_2 + 
 a \left( \ln \frac{\kappa_{23}}{\mu^2} -\frac{1}{\epsilon} \right) 
\right) \right.  \nonumber \\ \left. 
- 2 i \pi \int \frac{d \omega'_2}{2 \pi i} (e^{-i \pi} |s_2|)^{\omega'_2} 
V_{cut,reduced} \right]
\label{A2to4phaseexp_E} 
\eeqn
The one loop approximation of the Regge cut contribution, (\ref{Vcut0}), 
is infrared singular: 
\beqn
g^2 C(q_2,q_1) C(q_3,q_2) \int \frac{d \omega'_2}{2 \pi i} 
(e^{-i \pi} |s_2|)^{\omega'_2} V_{cut, reduced} \nonumber \\
= g^2 \frac{C(q_2,q_1) C(q_3,q_2)}{2} 
\left[ a \left( \ln \frac{\kappa_{123}\mu^2}{q_1^2 q_2^2} + \frac{1}{\epsilon}\right) 
+ ... \right],
\eeqn
whereas the two loop and higher order terms of $V_{cut, reduced}$ can been shown to be 
infrared finite~\cite{BLS}. As a consequence, in ({\ref{A2to4phaseexp_E}) 
the coefficient in the square brackets is infrared finite, and the singularities 
are collected in the overall phase factor $e^{-i \pi \omega_2}$.
    
For completeness we also list those energy discontinuities which do not vanish in this 
kinematic region, the discontinuity in the total energy $s$ and the discontinuity in $s_2$.
We again start from the 
analytic representation (\ref{24analyticgeneral_E}). 
After some algebra (which includes approximating phase factors by unity) 
we find\footnote{We use the 
definition $disc\, f(s) = \frac{1}{2i} \left( f(s+i\epsilon) - f(s-i\epsilon) \right)$.}: 
\beq
disc_s \frac{A_{2 \to 4, cut}}{\Gamma(t_1) \Gamma(t_3)}
\approx  \frac{- \pi s}{t_1t_2t_3} (|s_1|)^{\omega_1}
\int \frac{d \omega_2'}{2\pi i} (|s_2|)^{\omega_2'}
\,\, V_{cut}\,\,  (|s_3|)^{\omega_3}. 
\label{24discsleadinglog_E}
\eeq
In a similar way we compute the discontinuity in $s_2$:
\beq
disc_s \frac{A_{2 \to 4, cut}}{\Gamma(t_1) \Gamma(t_3)}
\approx  \frac{-\pi s}{t_1t_2t_3} (|s_1|)^{\omega_1}
\int \frac{d \omega'_2}{2\pi i} (|s_2|)^{\omega_2'}
\,\, \tilde{V}_{cut}\,\,  (|s_3|)^{\omega_3},
\eeq
where, instead of (\ref{Vcut}), 
\begin{eqnarray}
\tilde{V}_{\rm cut} &=& \frac{t_2 N_c}{8} g^4 \int \frac{d^2k d^2k'}{(2\pi)^6} 
        \big[C(q_2,q_1) - \frac{q_1^2}{(k+q_1)^2} C(k+q_2,k+q_1)\big] \nonumber\\
&\times&
G^{(8_A)}(k+q_2,-k;k'+q_2,-k';\omega_2') \nonumber \\
&\times&
\big[ C(q_3,q_2) -  C(k'+q_3,k'+q_2) \frac{q_3^2}{(k'+q_3)^2}\big]. 
\label{Vtildecut_E}
\end{eqnarray}
In lowest order we have:
\begin{equation}
\tilde{V}_{\rm cut}^{(0)} =g^2  \frac{C(q_2,q_1) C(q_3,q_2)}{2\omega'_2} 
a \left(\ln \frac{\kappa_{123} \mu^2} {\kappa_{12}\kappa_{23}}
  + \frac{1}{\epsilon}\right).
\label{Vtildecut0_E}
\end{equation}

A completely analogous discussion applies to the case $3 \to 3$ 
(Figs.~\ref{kin3to3},~\ref{Dispers3to3}) in the 
multi-Regge region (for a detailed 
discussion of the `analytic' representation see~\cite{JB2}). 
Our ansatz is:
\begin{eqnarray}
\frac{A_{3\to3}}{\Gamma(t_1) \Gamma( t_3)} &=& \frac{2s}{t_1 t_2 t_3}\int \frac{d \omega'_2}{2\pi i} \nonumber\\ 
&&\hspace{-3.6cm} \Big[\left(\frac{|s_1|}{\mu^2}\right)^{\omega_1-\omega'_2}
\left(\frac{|s_{02}|}{\mu^2}\right)^{\omega'_2-\omega_3}
\left(\frac{|s|}{\mu^2}\right)^{\omega_3}
\xi_{12'} \xi_{2'3} \xi_3 \,\,
\frac{U_1(t_1,t_2,t_3, \kappa_{12}, \kappa_{23};\omega'_2)}
{\sin \pi \omega_{12'} \sin \pi \omega_{2'3}}
\nonumber \\  
&&\hspace{-3.6cm}+
\left(\frac{|s_3|}{\mu^2}\right)^{\omega_3-\omega'_2}
\left(\frac{|s_{2}|}{\mu^2}\right)^{\omega'_2-\omega_1}
\left(\frac{|s|}{\mu^2}\right)^{\omega_1}
\xi_{32'} \xi_{2'1} \xi_1\,\, 
\frac{U_2(t_1,t_2,t_3,\kappa_{12},\kappa_{23};\omega'_2)}
{\sin \pi \omega_{32'} \sin \pi \omega_{2'1}}
\nonumber \\  
&&\hspace{-3.6cm}+
\left(\frac{|s_{02}|}{\mu^2}\right)^{\omega'_2-\omega_3}
\left(\frac{|s_{13}|}{\mu^2}\right)^{\omega'_2-\omega_1}
\left(\frac{|s|}{\mu^2}\right)^{\omega_1 + \omega_3 - \omega'_2}
\xi_{2'1} \xi_{2'3} \xi_{(1+3)2'}\,
\frac{U_3(t_1,t_2,t_3, \kappa_{12}, \kappa_{23},\kappa_{123};\omega'_2)}
{\sin \pi \omega_{2'1} \sin \pi \omega_{2'3} \sin \pi (\omega_1 +\omega_3 - \omega_{2'})}
\nonumber \\  
&&\hspace{-3.6cm}+
\left(\frac{|s_{02}|}{\mu^2}\right)^{\omega(t_1)}
\left(\frac{|s_{13}|}{\mu^2}\right)^{\omega(t_3)}
\left(\frac{|s_2|}{\mu^2}\right)^{\omega'_2-\omega_1 - \omega_3}
\xi_{3} \xi_{1} \xi_{2'(1+3)} \,\,
\frac{U_4(t_1,t_2,t_3, \kappa_{12}, \kappa_{23},\kappa_{123};\omega'_2)}
{\sin \pi (\omega_{2'} - \omega_1 -\omega_3)}
\nonumber  \\  
&&\hspace{-3.6cm}+
\left(\frac{|s_3|}{\mu^2}\right)^{\omega_3-\omega'_2}
\left(\frac{|s_1|}{\mu^2}\right)^{\omega_1-\omega'_2}
\left(\frac{|s|}{\mu^2}\right)^{\omega'_2}
\xi_{32'} \xi_{12'} \xi_{2'}\,\, 
\frac{U_5(t_1,t_2,t_3,\kappa_{12},\kappa_{23};\omega'_2)}
{\sin \pi \omega_{32'} \sin \pi \omega_{12'}}
 \Big],
\label{33analyticgeneral_E}
\end{eqnarray}
where 
\beq
\xi_{i(j+k)} = e^{-i \pi (\omega_i - (\omega_j+\omega_k)) } + 1,\;\; 
\xi_{(i+j)k)} = e^{-i \pi ((\omega_i + \omega_j) - \omega_k) } + 1.
\eeq
The pieces labelled by 1,2,5 are `normal' and contain only Regge poles. They coincide with 
those of the $2 \to 4$ amplitude:
\beq
U_i = W_i\,,\,\,\,i=1,2,5, 
\eeq
and they fit into the factorization pattern, 
The terms 3 and 4 have the extra Regge cut piece shown in Fig.~\ref{BFKLlad} (right figure), 
which is described in terms of the color octet BFKL equation. In analogy with 
(\ref{W3}), (\ref{W4}) one finds:
\beqn
U_3 &=& \frac{\sin \pi \omega_1 \sin \pi \omega_3}
{\sin\pi \omega'_2} V_2(t_1,t_2,\kappa_{12}) \frac{1}{\omega'_2 - \omega_2} 
V_1(t_2,t_3,\kappa_{23})\nonumber\\
& + & \sin \pi \omega_{2'1} \sin \pi \omega_{2'3}
\left(U_{\rm cut}-U_p \right),
\label{U3_E}\\
U_4 &=& \frac{1}{\sin \pi \omega'_2} V_2(t_1,t_2,\kappa_{12}) \frac{1}{\omega'_2 - \omega_2} 
V_1(t_2,t_3,\kappa_{23}) \nonumber\\
& + &  
\left(U_{\rm cut}-U_p \right).
\label{U4_E}
\eeqn
The cut piece has been given in (\ref{Vcut}), (\ref{Vcut0})):
\begin{eqnarray}
U_{\rm cut} &=& \frac{t_2 N_c}{8} g^4 \int \frac{d^2k d^2k'}{(2\pi)^6} 
        \frac{q_1^2}{(k-q_1)^2} C(q_2-k,q_1-k) \nonumber\\
&\times&
G^{(8_A)}(k,q_2-k;k',q_2-k';\omega_2') 
C(k'-k_2,k') \frac{q_3^2}{(k'-k_2)^2}, 
\label{Ucut_E}
\end{eqnarray}
with the lowest order approximation 
\begin{equation}
U_{\rm cut}^{(0)} = \frac{g^2 C(q_2,q_1) C(q_3,q_2)}{2\omega'_2}  
\;a \ln \frac{\kappa_{12}\kappa_{23}}{(\vec{q}_1+\vec{q}_3-\vec{q}_2)^2 q_2^2}.
\label{Ucut0_E}
\end{equation}
It contains Regge cut singularitities, and breaks the factorization.
Note that, in contrast to the $2 \to 4$ case, the one loop approximation  
of the Regge cut term, has no $1/\epsilon$ pole, i.e. it is infrared finite. 

In analogy with the $2 \to 4$ case, these Regge cut pieces 
does not show up in the physical region where all energies are positive.
It is, again, only in the other physical region $s,s_2>0$, 
$s_1,s_3,s_{13},s_{02}<0$ where these pieces become visible. 
Proceeding in the same fashion as before (\ref{A2to4_total_E}) we find for this 
region:
$$\frac{A_{3\to3}}{\Gamma(t_1) \Gamma( t_3)} = \frac{2s}{t_1 t_2 t_3} 
(|s_1|)^{\omega_1} (|s_3|)^{\omega_3} \cdot$$
\beqn
(|s_2|)^{\omega_2}\Big[ e^{-i \pi \omega_2} \left( \frac{V_1}{\sin \pi \omega_{12}} 
   + e^{-i \pi \omega_{12}} \frac{V_2}{\sin \pi \omega_{21}} \right)  
\left( e^{-i \pi \omega_{32}} \frac{V_1}{\sin \pi \omega_{23}} 
+  \frac{V_1}{\sin \pi \omega_{32}} \right) - 
\nonumber\\ 2 i \frac{V_1 V_2 }{\sin \pi \omega_2} \Big]
       + \frac{2 i s}{t_1 t_2 t_3} (|s_1|)^{\omega_1} (|s_3|)^{\omega_3} 
\int \frac{d \omega'_2}{2 \pi i}
(e^{-i \pi} |s_2|)^{\omega'_2} U_{cut}.
\eeqn
The last term can be written as 
\beq
\int \frac{d \omega`_2}{2 \pi i}
(e^{-i \pi} |s_2|)^{\omega'_2} U_{cut} 
= g^2 C(q_2,q_1) C(q_3,q_2) (e^{-i \pi} |s_2|)^{\omega_2} \int \frac{d \omega'_2}{2 \pi i}
(e^{-i \pi} |s_2|)^{\omega'_2} U_{cut,reduced}.
\eeq
It is important to note that the one loop approximation, $U_{cut}^{(0)}$, 
\beqn
 g^2 C(q_2,q_1) C(q_3,q_2) \int \frac{d \omega'_2}{2 \pi i}
(e^{-i \pi} |s_2|)^{\omega'_2} U_{cut,reduced} \nonumber \\ =
g^2 \frac{C(q_2,q_1) C(q_3,q_2)}{2} 
\left[ a \left(\ln \frac{\kappa_{12}\kappa_{23}}{(\vec{q}_1+\vec{q}_3-\vec{q}_2)^2 q_2^2}\right)
+ ...\right],
\eeqn
as well as the higher order terms are infrared finite. We therefore 
write $A_{3 \to 3}$ in the following form: 
$$\frac{A_{3\to3}}{\Gamma(t_1) \Gamma( t_3)} = \frac{2s}{t_1 t_2 t_3} 
(|s_1|)^{\omega_1} (|s_2|)^{\omega_2} (|s_3|)^{\omega_3} g^2 C(q_2,q_1) C(q_3,q_2)\cdot$$
\beqn
\left[ 1+ i \frac{\pi}{2} \left(
\omega_1+ 
 a \left( \ln \frac{\kappa_{12}}{\mu^2} -\frac{1}{\epsilon}\right) 
+ \omega_3 + 
 a \left( \ln \frac{\kappa_{23}}{\mu^2} -\frac{1}{\epsilon} \right) 
\right)\right.  \nonumber \\ \left.
- 2 i \pi \int \frac{d \omega'_2}{2 \pi i} (e^{-i \pi} |s_2|)^{\omega'_2} 
U_{cut,reduced}\right].
\label{A3to3phaseexp_E} 
\eeqn
On the rhs, the square bracket term is infrared finite.
This shows that the infrared structure of $A_{3 \to 3}$ is quite 
different from  $A_{2 \to 4}$.
  
We conclude this section by listing the discontinuities 
in the energies $s$ and $s_2$:
\beq
disc_s \frac{A_{3 \to 3, cut}}{\Gamma(t_1) \Gamma(t_3)}
\approx  \frac{-\pi s}{t_1t_2t_3} (|s_1|)^{\omega_1}
\int \frac{d \omega_2'}{2\pi i} (|s_2|)^{\omega_2'}
\,\, U_{cut}\,\,  (|s_3|)^{\omega_3}. 
\label{33discsleadinglog_E}
\eeq
In a similar way we compute the discontinuity in $s_2$:
\beq
disc_s \frac{A_{3 \to 3, cut}}{\Gamma(t_1) \Gamma(t_3)}
\approx  \frac{-\pi s}{t_1t_2t_3} (|s_1|)^{\omega_1}
\int \frac{d \omega'_2}{2\pi i} (|s_2|)^{\omega_2'}
\,\, \tilde{U}_{cut}\,\,  (|s_3|)^{\omega_3},
\eeq
where $\tilde{U}$ is obtained from $U$ in the same way as $\tilde{V}$ 
was obtained from $V$.

In order to compare these results with the BDS formula, we divide the scattering amplitudes by their 
Born approximation. For example, we obtain $M_{2 \to 3}$ by  
dividing $A_{2\to 3}$ by the Born approximation 
$$\frac{2s}{t_1t_2} g^2 \delta_{\lambda_A \lambda_{A'}} \delta_{\lambda_B \lambda_{B'}}
g C(q_2,q_1)$$ 
and $M_{2\to4}$ by dividing $A_{2\to 4}$ by
$$\frac{2s}{t_1t_2t_3} g^2 \delta_{\lambda_A \lambda_{A'}} 
\delta_{\lambda_B \lambda_{B'}} g^2 C(q_2,q_1)C(q_3,q_2).$$

\end{document}